\documentclass[longbibliography,twocolumn,amsmath,amssymb,superscriptaddress,aps,prapplied,a4paper,floatfix]{revtex4-2}
\pdfoutput=1
\usepackage{graphicx}
\usepackage[paperwidth=212mm,paperheight=297mm,centering,hmargin=1.65cm,vmargin=2.6cm]{geometry}
\usepackage{color} 
\usepackage{dcolumn}% Align table columns on decimal point
\usepackage{bm}% bold math
\usepackage{physics}
\usepackage[version=3]{mhchem}
\usepackage{float}
\usepackage[separate-uncertainty = true,multi-part-units=single]{siunitx}
\DeclareSIUnit{\molar}{M}
\DeclareSIUnit{\counts}{cnt}
\usepackage{amssymb}
\usepackage{gensymb}
\usepackage{physics}
\newcommand{\RomanNumeral}[1]{\uppercase\expandafter{\romannumeral#1\relax}}
\usepackage[capitalise]{cleveref}
\usepackage{jabbrv}
\usepackage{chemformula}
\usepackage{multirow}
\usepackage{dsfont}
\usepackage{comment}
\usepackage{makecell}

\newcommand{\TwoEightSi}{\ensuremath{^{28}\text{Si}}}
\newcommand{\TXo}{TX\ensuremath{_\mathrm{0}}}

\newcommand{\To}{T\ensuremath{_\mathrm{0}}}

\begin{document}

\title{Optical transition parameters of the silicon T centre}

\author{Chloe~Clear}
\email{cclear@photonic.com}
\affiliation{Department of Physics, Simon Fraser University, Burnaby, British Columbia, Canada} 
\affiliation{Photonic Inc., Coquitlam, British Columbia, Canada}

\author{Sara~Hosseini}
\affiliation{Department of Physics, Simon Fraser University, Burnaby, British Columbia, Canada} 
\affiliation{Photonic Inc., Coquitlam, British Columbia, Canada}

\author{Amirhossein~AlizadehKhaledi}
\affiliation{Department of Physics, Simon Fraser University, Burnaby, British Columbia, Canada} 
\affiliation{Photonic Inc., Coquitlam, British Columbia, Canada}

\author{Nicholas~Brunelle}
\affiliation{Department of Physics, Simon Fraser University, Burnaby, British Columbia, Canada} 
\affiliation{Photonic Inc., Coquitlam, British Columbia, Canada}

\author{Austin~Woolverton}
\affiliation{Department of Physics, Simon Fraser University, Burnaby, British Columbia, Canada} 
\affiliation{Photonic Inc., Coquitlam, British Columbia, Canada}

\author{Joshua~Kanaganayagam}
\affiliation{Department of Physics, Simon Fraser University, Burnaby, British Columbia, Canada} 
\affiliation{Photonic Inc., Coquitlam, British Columbia, Canada}

\author{Moein~Kazemi}
\affiliation{Department of Physics, Simon Fraser University, Burnaby, British Columbia, Canada} 
\affiliation{Photonic Inc., Coquitlam, British Columbia, Canada}

\author{Camille~Chartrand}
\affiliation{Department of Physics, Simon Fraser University, Burnaby, British Columbia, Canada} 
\affiliation{Photonic Inc., Coquitlam, British Columbia, Canada}

\author{Mehdi~Keshavarz}
\affiliation{Department of Physics, Simon Fraser University, Burnaby, British Columbia, Canada} 
\affiliation{Photonic Inc., Coquitlam, British Columbia, Canada}

\author{Yihuang~Xiong}
\affiliation{Thayer School of Engineering, Dartmouth College, 14 Engineering Drive, Hanover, New Hampshire 03755, USA}

\author{Louis~Alaerts}
\affiliation{Thayer School of Engineering, Dartmouth College, 14 Engineering Drive, Hanover, New Hampshire 03755, USA}

\author{Öney~O.~Soykal} 
\affiliation{Photonic Inc., Coquitlam, British Columbia, Canada}

\author{Geoffroy~Hautier}
\affiliation{Thayer School of Engineering, Dartmouth College, 14 Engineering Drive, Hanover, New Hampshire 03755, USA} 
\affiliation{Institute of Condensed Matter and Nanosciences (IMCN), Université catholique de Louvain, Chemin des Étoiles 8, B-1348 Louvain-la-Neuve, Belgium}

\author{Valentin~Karassiouk}
\affiliation{Department of Physics, Simon Fraser University, Burnaby, British Columbia, Canada} 

\author{Mike~Thewalt}
\affiliation{Department of Physics, Simon Fraser University, Burnaby, British Columbia, Canada} 

\author{Daniel~Higginbottom}
\affiliation{Department of Physics, Simon Fraser University, Burnaby, British Columbia, Canada} 
\affiliation{Photonic Inc., Coquitlam, British Columbia, Canada}

\author{Stephanie~Simmons}
\affiliation{Department of Physics, Simon Fraser University, Burnaby, British Columbia, Canada} 
\affiliation{Photonic Inc., Coquitlam, British Columbia, Canada}

\date{\today}

\begin{abstract}
The silicon T centre's narrow, telecommunications-band optical emission, long spin coherence, and direct photonic integration have spurred interest in this emitter as a spin-photon interface for distributed quantum computing and networking. However, key parameters of the T centre's spin-selective optical transitions remain undetermined or ambiguous in literature. In this paper we present a Hamiltonian of the T centre TX state and determine key parameters of the optical transition from \To{} to \TXo{} from a combined analysis of published results, density functional theory, and new spectroscopy. We resolve ambiguous values of the internal defect potential in the literature, and we present the first measurements of electrically tuned T centre emission. As a result, we provide a model of the T centre's optical and spin properties under strain, electric, and magnetic fields that can be utilized for realizing quantum technologies.
\end{abstract}

\maketitle

\section{Introduction}
The T centre \cite{Bergeron:2020_PRX,Dhaliah2022} is a well-known carbon-based colour center in silicon that can act as a spin-photon interface (SPI) with appealing properties for scalable quantum technologies based on distributed quantum entanglement \cite{Simmons2023}. T centres emit in the telecommunications (telecom) O-band, enabling low-loss transmission through fibre networks without wavelength conversion, and contain long-lived electron and nuclear spins that can be operated as local memory qubit registers \cite{Bergeron:2020_PRX}. As a host material, silicon permits spin coherence times up to several hours \cite{Saeedi2013}. SPIs with memory qubit registers have been proposed as a platform for quantum information technologies including distributed quantum computing \cite{Benjamin2009prospects,Monroe2014,Simmons2023,Li2024highrate}. Compared to state-of-the-art quantum network demonstrations with diamond colour centres \cite{Pompili2021a,Knaut2023entanglement} and trapped ions \cite{Drmota2023verifiable}, T centres can take full advantage of direct integration with silicon nanophotonics \cite{Higginbottom2022,DeAbreu2023Waveguide-integratedCentres} and co-packaging with existing silicon photonic and microelectronic components including optical switches, modulators and single-photon detectors \cite{Feng2022}.

%motivation for hamiltonian
Optimizing the crucial operations of a SPI entanglement distribution network \cite{Pompili2021a} such as state preparation, spectral tuning \cite{White_2020}, single-shot readout \cite{Anderson2022_singleshot} and spin-preserving photon emission \cite{Barrett2005} require detailed knowledge of the the system Hamiltonian. In atomic systems and even some well-established solid-state SPIs, such as nitrogen vacancy centres, the transition strengths, orientations and branching ratios are well characterized. Although research in silicon colour centres extends back many decades \cite{davies1989}, their potential application in quantum technologies has been realized only recently \cite{Chartrand2018,Bergeron:2020_PRX}. Because of this, there is a general lack of understanding of their electronic fine structure, and resulting optical and spin properties. Recent silicon colour centre studies have focused on the G \cite{Redjem2020,Udvarhelyi2021}, W \cite{Ivanov2022,Baron2022}, and T centres \cite{Bergeron:2020_PRX,Higginbottom2022,Dhaliah2022}. Of these emitters, the T centre is especially interesting for its capability as a SPI.

There are only a handful of studies concerning the T centre's Hamiltonian \cite{Dhaliah2022}, with some ambiguity in literature over key values \cite{Safonov1995,Safonov1999a}. In these works the anisotropic bound exciton state TX is modelled as an isoelectronic acceptor state. The spin components of TX$_0$ and TX$_1$ depend on the defect potential, and they are sensitive to the orientation of external fields and crystal strain. Knowing these eigenstates is essential to, for example, performing optical quantum non-demolition measurements of the T centre spin qubits. Published defect potential and Zeeman parameters determined from early studies disagree \cite{Safonov1995,Safonov1999a}. 

\begin{figure*}
\centering
    \includegraphics[width=1\textwidth]{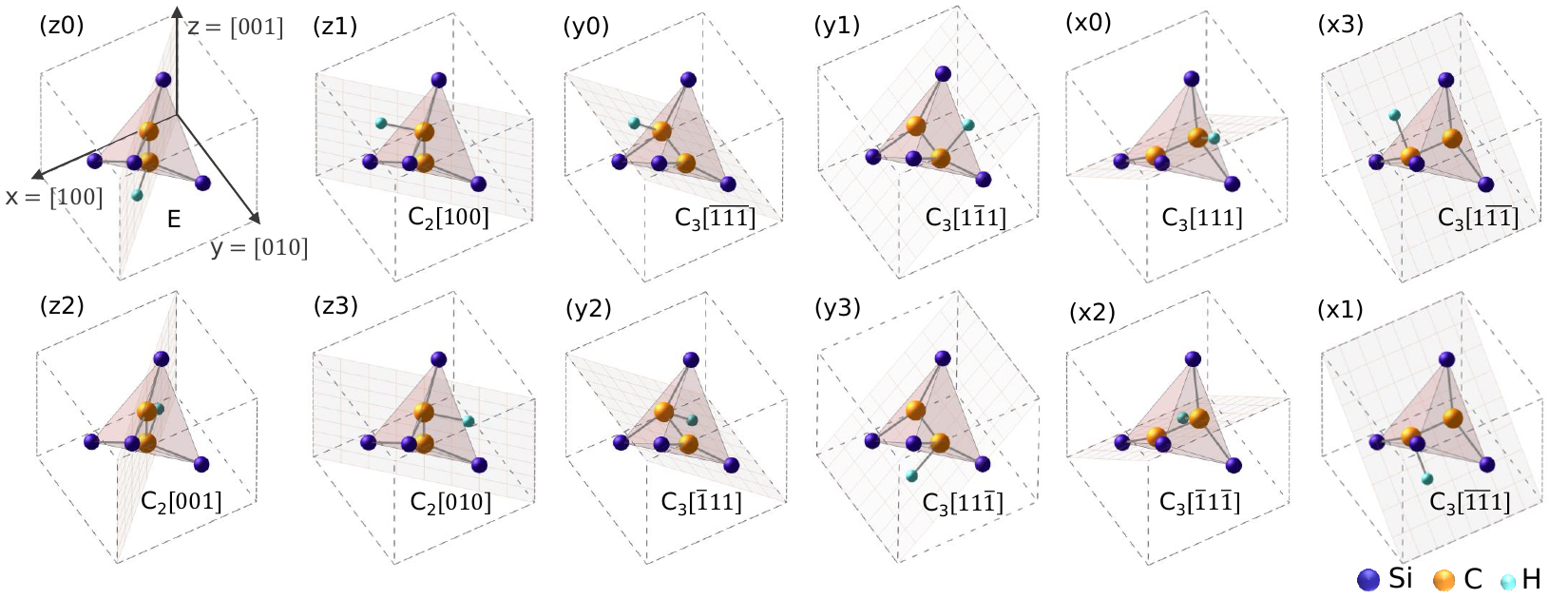}
    \caption{T centre atomic structure within the silicon crystal unit cell. For simplicity, only neighboring silicon atoms are included. Twelve orientations are shown, the identity \textit{E} and eleven orientations that are obtained by pure rotations about the labelled axis by either $C_2=180^\circ$ or $C_3=120^\circ$. Each orientation has a unique label $(ij)$ with $i=\{x,y,z\}$ and $j=\{0,1,2,3\}$, where $i$ is the direction of the carbon-carbon axis and $j$ represents the rotation of the hydrogen atom about the C-C axis. The defect plane of each orientation is shown. Inversion partners (z0' etc.) are not shown.}
    \label{fig:orientations}
\end{figure*}

In this manuscript, we characterize the T centre's optical transitions from the ground state T$_0$ to the lowest energy bound-exciton state TX$_0$, excluding the hyperfine structure. Due to the strong valence band character of TX, we model it as as a shallow acceptor state in a defect potential. We fit the Hamiltonian parameters sequentially, drawing upon new spectroscopy of T centre ensembles, new density functional theory (DFT) calculations, and holistic reanalysis of previously published data. Isotopically enriched \TwoEightSi{} crystal hosts enable high-resolution spectroscopy of the T centre at low fields. We perform photoluminescence excitation (PLE) spectroscopy for varying magnetic field and find excellent agreement with our model. We unambiguously determine the defect potential parameters from this data, resolving longstanding disagreements in literature \cite{Safonov1995, Safonov1999a}. Moreover, we report the first demonstration of electrically-tuned T centre emission. This enables T centres with inhomogeneous local environments to be spectrally matched for remote entanglement \cite{Kambs2018}, or to be tuned into resonance with optical cavities for emission enhancement \cite{Cavity_enhanced_QN,Simmons2023}. We measure the effect of electric fields oriented along the \ce{Si} crystallographic axes $~$[110] and $~$[001]. Under a field of 125 kV/m, we observe up to $1.5$~GHz shift, two orders of magnitude larger than the inhomogeneous linewidth. Our model captures the linear and the quadratic behavior of this shift by a DC Stark effect. From these results, we predict the electric fields required to correct inhomogeneous broadening in an integrated T centre device. This provides a roadmap for tuning T centre SPIs in an on-chip photonic network. 

In the following sections we first introduce the symmetry and Hamiltonian of the T centre, then determine the strain, Zeeman and electric field dependence of the transition energy.

\section{Structure and symmetry of the T centre}

The T centre is a carbon-carbon-hydrogen point defect within the tetrahedral silicon unit cell. The defect itself has monoclinic $(I)$~$C_{1h}$ symmetry. Each distinct orientation in the lattice can be found by applying the symmetry operations of $T_{h}$. This point group $T_{h}$ is the direct product between the group $T$, describing the identity (E) and 11 pure rotations ($3C_2$ and $8C_3$) from the identity that leave the tetrahedron unchanged, and the group $C_i$ which inverts the coordinates of these rotations, see \cref{fig:orientations}. This yields 24 distinct T centre orientations that divide into 12 inversion symmetric pairs \cite{Safonov1995,Safonov1996b}. For further details of the orientation coordinate transformations see Appendix~\ref{app:T centre orientation analysis}. We define here a numeric labelling system which groups the orientations by C-C axis. The three atomic components of the T centre form the defect plane. We choose identity orientation (z0), such that the defect lies in the $(1\overline{1}0)$ crystal plane, where C-C $\parallel [001]$.  

\begin{figure}
    \centering
    \includegraphics[width=1\columnwidth]{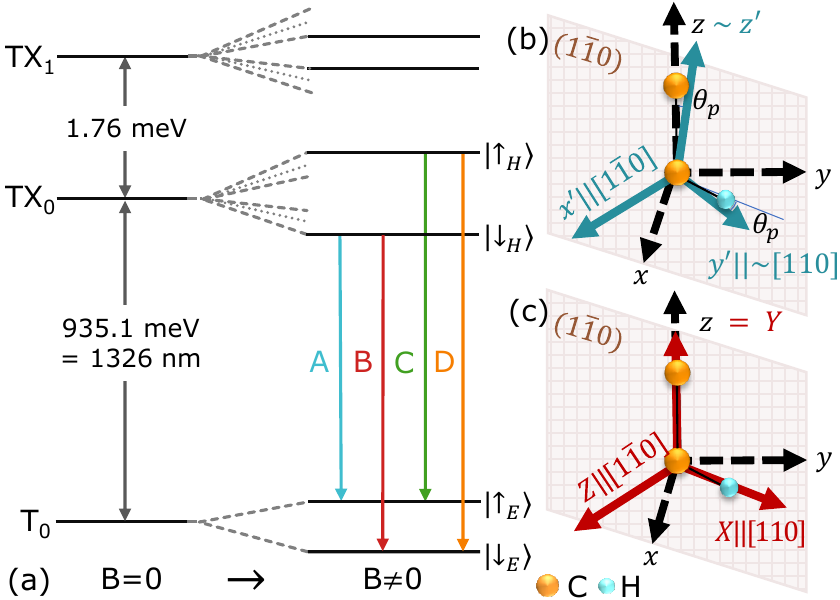}
    \caption{(a) Simplified energy level diagram of the T centre with and without a magnetic field. The effective internal strain splits TX into TX$_0$ and TX$_1$. Under a magnetic field, T$_0$, TX$_0$ and TX$_1$ Zeeman split with an orientation-dependent effective g factor. Spin labels for the Zeeman splitting of the T$_0$ and TX$_0$ are shown which reflect the hole (H) and electron (E).(b) Comparison between Si crystal coordinates (black) and defect potential coordinates (blue) for the identity orientation (z0). (c) Comparison between Si crystal coordinates (black) and identity (z0) defect dipole coordinates (red) for the identity orientation (z0).
    }
    \label{fig:introfig}
\end{figure}

It is known that the ground state T hosts an unpaired electron with spin $S=1/2$, whereas the TX state includes a bound exciton (BE). In the TX state the two bound electrons pair to form a singlet with $S=0$, leaving a four-fold hole state in the valence band with total angular momentum $J=3/2$ \cite{Bergeron:2020_PRX}. This is illustrated in \cref{fig:introfig}(a). The defect potential can be effectively modeled as an internal strain that causes TX to split into two Kramer's doublets: TX$_0$ and TX$_1$ \cite{Safonov1995,brink1962angular,bir1974symmetry} and mixes the $m_j$ = $(\pm 3/2)$, $(\pm 1/2)$ Zeeman components in each eigenstate. For $B \ll (E_{\text{TX}_1} - E_{\text{TX}_0})/\mu_\mathrm{B}$, where $\mu_B$ is the Bohr magneton, the dominant spin components of TX$_0$ (TX$_1$) are $m_j = \pm 1/2$ ($\pm 3/2$), see Appendix \ref{app:TX0 eigenvector element analysis}.

The defect potential coordinate frame diagonalizing the internal strain is shown in \cref{fig:introfig}(b). By symmetry, $\hat{x}' = [1\overline{1}0]$ (perpendicular to the defect plane). %$\hat{y}'$ and $\hat{z}'$ are a small rotation $\theta_p$ from $[110]$ and $[1\overline{1}0]$
$\hat{y}'$ and $\hat{z}'$ lie in the defect plane and are a small rotation $\theta_p$ from $[110]$ and $[001]$, respectively \cite{Safonov1995}. Safonov et. al. \cite{Safonov1995} determined $3\degree < \abs{\theta_p} < 4\degree$. We find $\theta_p=(-7.7\pm0.5)\degree$, see \ref{app:Defect potential and external strain parameters}.

\section{TX Hamiltonian}
To model TX, we use the Hamiltonian of a shallow acceptor state in a defect potential under external fields in a silicon lattice. The Hamiltonian is given by 
\begin{equation}\label{eq_accepH}
    H_\mathrm{a} = H_\mathrm{s}(\epsilon^c)+H_\mathrm{xs}(\epsilon^\mathrm{ext})+H_\mathrm{Z}(\mathbf{B}) +H_\mathrm{E}(\mathbf{E}) \,,
\end{equation}
where $H_\mathrm{s}$ describes the effective internal strain caused by the defect potential for a shallow acceptor \cite{bir1974symmetry, BIR1963_1}, $H_\mathrm{xs}$ an external strain field, $H_\mathrm{Z}$ an external magnetic field \cite{Bhattacharjee_1972}, and $H_\mathrm{E}$ an external electric field \cite{BIR1963_2}.

Following Bir and Pikus \cite{BIR1963_1}, the internal strain Hamiltonian is given by
\begin{equation}\label{eq_strainH}
    H_\mathrm{s}(\mathbf{\epsilon}^c)= b\sum _{i}(J_{i}^{2}-\mathds{1})\epsilon^c_{ii}+{\frac {d}{\sqrt {3}}}\sum _{i\neq j}[J_{i},J_{j}]\epsilon^c_{ij} \,,
\end{equation}
where the first term contains the hydrostatic strain components. These terms are characterised by the the deformation parameters $b$ and $d$. The elements $\epsilon^c_{ij}$ capture the defect potential represented in the Si crystal basis. In order to find the deformation parameters, the defect potential tensor can be found in its principle axes, \cref{fig:introfig}(b), with three parameters $\epsilon^p_{ii}$. The $J_i$ operators are the angular momentum operators for $J=3/2$, %represented in the Luttinger-Kohn basis \cite{luttinger1956},
\begin{comment}

\end{comment}
where $[J_{i},J_{j}]={\frac {1}{2}}(J_{i}J_{j}+J_{j}J_{i})$ and $i,j=\{x,y,z\}$. When an external strain field is present, the term takes the form $H_\mathrm{xs}(\epsilon^\mathrm{ext})=H_\mathrm{s}(\epsilon^\mathrm{ext})+\Delta_\mathrm{s}$, where the first term takes the same form as \cref{eq_strainH} with an external applied strain tensor $\epsilon^\mathrm{ext}$. In addition to this, a piezospectroscopic shift, $\Delta_\mathrm{s}$ is introduced \cite{AAKaplyanskii1964}. Depending on the direction of applied stress, the orientations group into subsets allowing for the identification of monoclinic defect symmetry \cite{Safonov1995}. For a defect with monoclinic symmetry with defect plane in $(1\overline{1}0)$, this shift takes the form
\begin{multline}
    \Delta_\mathrm{s} = A_1 \sigma^\mathrm{ext}_{zz} + A_2 (\sigma^\mathrm{ext}_{xx}+\sigma^\mathrm{ext}_{yy}) - 2A_3 \sigma^\mathrm{ext}_{xy} \\ + 2A_4 (\sigma^\mathrm{ext}_{yz}+\sigma^\mathrm{ext}_{zx}),
    \label{eqn:piezo}
\end{multline}
with piezospectroscopic tensor elements $A_1,A_2,A_3,A_4$. See Appendix \ref{app:Defect potential and external strain parameters} for further details.

The presence of an external homogeneous magnetic field can be introduced up to second order with
\begin{equation} \label{eq_zeeH}
    H_\mathrm{Z}({\mathbf {B}})=\mu _{B}\big(g_{1}\sum _{i}B_{i}J_{i}+g_{2}\sum _{i}B_{i}J_{i}^{3}\big)
    %+\\
    %q_1|\mathbf {B}|^2 +q_2\big(\sum_i B_i J_i\big)^2+q_3\big(\sum_i B_i^2 J_i^2 \big)
    \,
\end{equation}

where $g_1$, $g_2$ are the hole g-factors. 
 Diamagnetic shifts are evident at $B \gtrsim 1$~T \cite{bir1974symmetry,Bhattacharjee_1972,Bergeron:2020_PRX}. At the fields measured in this work the diamagnetic shifts are negligible.

Finally, we model the Stark shift of the electronic transition under the influence of a local electric field $\varepsilon$ up to the second order as \cite{PhysRevLett.97.083002}
\begin{equation}
H_\mathrm{E}(\mathbf{E}) =-\Delta\mu \cdot \varepsilon-\frac{1}{2}\varepsilon\cdot\Delta\alpha\cdot\varepsilon^T \,,
\label{eqn:stark_full}
\end{equation}
where $\Delta\mu$ is the change in the dipole moment and $\Delta\alpha$ is the change in the polarizability tensor of the defect due to the electric field effect.

Similar to the internal strain, we define an associated dipole coordinate system for the identity orientation, see \cref{fig:introfig}(c), where two axes $(X, Y)$ lie in the defect plane and $Z$ is perpendicular. This basis reflects the monoclinic symmetry of the defect which has a permanent dipole moment with a linear ($\pi$) component out of plane and elliptical ($\sigma$) component in the plane~\cite{AAKaplyanskii1964}. Moreover, it has been found in Ivanov et. al that both the ground and TX state host a $p_z$ like orbital that orients out of the defect plane which has $a^{''}$ symmetry~\cite{Ivanov2022}. For a monoclinic defect in this defect dipole basis we can therefore write the linear component as \cite{Kaplyanskii1967NONCUBICCI}
\begin{equation}
\begin{split} 
    \Delta E_L ~ &= -\Delta\mu\cdot \varepsilon \\
    &=~ A_X E^d_X +  ~ A_Y E^d_Y \ ,
\label{eqn:linear_efield}
\end{split}
\end{equation}
where $A_X$ and $A_Y$ are the non-zero components of the dipole moment and $E^d_X, E^d_Y$ are the components of the electric field in the defect dipole coordinate system. 

The quadratic Stark shift depends on the polarizability tensor. Considering that the only symmetry operation leaving the defect unchanged is reflection in the defect plane, the polarizability matrix can be simplified to \cite{Powell2010SymmetryGT}
 \begin{equation}
\Delta\alpha~=
\begin{bmatrix}
    \alpha_{XX} & \alpha_{XY} & 0  \\
    \alpha_{YX} & \alpha_{YY} & 0  \\
    0 & 0 & \alpha_{ZZ}  
\end{bmatrix}
\label{eqn:polamatrix_text}
\end{equation}
We assume  $\alpha_{XY}=\alpha_{YX}$ because the  
polarizability matrix in the case of static field is real and symmetric \cite{bonin1997electric}. Substituting the polarizability matrix \cref{eqn:polamatrix_text} into the second term of \cref{eqn:stark_full}, the quadratic term of the Stark shift in the defect dipole coordinate system is
%\begin{widetext}
\begin{multline}
    \Delta E_{Q}=-\frac{1}{2}[~\alpha_{XX}~{E_{X}^d}^2+2~\alpha_{XY}~E_X^d E_Y^d+ \\
    ~\alpha_{YY}~{E_{Y}^d}^2+~\alpha_{ZZ}~{E_Z^d}^2] \,.
    \label{eqn:quad_efield}
\end{multline}
To convert from the identity orientation's defect dipole to the crystal basis, $E^d_X = (E_x+E_y)/\sqrt{2}$, $E^d_Y = E_z$ and $E^d_Z = (E_x-E_y)/\sqrt{2}$.
\begin{table*}[htb!]
  \centering
 
\begin{tabular}{ p{4.5cm}p{1.5cm}p{2cm}p{1.5cm}p{4cm}p{2cm} } 
\hline
 Parameter & Symbol & Value & $\pm$ s.d. & Prior literature values & Unit \\
\hline 
\multirow{2}{4em}{Deformation} & $b$    & -1.72 &   0.12 & $^\dag$$0.8$, $^\ddag$$-1.8$ & eV \\ 
&  $d$  & -2.39   & 0.14 & $^\dag$ 2.7, $^\ddag$ $-5.0$ & eV \\
\\

\multirow{2}{4em}{\hbox{Defect potential}} & $\epsilon_{y' y'}^d$    & $-4.3\times 10^{-4}$ &    $3\times 10^{-5}$ & $^\dag$ $-6.5\times 10^{-4}$, $^\ddag$ $-3.5\times 10^{-4}$ &  \\ 
& $\epsilon_{z'z'}^d$  & $-6.3\times 10^{-4}$   & $2\times 10^{-5}$  & $^\dag$ $-2.6\times 10^{-4}$, $^\ddag$ $-2.9\times 10^{-4}$ & \\ 
\\
Defect offset angle & $\theta_p$  & $-7.7$  & 0.5 & $^\dag$ 3--4, $^\ddag$ 0.0 & degrees ($^\circ$) \\
\\
\multirow{4}{*}{\hbox{Piezospectroscopic terms}}
& $A_1$  &$-12.2\times 10^{-12}$  &$3\times 10^{-13}$ & $^\dag$ $-16.1\times 10^{-12}$  & eV/Pa \\
& $A_2$  & $16.2 \times 10^{-12}$ &$2\times 10^{-13}$ & $^\dag$ $14.9 \times 10^{-12}$  & eV/Pa \\ 
& $A_3$  & $0.9\times 10^{-12}$  &$2\times 10^{-13}$ & $^\dag$ $-1.55 \times 10^{-12}$ & eV/Pa \\ 
& $A_4$  & $-2.0\times 10^{-12}$   &$2\times 10^{-13}$ & $^\dag$ $1.86 \times 10^{-12}$  & eV/Pa \\ 
\\
\multirow{2}{4em}{\hbox{Hole g-factors}} & $|g_1|$    & 1.226 &    0.005 & $^\dag$ 1.3, $^\ddag$ 1.1 &  \\ 
 & $|g_2| $    &  $0.005$ &   $0.002$ & $^\dag$$-0.1$, $^\ddag$ 0.03 & \\ 
\\
\multirow{2}{4em}{\hbox{Linear electric field coupling}} & $A_X$   & $-3596$ &    $137$ & & Hz.m/V \\  
 & $A_Y$   & $7519$ &    $82$ &   & Hz.m/V \\ 
\\
\multirow{4}{4em}{\hbox{Quadratic electric field coupling}}
& $\alpha_{XX}$    & $0.123$ &    $0.005$ &   & Hz.m$^2$/V$^2$ \\ 
& $\alpha_{XY}$    & $0.000$ &    $0.002$ &  & Hz.m$^2$/V$^2$  \\ 
& $\alpha_{YY}$ & $0.106$ &    $0.003$ &  &  Hz.m$^2$/V$^2$ \\   
& $\alpha_{ZZ}$   & $0.002$ &    $0.003$ &  &  Hz.m$^2$/V$^2$ \\ 
 
\hline
\end{tabular}

\caption{Parameters of the TX Hamiltonian, \cref{eq_accepH,eq_strainH,eqn:piezo,eq_zeeH}, determined in this work. References for prior literature values are expressed with respect to the model Hamiltonian in this paper. Reference code is $^\dag$\cite{Safonov1995} and $^\ddag$\cite{Safonov1999a}.}
\label{tab:params}
\end{table*}

The Hamiltonian parameters are listed in \cref{tab:params}, along with the values that we determine in the following sections. This unified parameter set substantially revises the few values previously available in literature, and includes terms such as the electric field dependence that have never before been characterized.

\section{Deformation and defect parameters}\label{sec:internalstrain}

The parameters that characterize the defect potential in \cref{eq_accepH} are inconsistent in literature \cite{Safonov1995,Safonov1999a}. We freshly assess both the defect potential and external field terms. First, we extract the T centre spectra under applied strain data from Ref.~\cite{Safonov1995} and re-analyze using the Hamiltonian in \cref{eq_accepH} (details in Appendix \ref{app:Defect potential and external strain parameters}). The deformation potential parameters are found to be $b=-1.72\pm 0.12, d=-2.39\pm 0.14$. The defect potential tensor is found in the defect potential principle axes $(\mathbf{\epsilon}^p_{ij})$ adopted from Safonov et al. \cite{Safonov1995}. The non-zero tensor components are found to be $\epsilon_{y' y'}^p=(-4.3\pm 0.3) \times 10^{-4}$ and $\epsilon_{z'z'}^p=(-6.3\pm 0.2) \times 10^{-4}$.

It was not clear in prior work which orientation maps to the derived defect potential. To assign the defect potential to a particular orientation we combine external strain and high-field Zeeman measurements from literature. We consider the effective Landé g-factors of TX$_0$, $g_\mathrm{H}$, measured by Bergeron et. al. \cite{Bergeron:2020_PRX}. In that work the hole splitting of TX$_0$ was measured for $\hat{B} \sim[110]$, which gives maximal values of $g_\mathrm{H}\sim3.5$ for two orientations and minimal values of $g_\mathrm{H} \sim 1.1$ for another two. From our Hamiltonian model around $\hat{B}=[110]$, we find that the orientations with the smallest (largest) hole splitting are orientations z0 and z2 (z1 and z3). 

To fully determine the internal effective strain tensor, the sign of $\theta_p$ needs to be found. We fit the external strain and magnetic field spectra simultaneously and find that the axes tilt clockwise towards the hydrogen atom ($\theta_p < 0$) by $\theta_p = -7.7 \pm 0.5 \degree$. DFT calculations using a defect supercell model find $\theta_p = -8.10(^\circ)$, in close agreement to the fitted model. For details, see Appendix \ref{app:First-principles calculations of strain}.

We determine the remaining parameters in our model by spectroscopy with T centre ensembles in an isotopically enriched \TwoEightSi{} host, as described in the following sections.

\begin{figure*}

\includegraphics[]{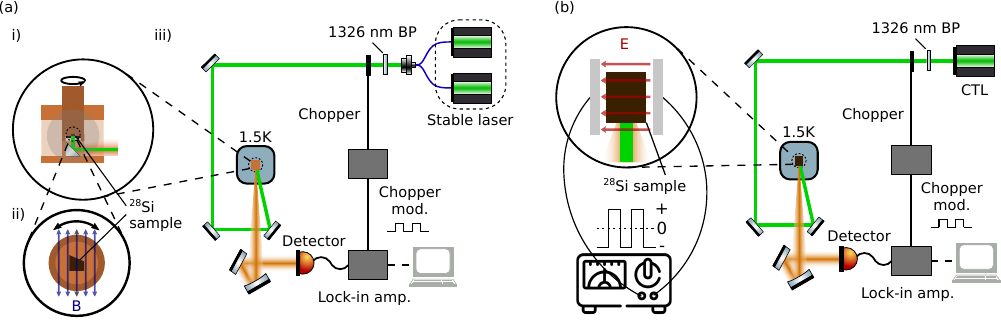}
    \caption{(a) Experimental layout for Zeeman spectroscopy. (i) A \TwoEightSi{} T centre sample is mounted on a cylinder facing downward and slotted between two permanent disk magnets. Lasers are aligned to the sample using a 45$^\circ$ mirror while the sample may be rotated within the magnetic field. (ii) View of the sample from below showing the magnetic field direction relative to the sample orientation and axis of rotation. (iii) Excitation and detection setup. Two narrow-linewidth tunable lasers are combined in fibre. The combined lasers are collimated and then pass through clean-up filters and a chopper, and then aligned to the sample held in a liquid He cryostat pumped to $1.5$~K. Emission from the sample is collimated and sent to a liquid nitrogen cooled Ge detector. Emission is spectrally filtered to select the T centre phonon sideband and reject the excitation laser. A lock-in amplifier removes background signals. (b) Experimental layout for TX$_0$ electric field spectroscopy. As before, except the sample is rigidly mounted and a pulsed electric field is applied by electrodes on opposite faces of the sample. A single tunable laser excites the T centre ensemble resonantly.
    }
    \label{fig:setup}
\end{figure*}

\section{Zeeman parameters}\label{sec:Zeeman_exp}

\subsection{Experiment} 
To determine the linear Zeeman Hamiltonian, we measure the energy of the optical transitions A,B,C,D between TX$_0$ and T$_0$ (see \cref{fig:introfig}(a)) as a function of magnetic field angle using two-laser photoluminescence excitation (PLE) spectroscopy. An experiment schematic is shown in \cref{fig:setup}(a). The sample is a mm-sized crystal obtained from the Avogadro project, enriched to 99.995\% \TwoEightSi{}~\cite{Chai_becker2010enrichment}, and irradiated and annealed to produce a T centre ensemble (as detailed in previous experiments with the same sample \cite{Bergeron:2020_PRX,Higginbottom2023memory}). The sample was rotated within a fixed magnetic field, equivalent to a varying $\vec{B}$ orientation between $[001]$ and $[\overline{1}10]$ in the reference frame of the sample. We choose a moderate field strength of $109.9\pm 0.1$~mT such that the quadratic terms can be safely neglected, isolating $g_1$ and $g_2$.

The sample was immersed in liquid He, pumped to $1.4$~K.Two stable, narrow-linewidth lasers were scanned in unison with a fixed offset equal to the ground state Zeeman splitting $\Delta_\mathrm{e} = g_e B \mu_\mathrm{B}$. A PLE resonance results when the $\Lambda$-system formed by the two lasers addresses one of the Zeeman-split TX$_0$ states. A single-laser PLE scan would hyperpolarize the sample and not drive continuous fluorescence. The two-laser scan finds two resonances per orientation, corresponding to transition pairs A, B and C, D. We plot all transitions A, B, C, D by superimposing the spectra as a function of each laser frequency.

\subsection{Results}
\subsubsection{Landé g-factor}

\begin{figure*}
    \includegraphics[width=1\textwidth]{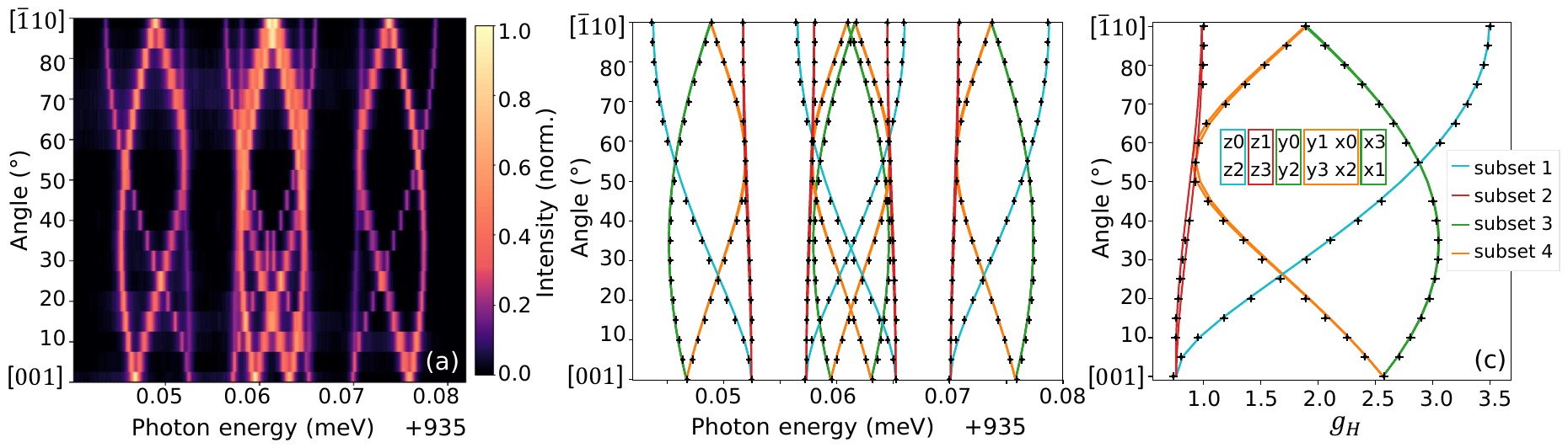}
    \caption{(a) Two-laser PLE spectra TX$_0$ as a function of magnetic field angle between $[001]$ and $[\bar{1}10]$ for $B=109.9$~mT. (b) Peak positions of transitions A,B,C,D from the PLE data (crosses) and the fitted Hamiltonian model. Crosses indicate $60$~MHz uncertainty from the photoluminescence resolution and a fixed angular uncertainty of $\pm1^\circ$. (c) Extracted $g_H$ factors for each subset. The orientation symmetry groups for this applied magnetic field angle are highlighted with the boxed legend.}
    \label{fig:PLEhole}
\end{figure*}

\Cref{fig:PLEhole}(a) shows the PLE spectra as a function of magnetic field angle between $[001]$ and $[\bar{1}10]$. The PLE linewidth ($\approx130$~MHz) is dominated by magnetic field variations across the sample and the zero-field inhomogeneous linewidth ($\approx40$~MHz). For $\vec{B}$ in this plane, we observe four orientation subsets as expected from the defect symmetry. We extract the peak positions and find excellent agreement with the fitted Hamiltonian model as shown in \cref{fig:PLEhole}(b). To fit the Landé $g$-factor parameters in the Zeeman Hamiltonian, the hole spin factor $g_\mathrm{H}=(B_i-D_i)/(\mu_B|\vec{B}|)$ has been found from the optical transitions $B_i$ and $D_i$ for orientations in the subset $i$, \cref{fig:PLEhole}(c). 

From simultaneously fitting the external stress data in \cite{Safonov1995} and the extracted $g_H$ data in \cref{fig:PLEhole}(c), we find the $g$-factor values for the T centre TX state Hamiltonian to be $|g_1|= 1.226 \pm 0.005$ and $|g_2|= 0.005\pm 0.002$. The high symmetry of the data is due to negligible diamagnetic shifts at this field. The color subsets of the optical transition energies represent the different orientation subgroups in point group $T$ (with their respective inversion set) that group energetically when the sample is rotated about a fixed magnetic field from $[001]$ to $[\overline{1}10]$. For subset (1), all the orientations have their defect plane along $(1\overline{1}0)$, the applied field sweeps from being parallel to the defect plane $\hat{B} =[001]$ to normal to the plane $\hat{B}= [\overline{1}10]$ which leads to significant dependence in transition energy. In contrast, subset 2, for which all orientations have their defect plane in $(110)$, experiences minimal variation.

In this analysis we treat $g_e$ as isotropic, and find $g_e = 2.005(3)$ consistent with $g_e = 2.005(8)$ measured in Ref.~\cite{Bergeron:2020_PRX}. In fact, additional pump-probe spectroscopy reveals evidence of $g_e$ anisotropy at the level of one part in a thousand, see Appendix \ref{app:Ground state Landé g-factor}.

\section{Radiative branching ratio}\label{sec:br}

In combination with selection rules imposed by the orbital character of T$_0$ and TX, the spin composition of their Zeeman levels predicted by our model determines the relative optical transition rates of transitions A--D, and therefore the optical branching ratios and optical transition cyclicity. T$_0$ has a local character and the TX state has Si valence band character according to DFT simulations \cite{Dhaliah2022,Ivanov2022}. Here, we approximate the branching ratios and cyclicities in the simple case of a T centre coupled to a single-mode waveguide or cavity that permits only linear dipole emission. The relative transition rates are 
\begin{equation}
\vec{T}_{uv} \propto \abs{\mel{\psi_u^\mathrm{e}}{D}{\psi_v^\mathrm{g}}}^2 \,,
\end{equation}
where $D$ is the linear dipole transition matrix (the identity in the  Luttinger-Kohn basis). To find $T$ for transitions A--D we first reduce TX to only TX$_0$ by selecting the two lowest energy eigenvectors. 

\begin{figure}[b!]
    \includegraphics[width=\linewidth]{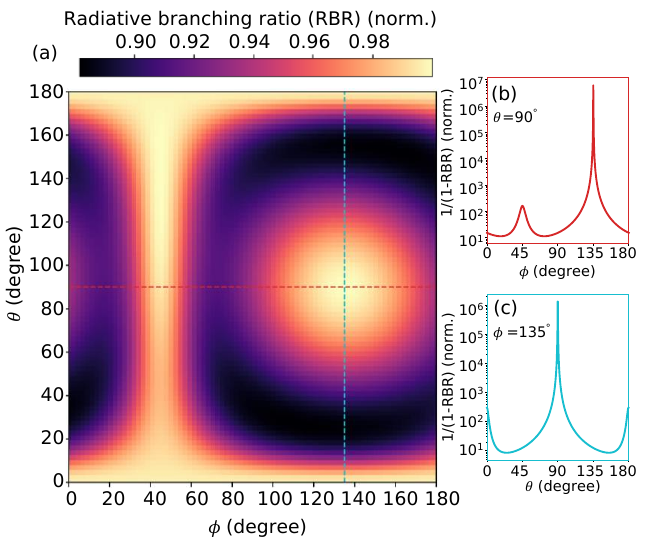}
    \caption{(a) Radiative branching ratio (RBR), for orientation (z0), simulated for different magnetic field orientations with initial population in the lowest TX eigenstate considering decay via optical transition B, the remaining decay probability follows optical transition A. (b) and (c) shows inverse of 1-RBR as a function of  magnetic field orientations cutting (a) horizontally and vertically, respectively.}
    \label{fig:Branching_ratio_identity}
\end{figure}

We analyze the results in terms of the radiative branching ratio (RBR) for $\ket{\downarrow_\mathrm{H}}$, \begin{equation}
    \mathrm{RBR_\downarrow} = T_\mathrm{B}/(T_\mathrm{A} + T_\mathrm{B}) \,.
\end{equation} 
RBR$_{\uparrow, \downarrow}$ are identical at low fields and diverge as $B$ increases. $1-\text{RBR}_\downarrow$ is the inverse of the optical cyclicity of B, under the conditions that that non-radiative and phonon-assisted decay is negligible (achievable by cavity coupling) and that direct relaxation ($T_1$ process) within TX has been frozen out ($k_\mathrm{B} T \ll 1.8$~meV). \Cref{fig:Branching_ratio_identity}(a) shows RBR as a function of magnetic field orientation at $B=250$~mT. \Cref{fig:Branching_ratio_identity}(b) and (c) show the the inverse of (1-RBR) (cyclicity) along slices through the RBR maxima. This indicates that the T centre's optical cyclicity is not limited by the intrinsic branching ratios in the linear dipole emission limit. The optical cyclicity can therefore sufficient for optical quantum non-demolition measurement so long as the non-radiative fraction can be made sufficiently small by Purcell enhancement. 

\section{Electric field parameters}

\subsection{Experiment}

To measure the transition energy under applied electric field, we pulsed voltage across another isotopically purified \TwoEightSi{} T centre sample mounted loosely between copper electrodes as shown in \cref{fig:setup}(b). Loose mounting was required to prevent external strain, but causes some ambiguity in field angle. The pulsed voltage alternated between $\pm V_0$ at $10.5$~kHz to prevent electric shielding by free charges in the sample. Electric fields were applied (approximately) along two crystal axes, up to $450$~V leading to ($E = 125$~kV/m) $\approx [110]$ and ($E = 60$~kV/m) $\approx [001]$. In contrast to an applied $\vec{B}$ field, the ground states remain degenerate and only a single laser is required to perform PLE. Once again, the phonon sideband fluorescence is collected onto a liquid nitrogen cooled detector.

\subsection{Results}

\begin{figure}[t!]
    \includegraphics[width=\linewidth]{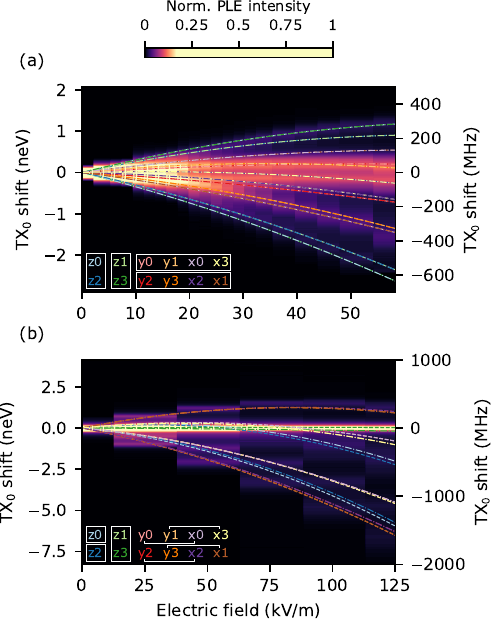}
    \caption{PLE signal versus the laser detuning and electric field while the electric field was applied approximately along (a) $~[001]$ and (b) $~[110]$. Dashed lines show the fitted model where each orientation is coloured according to the orientation keys at bottom left. Symmetry groups under $\hat{E} = [001]$ and $[110]$ are grouped by boxes and lines. Inversion partners are plotted in the same colour, with dot-dash lines. Inversion partners do not belong to the same symmetry groups in general.}
    \label{fig:Efield_fits}
\end{figure}

\Cref{fig:Efield_fits}(a and b) shows the measured PLE intensity as a function of laser detuning from the center of TX$_0$ at zero field, versus the applied electric field. This is the first observation the electric field shift of TX$_0$, and confirmation that T centre photon sources can be electrically tuned. Shifts of up to $1.5$~GHz, an order of magnitude larger than the inhomogeneous linewidth, are observed. In each dataset, an initial linear blue or red shift gives way to a strictly red-shifting quadratic term as the field is increased. 

Under electric fields applied along the symmetry axes, the orientations subdivide once again into orientation subgroups. The insets in \cref{fig:Efield_fits}(a) and (b) indicate the electric field subgroups along $\hat{E} = [001]$ and $[110]$. The linear and quadratic electric field terms separate the inversion partners, in contrast to the subgroups formed under a magnetic field. The linear and quadratic shift terms are included in Appendix \ref{app:Linear and quadratic terms of the stark shift}. 

The model predicts four distinct groups when $\vec{E} \approx [001]$. We observe an extra group which we attribute to the misalignment of the electric field. The model predicts seven distinct linear Stark shifts when the electric field is precisely along [110]. Two defect orientations (z1, z3) and their inversion partners have zero linear Stark shift, as this corresponds to the case of an electric field normal to the defect plane. Several of the groups are unresolved within the inhomogeneous linewidth of this ensemble. 

All data was fitted to \cref{eqn:linear_efield,eqn:quad_efield} to obtain the linear and quadratic electric field parameters. The fit explores plausible orientation assignments for each resolvable line on each data set, with further details provided in Appendix \ref{app:Fitting and angular misalignment}. The energy of each orientation under the applied field is shown as a dashed or dot-dashed line in \cref{fig:Efield_fits}. The dipole moment components estimated from fitting are listed in \cref{tab:params}, as are the nonzero elements of the polarizability tensor. The fit includes angular misalignment as a free parameter, as detailed in Appendix \ref{app:Fitting and angular misalignment}, to account for deviations between the applied electric field and the sample's crystallographic axes. We introduced two degrees of freedom to capture this misalignment, using spherical polar coordinates with $\theta$ the polar angle from the crystal $z$ axis and $\phi$ the azimuthal angle. The largest fitted misalignment found is for the $\hat{E}\sim[001]$ result, which was $\Delta \theta=(-13 \pm 1.0)^\circ$ and $\Delta \phi=(0.0\pm 4)^\circ$. This magnitude of misalignment is reasonable given the manual mounting process of the sample relative to the plates. Accounting for this angular misalignment, the fit captures the transition from linear to quadratic regimes within the precision of the PLE data. To determine the dipole moment to a unique orientation we performed density functional theory calculations on two molecular analogues of the T center (see Appendix \ref{app:First-principles calculation of Stark shift}). This simple model helps us assigning the correct sign for the linear  electric field coupling parameters while avoiding a full description of the exciton. Taking into account the effect of the exciton would require further calculations and is beyond the scope of this work. To confirm this assignment of dipole moment experimentally, orientation-specific Stark data (e.g. a simultaneous measurement of $g_\mathrm{H}$ and $\Delta E$) would be required.

%CONCLUSION--------------
\section{Discussion}

In this work we have introduced a Hamiltonian for the silicon T centre that captures the frequency dependence of TX$_0$ to applied strains, and electric and magnetic fields. The parameters of this Hamiltonian have been determined by reanalyzing previously published data, DFT simulations of the T centre, and new spectroscopy experiments. The values in \cref{tab:params} are the most complete summary of the T centre's optical transition properties available so far. This work substantially revises the value of some parameters determined by earlier work, and presents the first study of the T centre under applied electric fields. This Hamiltonian is instrumental to the development of T centre devices for distributed quantum computation and quantum communication \cite{Simmons2023} in three critical ways. 

First, the deformation, defect potential, and hole g-factors determine the spin composition of the TX state, and therefore the optical dipole orientation of the transitions from TX to T$_0$. These are also required to find the optical branching ratios, as illustrated in Sec.~\RomanNumeral{6} for a simple case of selection rules imposed by the polarization of a coupled photonic mode. The branching ratio is required to understand and optimize optical cyclicity for achieving single-shot, quantum non-demolition (QND) optical read-out of the T centre's electron spin. Due to the high non-radiative decay fraction (up to $76.6$\% \cite{Johnston2023cavitycoupled}) and phonon sideband fraction ($77$\% \cite{Bergeron:2020_PRX}) these potentially spin-scrambling pathways dominate the spin relaxation of T centres in bulk silicon under optical excitation. However, T centres can be enhanced in optical cavities such that the zero-phonon radiative pathways dominate \cite{Johnston2023cavitycoupled,Islam2023cavityenhanced}, and then the optical branching ratio determines the cyclicity. In an optical cavity selecting for linear dipole transitions and $\vec{B}$ along the defect potential axis $z'$, the B and C transitions become highly cyclic as required for QND measurement \cite{Raha2020opticalquantum}.

Second, both the Zeeman and electric field parameters determine how T centres may be tuned into spectral resonance with optical cavities and with each other. Remote entanglement schemes require two or more emitters producing photons that are nearly identical in every degree of freedom. Unlike atoms trapped in vacuum, individual solid-state emitters are generally not spectrally identical due to their inhomogeneous local environment. Even the very pure, isotopically enriched T centre samples studied here show a small degree of inhomogeneous spectral broadening across the sample. Consequently, remote entanglement schemes benefit from spectral tunability. We observed shifts of up to $1.5$~GHz under electric fields as low as $125$~kV/m. According to the fitted model, fields of $1$--$1.2$~MV/m applied along $[001]$ would be required to tune across the full spectral range of photonically integrated T centres as measured by Higginbottom et al. \cite{Higginbottom2022}. These local fields are routine in silicon microelectronics. Integrated electrodes for applying local fields of this magnitude have been demonstrated with other single emitters \cite{strak_integrated}, and PIN junctions can achieve much larger fields \cite{pin_anderson2019electrical, QD_pn}.

Finally, spectral wandering decreases the remote entanglement rates and fidelities feasible with solid-state SPIs. One can expect wandering caused by charge traps in proportion to the electric field sensitivity $\dv{f}{E}$. Integrated T centres already evidence spectral diffusion on the order of $1$~GHz, $100\times$ larger than the homogeneous linewidth \cite{Higginbottom2022,DeAbreu2023Waveguide-integratedCentres,Johnston2023cavitycoupled}. Our electric field measurements indicate that spectral diffusion should be no worse compared to $\vec{E}=0$ within the linear electric field regime $\Delta f < 0.5$~GHz, and that electric field `clock' transitions $\dv{f}{E} = 0$ exist for small applied fields, potentially reducing sensitivity to field in one axis. 

\section{Conclusion}

Silicon T centres combine uncommon advantages as a platform for distributed quantum technologies, not least of which is the advanced silicon semiconductor device industry. The design and operation of high-fidelity T centre devices requires precisely characterizing the T centre itself. In this work we have presented the most complete optical Hamiltonian of the T centre to date, observed electric field tuning, and indicated how a number of open questions surrounding T centre quantum technologies including QND single-shot readout and inhomogeneous broadening may be addressed in practice.

\begin{acknowledgements}

This work was supported by the Natural Sciences and Engineering Research Council of Canada (NSERC), the New Frontiers in Research Fund (NFRF), the Canada Research Chairs program (CRC), the Canada Foundation for Innovation (CFI), the B.C. Knowledge Development Fund (BCKDF), and the Canadian Institute for Advanced Research (CIFAR) Quantum Information Science program. 

The 28Si samples used in this study were prepared from
the Avo28 crystal produced by the International Avogadro Coordination (IAC) Project (2004–2011) in cooperation among the BIPM, the INRIM (Italy), the IRMM (EU), the NMIA (Australia), the NMIJ (Japan), the
NPL (UK), and the PTB (Germany).

Y.X. and G.H. acknowledge financial support by the U.S. Department of Energy, Office of Science, Basic Energy Sciences in Quantum Information Science under Award Number DE-SC0022289. This research used resources of the National Energy Research Scientific Computing Center, a DOE Office of Science User Facility supported by the Office of Science of the U.S. Department of Energy under Contract No. DE-AC02-05CH11231 using NERSC award BES-ERCAP0020966.

The authors thank Carlos Herranz and Matteo Pirro for feedback on this manuscript.
\end{acknowledgements}

%this is suppose to be rev-tex 4.2
\appendix

\section{T centre orientation analysis}
\label{app:T centre orientation analysis}

To find map to all 24 orientations of the T centre, $T_h$ symmetry operations need to be applied to the system coordinates, see \cref{table_ThOp}. The point group $T_h$ is made from the proper rotations from group $T$ and a single reflection which maps $T\rightarrow T_h$ about the mirror plane, corresponding to an inversion to each of the $T$ transformations. The $T_\mathrm{inv}$ subgroup gives the same physical properties as $T$ when applying an external stress or magnetic field for an uncoupled T centre, but not for an external electric field. Twelve orientations found from the proper rotations of the identity are shown in Fig.~1 in the main text (omitting the inversion partners). 

\begin{table}[h!]
\renewcommand{\arraystretch}{1}
\small\begin{tabular}{p{1.5cm}p{2.5cm}p{2.5cm}p{2.5cm} } 

\hline
 \multicolumn{4}{|c|}{$T_h$ coordinate transformations} \\
\hline

\hline
 \multicolumn{3}{|c|}{$T$ point group} & \multicolumn{1}{c|}{Inversion} \\
\hline

\hline
 \multicolumn{1}{|c|}{Rotation} & \multicolumn{1}{c|}{Vector of rotation} & \multicolumn{2}{c|}{Coordinate transformation} \\
\hline
\multicolumn{1}{|c|}{E} & \multicolumn{1}{c|}{-}    & \multicolumn{1}{c|}{[x,y,z]} & \multicolumn{1}{c|}{[-x,-y,-z]}\\ 

\multicolumn{1}{|c|}{$C_2[180^\circ]$} & \multicolumn{1}{c|}{[1,0,0]}    & \multicolumn{1}{c|}{[x,-y,-z]} & \multicolumn{1}{c|}{[-x,y,z]}\\

\multicolumn{1}{|c|}{$C_2[180^\circ]$} & \multicolumn{1}{c|}{[0,0,1]}    & \multicolumn{1}{c|}{[-x,-y,z]} & \multicolumn{1}{c|}{[x,y,-z]}\\

\multicolumn{1}{|c|}{$C_2[180^\circ]$} & \multicolumn{1}{c|}{[0,1,0]}    & \multicolumn{1}{c|}{[-x,y,-z]} & \multicolumn{1}{c|}{[x,-y,z]}\\

\multicolumn{1}{|c|}{$C_3[120^\circ]$} & \multicolumn{1}{c|}{[-1,-1,-1]}    & \multicolumn{1}{c|}{[y,z,x]} & \multicolumn{1}{c|}{[-y,-z,-x]}\\

\multicolumn{1}{|c|}{$C_3[120^\circ]$} & \multicolumn{1}{c|}{[1,-1,1]}    & \multicolumn{1}{c|}{[-y,-z,x]} & \multicolumn{1}{c|}{[y,z,-x]}\\

\multicolumn{1}{|c|}{$C_3[120^\circ]$} & \multicolumn{1}{c|}{[-1,1,1]}    & \multicolumn{1}{c|}{[-y,z,-x]} & \multicolumn{1}{c|}{[y,-z,x]}\\

\multicolumn{1}{|c|}{$C_3[120^\circ]$} & \multicolumn{1}{c|}{[1,1,-1]}    & \multicolumn{1}{c|}{[y,-z,-x]} & \multicolumn{1}{c|}{[-y,z,x]}\\

\multicolumn{1}{|c|}{$C_3[120^\circ]$} & \multicolumn{1}{c|}{[1,1,1]}    & \multicolumn{1}{c|}{[z,x,y]} & \multicolumn{1}{c|}{[-z,-x,-y]}\\

\multicolumn{1}{|c|}{$C_3[120^\circ]$} & \multicolumn{1}{c|}{[1,-1,-1]}    & \multicolumn{1}{c|}{[-z,-x,y]} & \multicolumn{1}{c|}{[z,x,-y]}\\

\multicolumn{1}{|c|}{$C_3[120^\circ]$} & \multicolumn{1}{c|}{[-1,1,-1]}    & \multicolumn{1}{c|}{[z,-x,-y]} & \multicolumn{1}{c|}{[-z,x,y]}\\

\multicolumn{1}{|c|}{$C_3[120^\circ]$} & \multicolumn{1}{c|}{[-1,-1,1]}    & \multicolumn{1}{c|}{[-z,x,-y]} & \multicolumn{1}{c|}{[z,-x,y]}\\
 \hline

\end{tabular}
\caption{T centre orientation transformations}
\label{table_ThOp}
\end{table}
\begin{table}[h!]
\renewcommand{\arraystretch}{1}
\small\begin{tabular}{p{1.5cm}p{2.5cm}p{2.5cm}p{2.5cm} } 

\hline
 \multicolumn{1}{|c|}{Rotation} & \multicolumn{1}{c|}{Vector of rotation} & \multicolumn{1}{c|}{Coordinate transformation} \\
\hline

\multicolumn{1}{|c|}{E} & \multicolumn{1}{c|}{-}    & \multicolumn{1}{c|}{[x,y,z]}\\ 

\multicolumn{1}{|c|}{$C_2[-180^\circ$]} & \multicolumn{1}{c|}{[1,0,0]}    & \multicolumn{1}{c|}{[x,-y,-z]} \\

\multicolumn{1}{|c|}{$C_2[-180^\circ]$} & \multicolumn{1}{c|}{[0,0,1]}    & \multicolumn{1}{c|}{[-x,-y,z]} \\

\multicolumn{1}{|c|}{$C_2[-180^\circ]$} & \multicolumn{1}{c|}{[0,1,0]}    & \multicolumn{1}{c|}{[-x,y,-z]} \\

\multicolumn{1}{|c|}{$C_3[-120^\circ]$} & \multicolumn{1}{c|}{[-1,-1,-1]}    & \multicolumn{1}{c|}{[z,x,y]} \\

\multicolumn{1}{|c|}{$C_3[-120^\circ]$} & \multicolumn{1}{c|}{[1,-1,1]}    & \multicolumn{1}{c|}{[z,-x,-y]} \\

\multicolumn{1}{|c|}{$C_3[-120^\circ]$} & \multicolumn{1}{c|}{[-1,1,1]}    & \multicolumn{1}{c|}{[-z,-x,y]} \\

\multicolumn{1}{|c|}{$C_3[-120^\circ]$} & \multicolumn{1}{c|}{[1,1,-1]}    & \multicolumn{1}{c|}{[-z,x,-y]} \\

\multicolumn{1}{|c|}{$C_3[-120^\circ]$} & \multicolumn{1}{c|}{[1,1,1]}    & \multicolumn{1}{c|}{[y,z,x]} \\

\multicolumn{1}{|c|}{$C_3[-120^\circ]$} & \multicolumn{1}{c|}{[1,-1,-1]}    & \multicolumn{1}{c|}{[-y,z,-x]} \\

\multicolumn{1}{|c|}{$C_3[-120^\circ]$} & \multicolumn{1}{c|}{[-1,1,-1]}    & \multicolumn{1}{c|}{[-y,-z,x]} \\

\multicolumn{1}{|c|}{$C_3[-120^\circ]$} & \multicolumn{1}{c|}{[-1,-1,1]}    & \multicolumn{1}{c|}{[y,-z,-x]} \\
 \hline

\end{tabular}
  \caption{T centre applied field transformations, equivalent to applying $C_2$ and $C_3$ rotations in Tab.~\ref{table_ThOp} in the opposite direction.}
  \label{table_ThOp_field}
\end{table}

\section{Hamiltonian fitting}
\label{app:Hamiltonian fittings}

\subsection{Defect potential and external strain parameters}
\label{app:Defect potential and external strain parameters}
The T centre defect potential parameters have been discussed previously in works by Safonov et al. \cite{Safonov1995,Safonov1996b}. The deformation potential parameters from Ref.~\cite{Safonov1995} (in which strain data was taken) %[Safanov, et al. 1995] 
are quoted as $b=-0.8$~eV, and $d=-2.7$~eV  and in Ref.~\cite{Safonov1996b} (where only magnetic field data was taken) they are $b=-1.8$~eV, and $d=-5$~eV. Given the unexplained difference, we repeated fits to the strain spectra extracted from Ref.~\cite{Safonov1995}. 

\begin{figure*}[t]
\centering
  \includegraphics[width=\linewidth]{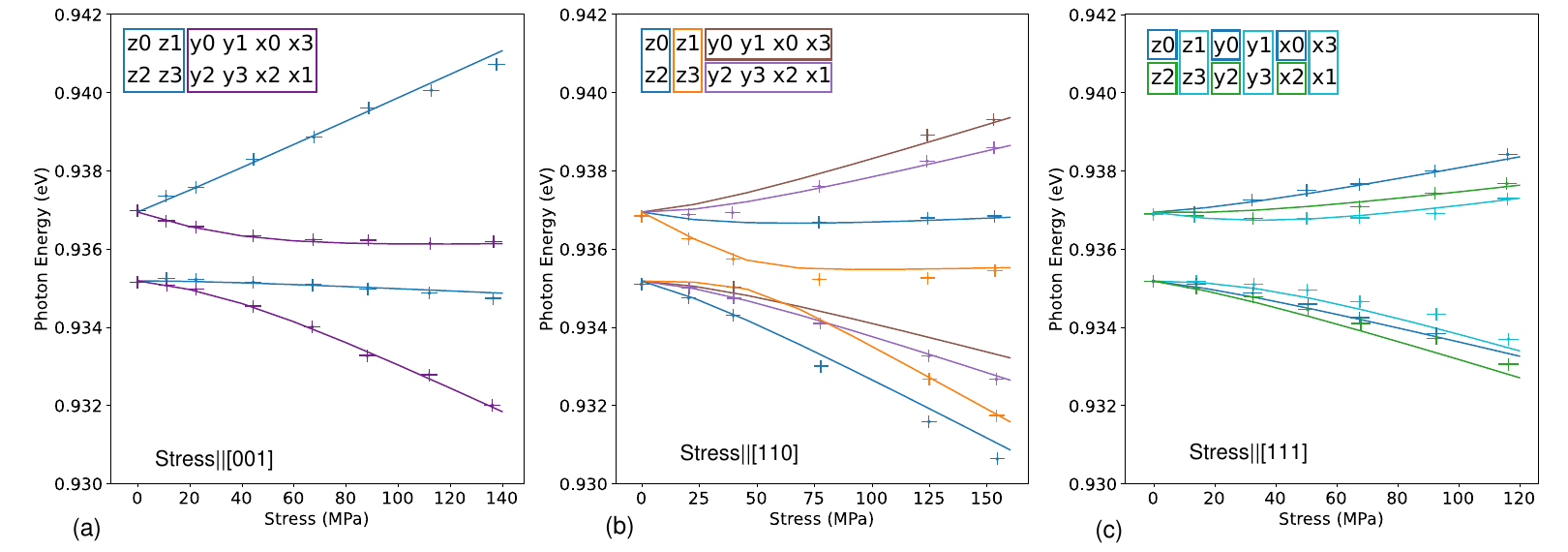}
\caption{TX$_0$ and TX$_1$ spectra under applied stress, experiment data (crosses) extracted from Ref.~\cite{Safonov1995}. Solid lines are the fitted model (Eqn.~1). The colored lines represent the different orientation subgroups, with labels shown on the figure.}
\label{fig_strainfits}
\end{figure*}

To fit this data, the Hamiltonian model needs to be modified to account for an external applied strain. The external strain tensor can be expressed as the symmetric matrix with the elements $\epsilon^\mathrm{ext}_{ij}= s_{ijkl} \sigma_{kl}$, where $s_{ijkl}$ is the compliance tensor of Silicon with the form
\begin{equation}
    \begin{pmatrix}
    \epsilon^\mathrm{ext}_{xx} \\
    \epsilon^\mathrm{ext}_{yy}\\
    \epsilon^\mathrm{ext}_{zz} \\
    \epsilon^\mathrm{ext}_{yz}\\
    \epsilon^\mathrm{ext}_{zx} \\
    \epsilon^\mathrm{ext}_{xy} \\
    \end{pmatrix} = 
    \begin{pmatrix}
    s_{11} & s_{12} & s_{12} & 0 & 0 & 0 \\
    s_{12} & s_{11} & s_{12} & 0 & 0 & 0\\
    s_{12} & s_{12} & s_{11} & 0 & 0 & 0 \\
    0 & 0 & 0 & s_{44} & 0 & 0 \\
    0 & 0 & 0 & 0 & s_{44} & 0 \\
    0 & 0 & 0 & 0 & 0 & s_{44} \\
    \end{pmatrix} 
    \begin{pmatrix}
    \sigma^\mathrm{ext}_{xx} \\
    \sigma^\mathrm{ext}_{yy}\\
    \sigma^\mathrm{ext}_{zz} \\
    \sigma^\mathrm{ext}_{yz}\\
    \sigma^\mathrm{ext}_{zx} \\
    \sigma^\mathrm{ext}_{xy} \\
    \end{pmatrix} \,.
\end{equation}
At 4.2~K the compliance parameters are $s_{11}=7.61736\times 10^{-12}$, $s_{12}=-2.12733\times 10^{-12}$ and $s_{44}=12.4626\times 10^{-12}$ \cite{Hall_1967}. The tensor $\sigma^\mathrm{ext}$ is the applied stress tensor.

To account for the applied strain, we include an additional term in the acceptor Hamiltonian
\begin{equation}\label{eq_straingen}
\begin{split} 
    H_\mathrm{xs}(\epsilon^\mathrm{ext}) = -a\sum _{i}\epsilon^\mathrm{ext} _{ii} \mathds{1} + b\sum _{i}(J_{i}^{2}-\textbf{J}^2/3)\epsilon^\mathrm{ext} _{ii} \\ +{\frac {d}{\sqrt {3}}}\sum _{i\neq j}[J_{i},J_{j}]\epsilon^\mathrm{ext}_{ij} \,.
\end{split}
\end{equation}
Setting $a=-b/4$ similar to the main text,
\begin{equation}\label{eq_straingen_2}
    H_\mathrm{xs}(\epsilon^\mathrm{ext})= b\sum _{i}(J_{i}^{2}-\mathds{1})\epsilon^\mathrm{ext} _{ii}+{\frac {d}{\sqrt {3}}}\sum _{i\neq j}[J_{i},J_{j}]\epsilon^\mathrm{ext}_{ij}\,.
\end{equation}
This has the same form as the internal defect potential term, Eqn.~2 in the main text, with the same deformation parameters but different strain tensor elements $\epsilon^\mathrm{ext}_{ij}$. To model the experiment in Ref.~\cite{Safonov1995}, external strain with compression $T<0$ has been applied in the directions [001], [110] and [111]. The corresponding stress tensors for applied stress along the directions [001] to [110] through [111], parameterised by $\theta$ (the polar between the directions) is
\begin{equation}
    \sigma^\mathrm{ext}(\theta) = 
    T\begin{pmatrix}
    \frac{1}{2} \sin^2{\theta}\\
    \sigma^\mathrm{ext}_{xx}\\
    \cos^2{\theta} \\
    \frac{1}{\sqrt{2}}\sin{\theta} \cos{\theta} \\
    \sigma^\mathrm{ext}_{yz} \\
    \sigma^\mathrm{ext}_{xx} \\
    \end{pmatrix} \,,
\end{equation}
and the external applied strain tensor is
\begin{equation}
    \epsilon^\mathrm{ext}(\theta) = T
    \begin{pmatrix}
    s_{12} \cos^2{\theta} + \frac{1}{2}\sin^2{\theta}(s_{11}+s_{12}) \\
    \epsilon^\mathrm{ext}_{xx}\\
    s_{12} \sin^2{\theta} + s_{11}\cos^2{\theta}  \\
    \frac{1}{\sqrt{2}}s_{44}\sin{\theta}\cos{\theta}\\
    \epsilon^\mathrm{ext}_{yz} \\
    \frac{1}{2}s_{44}\sin^2{\theta} \\
    \end{pmatrix}.
\end{equation}

For an external strain along [001], $\theta =0$ and $\sigma^\mathrm{ext}_{zz}=T$. The strain tensor has $\epsilon^\mathrm{ext}_{xx}=\epsilon^\mathrm{ext}_{yy}=T s_{12}$ and $\epsilon^\mathrm{ext}_{zz}=T  s_{11}$. 

For an external strain along [110], $\theta=\pi /2$ which gives the non-zero stress tensor elements $\sigma^\mathrm{ext}_{xx}=\sigma^\mathrm{ext}_{yy}=\sigma^\mathrm{ext}_{xy}=\frac{T}{2}$ and for the stress tensor $\epsilon^\mathrm{ext}_{xx}=\epsilon^\mathrm{ext}_{yy}=T/2 (s_{11}+s_{12})$, $\epsilon^\mathrm{ext}_{zz}=T s_{12}$, and $\epsilon^\mathrm{ext}_{xy}=\frac{T}{2} s_{44}$. 

Finally, for the external field applied along $[ 111 ]$, $\theta=\arccos{1/\sqrt{3}}$ all the stress tensor elements are $\sigma^\mathrm{ext}_{ij}=\frac{T}{3}$, with the strain elements $\epsilon^\mathrm{ext}_{xx}=\epsilon^\mathrm{ext}_{yy}=\epsilon^\mathrm{ext}_{zz}=\frac{T}{3} (s_{11}+2s_{12})$, $\epsilon^\mathrm{ext}_{xy}=\epsilon^\mathrm{ext}_{zx}=\epsilon^\mathrm{ext}_{yz}=\frac{T}{3} s_{44}$.

When accounting for applied stress on the defect, the piezospectroscopic shift also needs to be considered. For a defect with monoclinic symmetry using \cite{AAKaplyanskii1964} we find that the piezospectroscopic tensor for a defect with plane (1-10) has the form
\begin{equation}
    A_p = \begin{pmatrix}
A_2 & -A_3 & A_4 \\
 & A_2 & A_4 \\
 &  & A_1 \\
\end{pmatrix}.
\end{equation}
When applying this piezospectroscopic tensor along with the stress tensor the shift to the Hamiltonian takes the form
\begin{equation}\label{eq:peizo}
    \Delta = A_1 \sigma^\mathrm{ext}_{zz} + A_2 (\sigma^\mathrm{ext}_{xx}+\sigma^\mathrm{ext}_{yy}) - 2A_3 \sigma^\mathrm{ext}_{xy} + 2A_4 (\sigma^\mathrm{ext}_{yz}+\sigma^\mathrm{ext}_{zx}).
\end{equation}
Where this corresponds to the shift for the identity orientation in the crystal coordinate system. To find all other orientations the coordinate transformations shown in \cref{table_ThOp_field} needs to be applied, as instead of transforming the orientation we transform the applied stress in the opposite direction. We note that when finding the final transition energies for each orientation, as well as considering the stress piezospectroscopic shift, the internal strain defect potential basis needs to be transformed using the direct coordinate transformations shown in \cref{table_ThOp} keeping the external strain constant. The reason this transformation \textit{isn't} the inverse is we are actually transforming the atomic coordinate system and not the effective stress direction that each orientation sees.

The deformation parameters found from this re-analysis are $b=(-1.72\pm 0.12$)~eV, and $d=(-2.39 \pm 0.14)$~eV. The effective internal strain parameters are $\epsilon_{y'y'}^d=(-0.43\pm 0.03)\times 10^{-3}$, and $ \epsilon_{z'z'}^d=(-0.63\pm 0.02)\times 10^{-3}$. These components are in the C-C-H defect plane, the component out of the plane has been found to be zero. The splitting between the ground and TX is found to be $E_X = (0.93560\pm 0.00004)$eV. The defect potential offset angle is $\theta_p = (-7.7 \pm 0.5)^\circ$ (tilting clockwise towards the hydrogen atom) see \cref{fig_strainfits}. Finally, the piezospectroscopic tensor parameters are $A_1= (-12.2\pm 0.3)\times 10^{-12} $, $A_2=( 16.2\pm 0.2)\times 10^{-12}$, $A_3= (0.9\times \pm 0.2)\times10^{-12} $ and $A_4= (-2.0\pm 0.2)\times 10^{-12} $.

%DFT computations of strain tensor
\subsection{First-principles calculations of strain}
\label{app:First-principles calculations of strain}
\begin{figure}[b]
\centering
  \includegraphics[width=\linewidth]{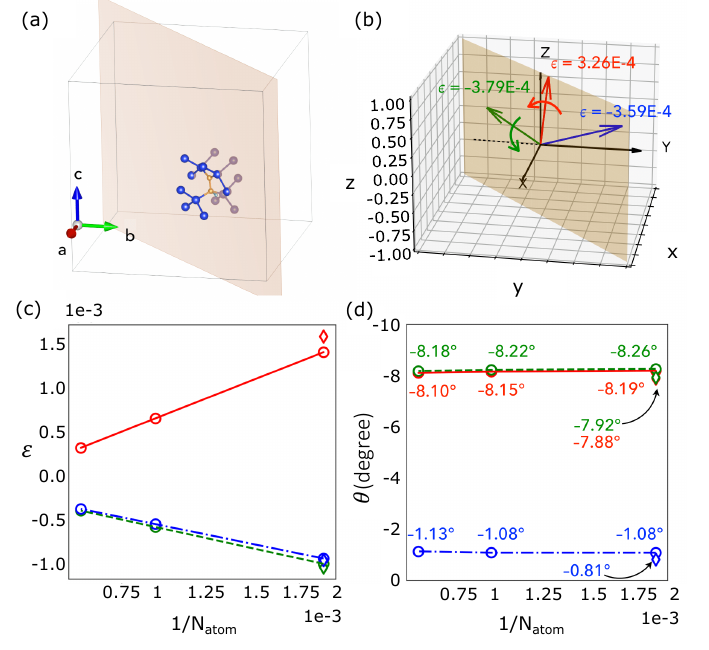}
\caption{Strain computations of the T centre in silicon. (a) Optimized atomic structure of the T centre lies in the (1-10) plane. (b) Two in-plane principal strains (red and green) and the out-of-plane principal strain (blue) evaluated in the 6$\times$6$\times$6 supercell. (c) The magnitude of the principal strain with respect to the inverse of the number of atoms in the supercell. (d) The tilting angles of the principal strains with respect to the inverse of the number of atoms computed at PBE (circle) and HSE06 level (diamond), respectively.}
\label{fig:dft_strain}
\end{figure}

First-principles calculations were performed using VASP \cite{G.Kresse-PRB96,G.Kresse-CMS96} and the projector-augmented wave (PAW) framework \cite{P.E.Blochl-PRB94}. Structural optimizations were performed with spin-polarized computations using Perdew-Burke-Ernzerhof (PBE) functional \cite{J.Perdew-PRL96}. The supercell volumes are fixed during the ionic relaxation and the atoms are relaxed until the forces are smaller than 0.001~eV/\r{A}. To account for the finite-size effect, we evaluated the stress tensor of the T centre in the 4$\times$4$\times$4 (512 atoms), 5$\times$5$\times$5  (1000 atoms), and 6$\times$6$\times$6 (1728 atoms) supercells of silicon. Using the finite difference method, the compliance tensor was evaluated in a 4$\times$4$\times$4 supercell that contains the T centre. Our simulation setup corresponds to the defect orientation $(z0)$, as shown in ~\cref{fig:dft_strain}(a) below.

The principal strains are obtained by diagonalizing the strain tensor as shown in \cref{fig:dft_strain}(b). Two of the principal strains residing in the defect plane $(1\overline{1}0)$ are highlighted by the red and green arrows, while the strain perpendicular to the defect plane is highlighted by the blue arrow. For the in-plane strain, the tilting angle is measured between the red arrow and the z-axis, and between the green arrow and the y-axis, as shown in \cref{fig:dft_strain}(b). We also quantify the tilt of the blue arrow away from the $[1\overline{1}0]$ direction. Our results show that the magnitude of the principal strain is slow to converge (\cref{fig:dft_strain}(c)). However, the tilting angles of the principal strains do not vary significantly, as shown in ~\cref{fig:dft_strain}(d). In addition, we investigated the effect of the functional by evaluating the strain tensor in a 4$\times$4$\times$4 supercell using the HSE06 hybrid functional \cite{Heyd2003}, as shown in the diamond marker in \cref{fig:dft_strain}(c,d). No significant effect of the functionals has been observed. Thus, the PBE strain tensor from 6$\times$6$\times$6 is used to guide our interpretation. The result suggests that one of the in-plane principal strains (red arrow) tilts clockwise towards the hydrogen atom with a tilting angle of $-8.10^\circ$, in close agreement with experimental measurement ($-7.7 \pm 0.5 (^\circ)$).

\subsection{Ground state Landé g-factor}
\label{app:Ground state Landé g-factor}
\begin{figure}[b]
  \includegraphics[width=\linewidth]{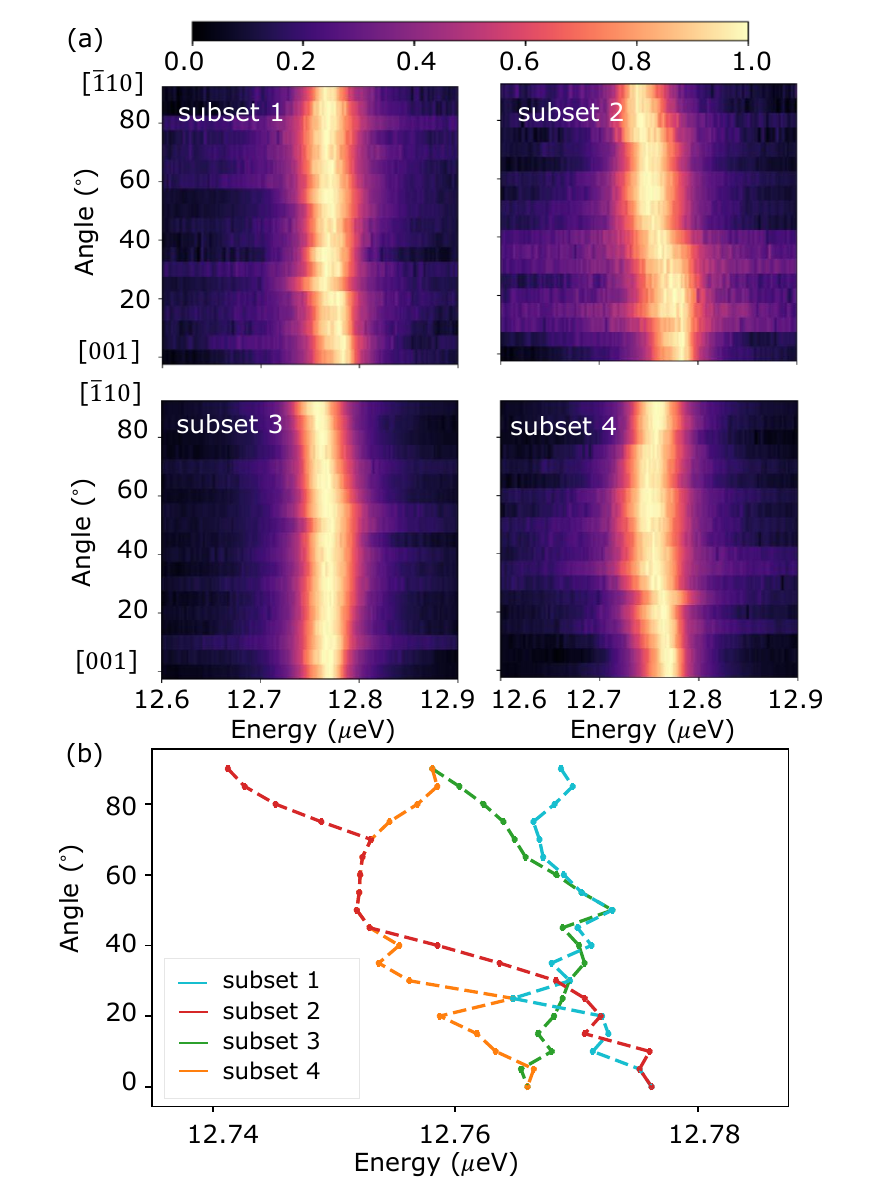}
    \caption{(a) Ground-state Zeeman splitting measured by pump-probe PLE for the 4 different orientation subsets as a function magnetic field angle. (b) Extracted peak positions from ground state splitting values for all subsets. Crosses indicate uncertainties for energies from Gaussian peak fitting (with one sigma confidence) of (a), and a fixed angular uncertainty of $\pm1^\circ$.}

    \label{fig:gEdat}
\end{figure}

Subsequent to the measurement of $g_\mathrm{H}$ in Sec.~V, we performed pump-probe PLE spectroscopy and resolve previously unobserved anisotropy in the electron g-factor. The same two lasers were configured to run pump-probe PLE scans on each of the A/B $\Lambda$-configurations. With the pump laser set to A, the corresponding electron Zeeman splitting was found from the PLE resonance when scanning the probe laser over B. In this configuration, we measure narrower linewidths of $\approx14$~MHz because this measurement is insensitive to inhomogeneity in the un-split TX$_0$ transition energy. At $1.5$~K the homogeneous linewidth of TX$_0$ is $1.1$~MHz, 5$\times$ larger than the lifetime limit, due to thermal excitation between TX$_0$ and TX$_1$ \cite{Bergeron:2020_PRX}. Magnetic field inhomogeneity across the sample accounts for the remaining linewidth.

The ground state splittings measured as a function of field direction, \cref{fig:gEdat}(a), show anisotropy in $g_\mathrm{e}$. It can be seen that the ground state splitting varies by $\approx 0.04$~ $\mu$eV ($9.6$~MHz). This variation corresponds to a Landé g-factor magnitude in the range of $g_\mathrm{e}=2.005\pm0.003$. The sample position may shift slightly during the rotation of the static magnetic field, varying $B$ and therefore the electron splitting. However, \cref{fig:gEdat}(b) shows the electron splitting of each subset recorded simultaneously, for a fixed angle and $B$. It is clear that the subsets have distinct splittings at each angle. An overall magnetic field variation is also evident, which prevents a precise determination of the $g_\mathrm{e}$ tensor from this data.

\subsection{TX$_0$ eigenvector element analysis}
\label{app:TX0 eigenvector element analysis}
\begin{figure}[t]
\centering
    \includegraphics[width=\linewidth]{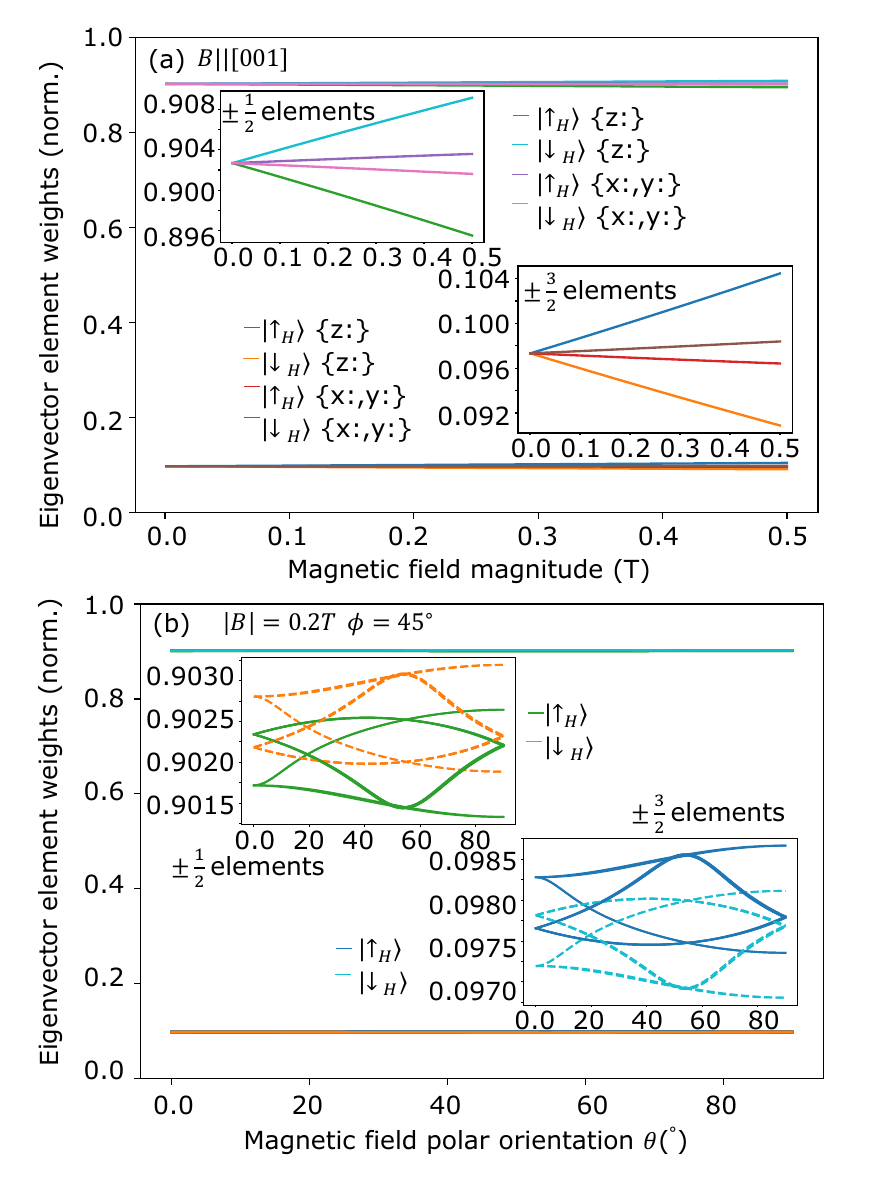}
\caption{Spin components of TX$_0$ under changing magnetic field strength (a) and orientation (b). The upper(lower) lines in the main plots are shown in the left (right) inset and correspond to the sum of the $\pm1/2$ ($\pm3/2$) basis terms for both TX$_0$ levels. Orientation subgroups are indicated by colour. In (b) all orientations are shown, but without labels.}
\label{fig:TX_eigenvector_elements}
\end{figure}

The TX Hamiltonian has four Zeeman eigenstates, two in TX$_0$ and two in TX$_1$. Considering the TX$_0$ only, each eigenvector has four elements corresponding to the Luttinger-kohn basis with the $m_j$ labels $\{+3/2,+1/2,-1/2,-3/2\}$. \Cref{fig:TX_eigenvector_elements} shows the components in this basis as a function of magnetic field magnitude (a) and orientation (b). It can be seen that the dominant $m_j$ component for both the TX$_0$ states is $\pm1/2$, with $\expval{\pm 1/2} \sim0.9$, which is stable with $1$\% over this magnetic field range.

\subsection{Stark shift Hamiltonian fitting}
\label{app:Stark shift Hamiltonian fitting}

\subsubsection{Linear and quadratic terms of the stark shift}
\label{app:Linear and quadratic terms of the stark shift}
To find the stark shift for all 24 orientations instead of rotating the defect, the shifts were calculated by equivalently rotating the field in the opposite direction using table~\ref{table_ThOp_field}. The results of these calculations for the 12 proper rotations of (T from the identity) for the linear and quadratic terms are summarized in the tables \ref{table:Eterms_Linear} and \ref{table:Eterms_Quadratic}, respectively. Orientations (13 to 24) (the inversion partners to (1 to 12)) are obtained by applying coordinate inversion $(x \rightarrow -x, y \rightarrow -y, z \rightarrow -z)$. Coordinates shown in both tables are in the crystal basis.

\begin{table*}[t]
\small
\renewcommand{\arraystretch}{}
 \begin{tabular}{ |p{1.3cm}||p{1.6cm}|p{5cm}|p{3.6cm}|p{2.6cm}|p{1.2cm}|}
 
\hline
\centering{Orient.} & \centering{Rot.} & \centering{Field components transformation relative to the defect} & \centering{Linear Stark shift}  
&\centering{$\Vec{\varepsilon} || [110] $} & { $ \vec{\epsilon} || [001] $} \\
 \hline

\[z0\]& \[ E \] & \[\varepsilon_x \to \varepsilon_x  \varepsilon_y\to \varepsilon_y  \varepsilon_z\to \varepsilon_z \] &\[\frac{A_X(\varepsilon_x+\varepsilon_y)}{\sqrt{2}} + A_Y \varepsilon_z \] &\[A_{X} \varepsilon \] &\[ A_Y \varepsilon\]\\
\hline

\[z1\]&  \[ C_2^{[1,0,0]} \]& \[\varepsilon_x\to \varepsilon_x, \varepsilon_y\to -\varepsilon_y,\varepsilon_z\to -\varepsilon_z \] &\[\frac{A_X(-\varepsilon_x+\varepsilon_y)}{\sqrt{2}} - A_Y \varepsilon_z \] & \[0 \]& \[-A_{Y} \varepsilon \]\\ \hline

\[z2\]& \[ C_2^{[0,0,1]} \] &  \[\varepsilon_x \to -\varepsilon_x, \varepsilon_y\to -\varepsilon_y,\varepsilon_z\to \varepsilon_z \] &\[\frac{A_X(-\varepsilon_x-\varepsilon_y)}{\sqrt{2}} + A_Y \varepsilon_z \]  &\[-A_{X} \varepsilon \] &\[ A_Y \varepsilon\]\\
\hline

\[z3\]&  \[ C_2^{[0,1,0]} \]& \[ \varepsilon_x \to -\varepsilon_x, \varepsilon_y\to \varepsilon_y,\varepsilon_z\to -\varepsilon_z \] & \[\frac{A_X(\varepsilon_x-\varepsilon_y)}{\sqrt{2}} - A_Y \varepsilon_z \] & \[0 \]& \[-A_{Y} \varepsilon \]\\ \hline     

\[y0\]&  \[ C_3^{[-1,-1,-1]} \] & \[\varepsilon_x\to \varepsilon_z, \varepsilon_y\to \varepsilon_x,\varepsilon_z\to \varepsilon_y \] &\[\frac{A_X(\varepsilon_x+\varepsilon_z)}{\sqrt{2}} + A_Y \varepsilon_y \]  &\[(A_X + \sqrt{2}A_Y)\frac{\varepsilon}{2}\]&  \[\frac{A_{X}}{\sqrt{2}} \epsilon\]\\ \hline

\[y1\]&  \[ C_3^{[1,-1,1]} \] & \[\varepsilon_x\to \varepsilon_z, \varepsilon_y\to -\varepsilon_x,\varepsilon_z\to -\varepsilon_y \] &\[\frac{A_X(-\varepsilon_x+\varepsilon_z)}{\sqrt{2}} - A_Y \varepsilon_y \] &\[-(A_X + \sqrt{2}A_Y)\frac{\varepsilon}{2}\]& \[\frac{A_{X} }{\sqrt{2}}\varepsilon\]\\ \hline

\[y2\]&  \[ C_3^{[-1,1,1]} \] & \[\varepsilon_x\to -\varepsilon_z, \varepsilon_y\to -\varepsilon_x,\varepsilon_z\to \varepsilon_y \] &\[\frac{A_X(-\varepsilon_x-\varepsilon_z)}{\sqrt{2}} + A_Y \varepsilon_y \] &\[(-A_X + \sqrt{2}A_Y)\frac{\varepsilon}{2}\]& \[\frac{-A_{X}}{\sqrt{2}} \varepsilon\]\\ \hline

\[y3\]&  \[ C_3^{[1,1,-1]} \] & \[\varepsilon_x\to -\varepsilon_z, \varepsilon_y\to \varepsilon_x,\varepsilon_z\to -\varepsilon_y \]&\[\frac{A_X(\varepsilon_x-\varepsilon_z)}{\sqrt{2}} - A_Y \varepsilon_y \]  &  \[(A_X - \sqrt{2}A_Y)\frac{\varepsilon}{2} \]&  \[ -\frac{A_{X}}{\sqrt{2} }\varepsilon\]\\ \hline

\[x0\]&  \[ C_3^{[1,1,1]} \] & \[\varepsilon_x\to \varepsilon_y, \varepsilon_y\to \varepsilon_z,\varepsilon_z\to \varepsilon_x \] &\[\frac{A_X(\varepsilon_y+\varepsilon_z)}{\sqrt{2}} + A_Y \varepsilon_x \]  &  \[(A_X + \sqrt{2}A_Y)\frac{\varepsilon}{2} \]&  \[ \frac{A_{X}}{\sqrt{2} }\varepsilon\]\\ \hline

\[x3\]&  \[ C_3^{[1,-1,-1]} \] &  \[\varepsilon_x\to -\varepsilon_y, \varepsilon_y\to \varepsilon_z,\varepsilon_z\to -\varepsilon_x \] &\[\frac{A_X(-\varepsilon_y+\varepsilon_z)}{\sqrt{2}} - A_Y \varepsilon_x \]  &  \[-(A_X + \sqrt{2}A_Y)\frac{\varepsilon}{2} \]&  \[ \frac{A_{X}}{\sqrt{2} }\varepsilon\]\\ \hline

\[x2\]& \[ C_3^{[-1,1,-1]} \] & \[\varepsilon_x \to -\varepsilon_y, \varepsilon_y\to -\varepsilon_z,\varepsilon_z\to \varepsilon_x \] &\[\frac{A_X(-\varepsilon_y-\varepsilon_z)}{\sqrt{2}} + A_Y \varepsilon_x \]  &  \[(-A_X + \sqrt{2}A_Y)\frac{\varepsilon}{2} \]&  \[ -\frac{A_{X}}{\sqrt{2} }\varepsilon\]\\ \hline

\[x1\] &  \[ C_3^{[-1,-1,1]} \] & \[\varepsilon_x\to \varepsilon_y, \varepsilon_y\to -\varepsilon_z,\varepsilon_z\to -\varepsilon_x \] &\[\frac{A_X(\varepsilon_y-\varepsilon_z)}{\sqrt{2}} - A_Y \varepsilon_x \]  &  \[(A_X - \sqrt{2}A_Y)\frac{\varepsilon}{2} \]&  \[ -\frac{A_{X}}{\sqrt{2} }\varepsilon\]\\ \hline

\hline
\end{tabular}
\caption{Linear term of the TX$_0 \leftrightarrow$ T$_0$ Stark shift for each T centre orientation. The two rightmost columns show the shift for fields along the crystal axes approximately corresponding to the measurements in the manuscript.}
\label{table:Eterms_Linear}
\end{table*}

\begin{table*}[!ht]
\renewcommand{\arraystretch}{1}
\small\begin{tabular}{ |p{1.5cm}||p{2.1cm}|p{6.2cm}|p{5cm}|}
 
 \hline
\centering{Orient.} &  \centering{Rot.} & \centering{Quadratic Stark shift for  $\vec{\varepsilon}~||~[110]$} & {Quadratic Stark shift for  $\vec{\varepsilon}~||~[001]$} \\
 \hline

\[z0\]& \[ E \] & \[-\alpha_{XX}~\frac{\varepsilon^2}{2}\]& \[-\alpha_{YY}~ \frac{\varepsilon^2}{2}\]\\
 \hline

\[z1\]&  \[ C_2^{[1,0,0]} \]& \[-\alpha_{ZZ}~\frac{\varepsilon^2}{2}\]& \[-\alpha_{YY}~ \frac{\varepsilon^2}{2}\]\\ \hline
 
\[z2\]& \[ C_2^{[0,0,1]} \] &\[-\alpha_{XX}~\frac{\varepsilon^2}{2}\]& \[-\alpha_{YY}~ \frac{\varepsilon^2}{2}\]\\ \hline

\[z3\]&  \[ C_2^{[0,1,0]} \]&  \[-\alpha_{ZZ}~\frac{\varepsilon^2}{2}\]& \[-\alpha_{YY}~ \frac{\varepsilon^2}{2}\]\\ \hline

\[y0\]&  \[ C_3^{[-1,-1,-1]} \] & \[- \left(\alpha_{YY} + \sqrt{2}~\alpha_{XY} + \frac{1}{2}(\alpha_{XX}+ \alpha_{ZZ})\right) ~\frac{\varepsilon^2}{4}\]&   \[-(\alpha_{XX}+\alpha_{ZZ})~ \frac{\varepsilon^2}{4}\]\\ \hline

\[y1\]&  \[ C_3^{[1,-1,1]} \] & \[-\left(\alpha_{YY} +\sqrt{2}~\alpha_{XY} + \frac{1}{2}(\alpha_{XX}+\alpha_{ZZ})\right) ~\frac{\varepsilon^2}{4}\]&   \[-(\alpha_{XX}+\alpha_{ZZ})~\frac{\varepsilon^2}{4}\]\\ \hline

\[y2\]&  \[ C_3^{[-1,1,1]} \] & \[-\left(\alpha_{YY} -\sqrt{2}~\alpha_{XY}+ \frac{1}{2}(\alpha_{XX}+\alpha_{ZZ})\right) ~\frac{\varepsilon^2}{4}\]&   \[-(\alpha_{XX}+\alpha_{ZZ})~\frac{\varepsilon^2}{4}\]\\ \hline

\[y3\]&  \[ C_3^{[1,1,-1]} \] & \[-\left(\alpha_{YY} -\sqrt{2}~\alpha_{XY} + \frac{1}{2}(\alpha_{XX}+\alpha_{ZZ})\right) ~\frac{\varepsilon^2}{4}\]&   \[-(\alpha_{XX}+\alpha_{ZZ})~\frac{\varepsilon^2}{4}\]\\ \hline

\[x0\]&  \[ C_3^{[1,1,1]} \] & \[-\left(\alpha_{YY} + \sqrt{2}~\alpha_{XY}+ \frac{1}{2} (\alpha_{XX}+\alpha_{ZZ})\right) ~\frac{\varepsilon^2}{4}\]&   \[-(\alpha_{XX}+\alpha_{ZZ})~\frac{\varepsilon^2}{4}\]\\ \hline

\[x3\]&  \[ C_3^{[1,-1,-1]} \] &  \[-\left(\alpha_{YY} + \sqrt{2}~\alpha_{XY}+ \frac{1}{2} (\alpha_{XX}+\alpha_{ZZ})\right) ~\frac{\varepsilon^2}{4}\]&   \[-(\alpha_{XX}+\alpha_{ZZ})~\frac{\varepsilon^2}{4}\]\\ \hline

\[x2\]& \[ C_3^{[-1,1,-1]} \] & \[-\left(\alpha_{YY} - \sqrt{2}~\alpha_{XY}+ \frac{1}{2} (\alpha_{XX}+\alpha_{ZZ})\right) ~\frac{\varepsilon^2}{4}\]&   \[-(\alpha_{XX}+\alpha_{ZZ})~\frac{\varepsilon^2}{4}\]\\ \hline

\[x1\]&  \[ C_3^{[-1,-1,1]} \] & \[ -\left(\alpha_{XX} - \sqrt{2}~\alpha_{XY}+\frac{1}{2} (\alpha_{YY} +\alpha_{ZZ})\right)~ \frac{\varepsilon^2}{4}\]  & \[-(\alpha_{XX}+\alpha_{ZZ})~ \frac{\varepsilon^2}{4}\]\\ \hline
\end{tabular}
\caption{Quadratic term of the TX$_0 \leftrightarrow$ T$_0$ Stark shift for each T centre orientation and for fields along the crystal axes approximately corresponding to the measurements in the manuscript.}
\label{table:Eterms_Quadratic}
\end{table*}

\subsubsection{Fitting and angular misalignment}
\label{app:Fitting and angular misalignment}
When fitting the Stark data, we resolve more PLE lines than expected based on the monoclinic symmetry of the defect. Specifically, for an external electric field along the [001] direction (where 4 lines are expected), we resolve 7 lines. Similarly, for the [110] direction (with 7 expected lines), we resolve 8. This apparent reduction in degeneracy may be due to misalignment between the applied electric field and the crystallographic symmetry directions. To account for potential misalignment, we fit the Stark model (Eqn.~(5)) including two free misalignment parameters. We use spherical polar coordinates: $\theta$ the polar angle from the crystal $z$ axis and $\phi$ the azimuthal angle. The fitted misalignment from the [110] ([001]) dataset is
$\Delta \theta=(1.6 \pm 0.6) ^\circ $, and $\Delta  \phi=(0.0\pm 0.7) ^\circ$ ($\Delta \theta=(-13 \pm 1.0)^\circ$ and $\Delta \phi=(0.0\pm 4)^\circ$).

\begin{figure}
\centering
\includegraphics[width=\linewidth]{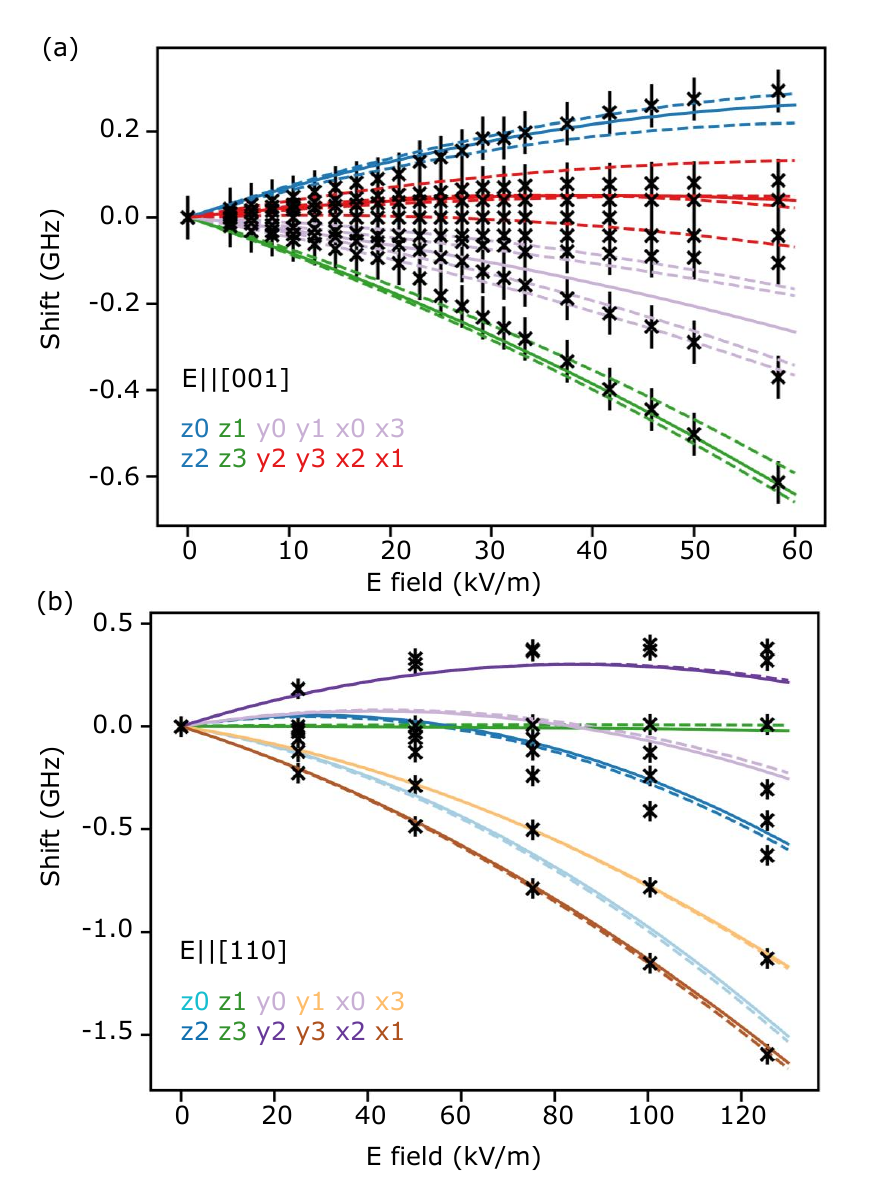}
\caption{Simulated stark shift with the extracted PLE data peaks for (a) $\hat{E} \approx [001]$, (b) $\hat{E} \approx [110]$. The dashed lines show the model fitted with the misalignment as shown in the main text, the solid lines show the model with the same parameters along the symmetry angles [001] (a) and [110] (b). To highlight the symmetry groups, we plot the 12 proper rotations of the T orientation, omitting inversion partners, as these produce degenerate lines at different positions.}
\label{fig:efield_miss_vs_ideal}
\end{figure}

The model is fitted simultaneously to the two datasets using a two-step minimization process that considers plausible assignments between each resolvable line and orientations in the model to calculate uncertainties. The fitting routine calculates fit parameter uncertainties considering the measurement uncertainty on each PLE peak frequency, as shown in \cref{fig:efield_miss_vs_ideal}. %For each data line, we averaged the Stark parameters over the corresponding plausible orientation subsets to ensure robustness in the final parameter estimates. 
We first implement an initial cost function in which each data line is assigned to the closest simulated orientation line and each simulated orientation line is assigned to the closest data line. This cost function weighs the [110] and [001] datasets equally. The total cost is the sum of the average per-point cost of each dataset. For each dataset, the cost function first totals the Euclidean distance from each orientation's PLE resonance expected by the model to the nearest observed PLE peak data point. This `model-data' cost is divided by the total number of orientations (24). The cost function then totals the Euclidean distance from each data point to the nearest orientation in the model. This `data-model' cost is divided by the number of observed data lines (7 and 8 respectively for $\hat{E} \approx [001]$ and $[110]$). The total cost is minimized with a Nelder-Meade fitting routine.

\begin{figure}[b]
\centering
\includegraphics[width=\linewidth]{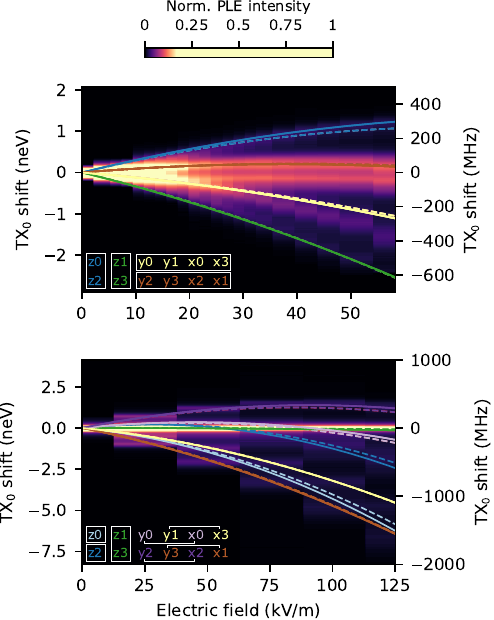}
\caption{Simulated stark shift for electric field along angles [001] (a) and [110] (b) superposed with the PLE data. The solid lines show the model with fitted parameters including misalignment, and the dashed lined show the model with fitted parameters with no assumed misalignment. We plot the 12 proper rotations of the T orientations, excluding inversion partners that result in degenerate lines at different positions.}
\label{fig:efield_ideal}
\end{figure}

This two-step assignment, considering both the data-model and model-data costs, is crucial when fitting with four free misalignment parameters. If we only include data-model cost, the fitting process can erroneously optimize for a large misalignment angle, ignoring extreme orientations in the model that don't match any data line. Conversely, if we only include model-data costs, some data lines may not be fit at all. A single data-model assignment is sufficient when fitting the data to the Stark model (Eqn.~(5)) without the four misalignment parameters (see  \cref{fig:efield_ideal}). 

%To assign orientation lines to data lines, we implemented an initial cost function designed to optimize the Stark parameters and misalignment angles. The process involves first assigning each data line to the closest simulated orientation line, followed by assigning each simulated orientation line to the closest data line. This two-step assignment is crucial. If we only assign each data line to the nearest orientation line, the fitting process can erroneously optimize for a large misalignment angle, ignoring other orientations that don't match well with any data line. Conversely, if we only assign each orientation line to the nearest data line, some data lines may not be properly accounted for in the fit.  

From the output of the cost model we assign each data line a subset of plausible orientation lines, such that a least squares fitting can be performed. To address fit ambiguity stemming from data linewidth and possible orientation assignments, we also consider four additional plausible orientation assignments to the data sets. We perform joint fits of the $[001]$ and $[110]$ data sets for each orientation assignment using the least squares method `curve fit' from SciPy, which provides fit uncertainties for each parameter based on the covariance matrix. The reported parameters are the average of these five fits, and the reported uncertainty is the larger of the standard deviation of the five fitted parameter sets or mean of the fit uncertainty.

%From this starting point, we consider fit ambiguity arising from the data linewidth and possible orientation assignments. We repeat the process with four additional plausible permutations of the initial orientation assignments. Each of these permutations produce equally good final fits, within the linewidth of the dataset. The remaining assignments are significantly worse, and rejected manually. The reported parameters are the average of these five fits, and the reported uncertainty is the larger of the largest fit uncertainty or the standard deviation of the five fitted parameter sets.

\Cref{fig:efield_miss_vs_ideal} shows the degree of misalignment fitted. In this figure, we plot the extracted PLE peaks against the modeled frequency shifts for each orientation. We compare the model fitted including misalignment (as in the main text), dashed lines, to the model with the same fitted parameters but no misalignment, solid lines. The small alignment fitted to the $\vec{E} \parallel [110]$ makes a negligible difference, but the $13(1)^\circ$ misalignment fitted to the $\vec{E} \parallel [001]$ dataset is required to capture the observed structure. In \cref{fig:efield_miss_vs_ideal} and \cref{fig:efield_ideal}, we display only the 12 proper rotations of the identity orientation to better emphasize the symmetry groups. The inverse orientations, while generating degenerate lines, appear in positions distinct from their inversion partners.

For comparison, we also consider the model fitted without misalignment parameters. We perform a single curve fitting routine where each data line for both sets is compared against the closest orientation line (data-model cost). Since there are fewer orientation subsets in this high-degeneracy model, and fewer free parameters, a single-stage fit is adequate. The parameters found from this fitting are $A_X=-3681 \pm 170$ Hz.m/V, $A_Y=7817 \pm 152$ Hz.m/V, $\alpha_{XX}=0.134 \pm 0.004$ Hz.m$^2/$V$^2$, $\alpha_{XY}=0 \pm 0.003 $ Hz.m$^2/$V$^2$, $\alpha_{YY}=0.092 \pm 0.005 $ Hz.m$^2/$V$^2$, $\alpha_{ZZ}=0.004 \pm 0.004$ Hz.m$^2/$V$^2$. Where all parameters lie within one or two standard deviations of the parameter set fitted with misalignment. %following parameters lie within one standard deviation (std) of the parameter set fitted with misalignment, $A_X$, $\alpha_{XY}$ and $\alpha_{ZZ}$, and all remaining parameters lie within two std. 
In \cref{fig:efield_ideal} we compare this parameter set to the parameter set fitted using free misalignment parameters. We plot the modelled energy shift for (a) $\vec{E} \parallel [001]$, (b) $\vec{E} \parallel [110]$, without any misalignment. Dashed lines are the parameter set fitted with ideal alignment, and solid lines show the parameter set fitted with misalignment, but plotted along the crystal axes (misalignment removed).

\subsubsection{Quadratic term of Stark shift for a shallow acceptor}
\label{Quadratic term of Stark shift for a shallow acceptor}
We can extend our treatment of TX as a shallow acceptor to provide a Hamiltonian for the T centre under an electric field. In this approach, the quadratic component of the shift is attributed to the loosely-bound hole following the model derived by Bir and Pikus for shallow acceptors \cite{BIR1963_2}. This replaces the more general polarizability tensor fitted in the main text. As the quadratic shift is proportional to the wavefunction's volume it can be assumed that the quadratic contribution from the loosely-bound hole dominates the tightly-bound electrons \cite{Jackson_1975}. 
To capture this, the quadratic term in $H_\mathrm{E}$ is expressed in a similar form to the effective strain as in the crystal basis as
\begin{equation} \label{eq_efeild_quad}
    H_\mathrm{E_Q}(\mathbf{E})=  \alpha E^2 \mathds{1} + \beta \sum _{i}(J_{i}^{2}-\frac{5}{4} \mathds{1} )E_{i}^2 +
    {\frac {\gamma }{\sqrt {3}}}\sum _{i\neq j}[J_{i},J_{j}]E_{i}E_{j} \,,
\end{equation}
where the terms $\alpha$, $\beta$ and $\gamma$ are the cubic coupling parameters.
Furthermore, although the electrons see an asymmetric central cell potential, the hole wavefunction sees a more symmetric potential averaged over many cells. A symmetric potential cannot give rise to a linear Stark shift, so we can attribute the linear shift to the electrons.

\subsubsection{Effective electric field across the sample}
\label{app:Effective electric field across the sample}
In the experiment the sample was mounted loosely between two copper plates. In order to estimate the value of the electric field across the sample, we took into account the effect of the Kapton and helium layers surrounding the sample. %as shown in \cref{fig:cavity}. 
The voltage across the sample is
 \begin{equation}
\frac{V_\mathrm{Si}}{V_\mathrm{tot}}=\frac{C_\mathrm{tot}}{C_\mathrm{Si}} \,, 
%&=0.53 V_\mathrm{tot} \text{ for } E||110
\end{equation}
where
\begin{equation}
C_\mathrm{tot}=\left(\frac{1}{C_\mathrm{lHe}}+\frac{1}{C_\mathrm{Kap}}+\frac{1}{C_\mathrm{Si}}\right)^{-1} \,.
\end{equation}
The dielectric constants of silicon, liquid helium and Kapton are, respectively, (11.7, 1.057 and 3.4) F/m, and the thickness of these materials when the sample was mounted for $E||[110]$([001]), were $d_\mathrm{lHe}=0.1$(0.1)~mm, $d_\mathrm{Si}=1.9$ (4.3)~mm, 
and $d_\mathrm{Kap}=0.17(0.54)$~mm, resulting in $C_\mathrm{tot}=3.26(1.61)$~F 
and $V_\mathrm{Si}=0.53(0.59) V_\mathrm{tot}$.

\subsection{First-principles calculation of Stark shift}
\label{app:First-principles calculation of Stark shift}
In order to support the experimental Stark shift, we performed density functional theory calculations on two molecular analogues of the T center using the Gaussian software \cite{Gaussian16}. In the ground state, the T center hosts an unpaired electron. Upon electronic excitation to the TX state, the defect becomes negatively charged and a bound-exciton is formed from the valence band hole. As a result of the large spatial extent, excitons are difficult to model using first-principles calculations based on periodic boundary conditions. Instead, we used two simplified molecular models, the ethyl radical ($C_2H_5$) and an ethyl surrounded by two layers of silicon terminated by hydrogens ($C_2HSi_{16}H_{36}$)) which we will refer to as the silicon-ethyl cluster. In order to model the TX transition, we assume a transition from the radical state (S=1/2) to the negatively charged state (S=0). This means we entirely neglect the effect of the exciton. The dipole moment of each state is then calculated and the first-order linear Stark shift is obtained by taking the difference between the two. 

We used a 6311G+ basis set in combination with the PBE functional \cite{J.Perdew-PRL96}. The symmetry was constrained to the \textit{Cs} point group. In the case of the ethyl radical, we directly optimized the geometry of the molecule using the Berny optimization algorithm implemented in Gaussian. For the silicon-ethyl cluster, we tested two geometries. In the first case, we also fully relaxed all the degree of freedoms using Gaussian. In the second case, we used the structure obtained from a VASP calculation (see Appendix \ref{app:First-principles calculations of strain}), relaxed the Si-H termination while keeping every other atoms fixed. Both approaches give very similar results and we only report the dipole moment changes obtained using the Gaussian geometry. 

For the ethyl molecule, the calculated linear electric field coupling coefficient are $A_X$ = 0.06D and $A_Y$ = 3.44D which translates to $A_X$ = 302 Hz.m/V and $A_Y$ = 17317 Hz.m/V, respectively. For the silicon-ethyl cluster, we obtain $A_X$ = -0.61D and $A_Y$ = 1.85D or $A_X$ = -3070 Hz.m/V and $A_Y$ = 9313 Hz.m/V. Considering the simplicity of the model, these results are in good agreement with the experimentally observed linear Stark shift ($A_X$ = -3596 $\pm$ 137 Hz.m/V, $A_Y$ = 7519 $\pm$ 82 Hz.m/V) and help clarifying the orientation of the dipole moment change.

\section{Multi-Channel Ultra-Stable Laser}
\label{app:Multi-Channel Ultra-Stable Laser}
In order to probe the narrow spectral peaks of the different T centre orientations and their transitions, we utilized a cavity-stabilized diode laser with linewidth$<2.6$~kHz. Our seed laser is a Toptica DL100 Pro with Pound-Drever-Hall (PDH) locked to Fabry-Perot cavity from Stable Laser Systems. The laser signal is split into 2 channels for independent control of frequency and power. IQ modulators (iXblue MXIQER-LN-30) allow a tunable range sufficient to probe all of the A and B transitions of the different T centre orientations at 100~mT. The optical signals are amplified using solid-state amplifiers (Aerodiode SOM-HPP-S1310) up to a maximum of $\sim$30~mW as measured at the input to the cryostat. 

\section{Electric field frequency}
\label{app:Electric field frequency}

\begin{figure}[!b]
\centering
\includegraphics[scale=0.7]{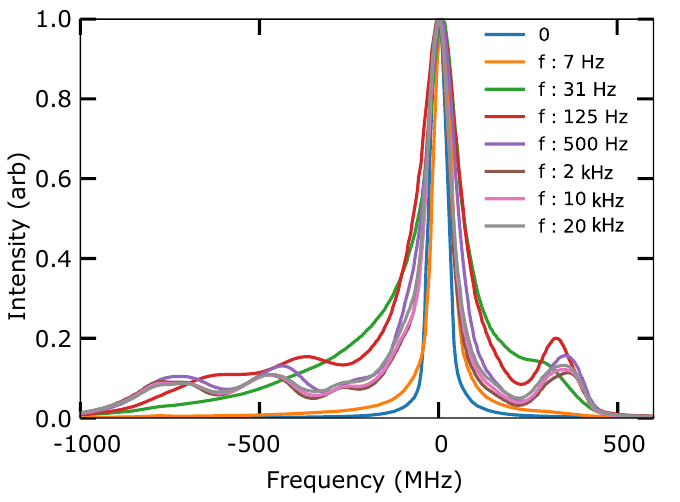}
	\caption{PLE spectrum for various pulsed voltage frequencies. Electric field is applied along the [110] axis. For switching frequency less than 30~Hz no obvious stark shift was observed.    }
	\label{fig:110_frequency}
\end{figure}

 With pulse frequency lower than 30~Hz, no obvious Stark shift was observed. We attribute this to shielding by accumulated charges. \Cref{fig:110_frequency} shows a PLE spectrum repeated with different electric field pulse frequencies. Below $30$~Hz, only a single unshifted peak is observed corresponding to zero total field. Above 500~Hz, the PLE signal stabilizes, indicating no significant screening effect happening in this regime.

\bibliography{references}

\begin{thebibliography}{53}%
\makeatletter
\providecommand \@ifxundefined [1]{%
 \@ifx{#1\undefined}
}%
\providecommand \@ifnum [1]{%
 \ifnum #1\expandafter \@firstoftwo
 \else \expandafter \@secondoftwo
 \fi
}%
\providecommand \@ifx [1]{%
 \ifx #1\expandafter \@firstoftwo
 \else \expandafter \@secondoftwo
 \fi
}%
\providecommand \natexlab [1]{#1}%
\providecommand \enquote  [1]{``#1''}%
\providecommand \bibnamefont  [1]{#1}%
\providecommand \bibfnamefont [1]{#1}%
\providecommand \citenamefont [1]{#1}%
\providecommand \href@noop [0]{\@secondoftwo}%
\providecommand \href [0]{\begingroup \@sanitize@url \@href}%
\providecommand \@href[1]{\@@startlink{#1}\@@href}%
\providecommand \@@href[1]{\endgroup#1\@@endlink}%
\providecommand \@sanitize@url [0]{\catcode `\\12\catcode `\$12\catcode `\&12\catcode `\#12\catcode `\^12\catcode `\_12\catcode `\%12\relax}%
\providecommand \@@startlink[1]{}%
\providecommand \@@endlink[0]{}%
\providecommand \url  [0]{\begingroup\@sanitize@url \@url }%
\providecommand \@url [1]{\endgroup\@href {#1}{\urlprefix }}%
\providecommand \urlprefix  [0]{URL }%
\providecommand \Eprint [0]{\href }%
\providecommand \doibase [0]{https://doi.org/}%
\providecommand \selectlanguage [0]{\@gobble}%
\providecommand \bibinfo  [0]{\@secondoftwo}%
\providecommand \bibfield  [0]{\@secondoftwo}%
\providecommand \translation [1]{[#1]}%
\providecommand \BibitemOpen [0]{}%
\providecommand \bibitemStop [0]{}%
\providecommand \bibitemNoStop [0]{.\EOS\space}%
\providecommand \EOS [0]{\spacefactor3000\relax}%
\providecommand \BibitemShut  [1]{\csname bibitem#1\endcsname}%
\let\auto@bib@innerbib\@empty
%</preamble>
\bibitem [{\citenamefont {Bergeron}\ \emph {et~al.}(2020)\citenamefont {Bergeron}, \citenamefont {Chartrand}, \citenamefont {Kurkjian}, \citenamefont {Morse}, \citenamefont {Riemann}, \citenamefont {Abrosimov}, \citenamefont {Becker}, \citenamefont {Pohl}, \citenamefont {Thewalt},\ and\ \citenamefont {Simmons}}]{Bergeron:2020_PRX}%
  \BibitemOpen
  \bibfield  {author} {\bibinfo {author} {\bibfnamefont {L.}~\bibnamefont {Bergeron}}, \bibinfo {author} {\bibfnamefont {C.}~\bibnamefont {Chartrand}}, \bibinfo {author} {\bibfnamefont {A.~T.~K.}\ \bibnamefont {Kurkjian}}, \bibinfo {author} {\bibfnamefont {K.~J.}\ \bibnamefont {Morse}}, \bibinfo {author} {\bibfnamefont {H.}~\bibnamefont {Riemann}}, \bibinfo {author} {\bibfnamefont {N.~V.}\ \bibnamefont {Abrosimov}}, \bibinfo {author} {\bibfnamefont {P.}~\bibnamefont {Becker}}, \bibinfo {author} {\bibfnamefont {H.-J.}\ \bibnamefont {Pohl}}, \bibinfo {author} {\bibfnamefont {M.~L.~W.}\ \bibnamefont {Thewalt}},\ and\ \bibinfo {author} {\bibfnamefont {S.}~\bibnamefont {Simmons}},\ }\bibfield  {title} {\bibinfo {title} {{Silicon-Integrated Telecommunications Photon-Spin Interface}},\ }\href {https://doi.org/10.1103/prxquantum.1.020301} {\bibfield  {journal} {\bibinfo  {journal} {PRX Quantum}\ }\textbf {\bibinfo {volume} {1}},\ \bibinfo {pages} {20301} (\bibinfo {year} {2020})}\BibitemShut {NoStop}%
\bibitem [{\citenamefont {Dhaliah}\ \emph {et~al.}(2022)\citenamefont {Dhaliah}, \citenamefont {Xiong}, \citenamefont {Sipahigil}, \citenamefont {Griffin},\ and\ \citenamefont {Hautier}}]{Dhaliah2022}%
  \BibitemOpen
  \bibfield  {author} {\bibinfo {author} {\bibfnamefont {D.}~\bibnamefont {Dhaliah}}, \bibinfo {author} {\bibfnamefont {Y.}~\bibnamefont {Xiong}}, \bibinfo {author} {\bibfnamefont {A.}~\bibnamefont {Sipahigil}}, \bibinfo {author} {\bibfnamefont {S.~M.}\ \bibnamefont {Griffin}},\ and\ \bibinfo {author} {\bibfnamefont {G.}~\bibnamefont {Hautier}},\ }\bibfield  {title} {\bibinfo {title} {{First-principles study of the T center in silicon}},\ }\href {https://doi.org/10.1103/PhysRevMaterials.6.L053201} {\bibfield  {journal} {\bibinfo  {journal} {Physical Review Materials}\ }\textbf {\bibinfo {volume} {6}},\ \bibinfo {pages} {L053201} (\bibinfo {year} {2022})}\BibitemShut {NoStop}%
\bibitem [{\citenamefont {Simmons}(2023)}]{Simmons2023}%
  \BibitemOpen
  \bibfield  {author} {\bibinfo {author} {\bibfnamefont {S.}~\bibnamefont {Simmons}},\ }\href {https://arxiv.org/abs/2311.04858} {\bibinfo {title} {Scalable fault-tolerant quantum technologies with silicon colour centres}} (\bibinfo {year} {2023}),\ \Eprint {https://arxiv.org/abs/2311.04858} {arXiv:2311.04858 [quant-ph]} \BibitemShut {NoStop}%
\bibitem [{\citenamefont {Saeedi}\ \emph {et~al.}(2013)\citenamefont {Saeedi}, \citenamefont {Simmons}, \citenamefont {Salvail}, \citenamefont {Dluhy}, \citenamefont {Riemann}, \citenamefont {Abrosimov}, \citenamefont {Becker}, \citenamefont {Pohl}, \citenamefont {Morton},\ and\ \citenamefont {Thewalt}}]{Saeedi2013}%
  \BibitemOpen
  \bibfield  {author} {\bibinfo {author} {\bibfnamefont {K.}~\bibnamefont {Saeedi}}, \bibinfo {author} {\bibfnamefont {S.}~\bibnamefont {Simmons}}, \bibinfo {author} {\bibfnamefont {J.~Z.}\ \bibnamefont {Salvail}}, \bibinfo {author} {\bibfnamefont {P.}~\bibnamefont {Dluhy}}, \bibinfo {author} {\bibfnamefont {H.}~\bibnamefont {Riemann}}, \bibinfo {author} {\bibfnamefont {N.~V.}\ \bibnamefont {Abrosimov}}, \bibinfo {author} {\bibfnamefont {P.}~\bibnamefont {Becker}}, \bibinfo {author} {\bibfnamefont {H.-J.}\ \bibnamefont {Pohl}}, \bibinfo {author} {\bibfnamefont {J.~J.~L.}\ \bibnamefont {Morton}},\ and\ \bibinfo {author} {\bibfnamefont {M.~L.~W.}\ \bibnamefont {Thewalt}},\ }\bibfield  {title} {\bibinfo {title} {Room-temperature quantum bit storage exceeding 39 minutes using ionized donors in silicon-28},\ }\href {https://doi.org/10.1126/science.1239584} {\bibfield  {journal} {\bibinfo  {journal} {Science}\ }\textbf {\bibinfo {volume} {342}},\ \bibinfo {pages} {830} (\bibinfo {year} {2013})}\BibitemShut
  {NoStop}%
\bibitem [{\citenamefont {Benjamin}\ \emph {et~al.}(2009)\citenamefont {Benjamin}, \citenamefont {Lovett},\ and\ \citenamefont {Smith}}]{Benjamin2009prospects}%
  \BibitemOpen
  \bibfield  {author} {\bibinfo {author} {\bibfnamefont {S.~C.}\ \bibnamefont {Benjamin}}, \bibinfo {author} {\bibfnamefont {B.~W.}\ \bibnamefont {Lovett}},\ and\ \bibinfo {author} {\bibfnamefont {J.~M.}\ \bibnamefont {Smith}},\ }\bibfield  {title} {\bibinfo {title} {{Prospects for measurement-based quantum computing with solid state spins}},\ }\href {https://doi.org/10.1002/LPOR.200810051} {\bibfield  {journal} {\bibinfo  {journal} {Laser {\&} Photonics Reviews}\ }\textbf {\bibinfo {volume} {3}},\ \bibinfo {pages} {556} (\bibinfo {year} {2009})}\BibitemShut {NoStop}%
\bibitem [{\citenamefont {Monroe}\ \emph {et~al.}(2014)\citenamefont {Monroe}, \citenamefont {Raussendorf}, \citenamefont {Ruthven}, \citenamefont {Brown}, \citenamefont {Maunz}, \citenamefont {Duan},\ and\ \citenamefont {Kim}}]{Monroe2014}%
  \BibitemOpen
  \bibfield  {author} {\bibinfo {author} {\bibfnamefont {C.}~\bibnamefont {Monroe}}, \bibinfo {author} {\bibfnamefont {R.}~\bibnamefont {Raussendorf}}, \bibinfo {author} {\bibfnamefont {A.}~\bibnamefont {Ruthven}}, \bibinfo {author} {\bibfnamefont {K.~R.}\ \bibnamefont {Brown}}, \bibinfo {author} {\bibfnamefont {P.}~\bibnamefont {Maunz}}, \bibinfo {author} {\bibfnamefont {L.-M.}\ \bibnamefont {Duan}},\ and\ \bibinfo {author} {\bibfnamefont {J.}~\bibnamefont {Kim}},\ }\bibfield  {title} {\bibinfo {title} {Large-scale modular quantum-computer architecture with atomic memory and photonic interconnects},\ }\href {https://doi.org/10.1103/PhysRevA.89.022317} {\bibfield  {journal} {\bibinfo  {journal} {Phys. Rev. A}\ }\textbf {\bibinfo {volume} {89}},\ \bibinfo {pages} {022317} (\bibinfo {year} {2014})}\BibitemShut {NoStop}%
\bibitem [{\citenamefont {Li}\ and\ \citenamefont {Thompson}(2024)}]{Li2024highrate}%
  \BibitemOpen
  \bibfield  {author} {\bibinfo {author} {\bibfnamefont {Y.}~\bibnamefont {Li}}\ and\ \bibinfo {author} {\bibfnamefont {J.}~\bibnamefont {Thompson}},\ }\href {https://arxiv.org/abs/2401.04075} {\bibinfo {title} {High-rate and high-fidelity modular interconnects between neutral atom quantum processors}} (\bibinfo {year} {2024}),\ \Eprint {https://arxiv.org/abs/2401.04075} {arXiv:2401.04075 [quant-ph]} \BibitemShut {NoStop}%
\bibitem [{\citenamefont {Pompili}\ \emph {et~al.}(2021)\citenamefont {Pompili}, \citenamefont {Hermans}, \citenamefont {Baier}, \citenamefont {Beukers}, \citenamefont {Humphreys}, \citenamefont {Schouten}, \citenamefont {Vermeulen}, \citenamefont {Tiggelman}, \citenamefont {dos Santos~Martins}, \citenamefont {Dirkse}, \citenamefont {Wehner},\ and\ \citenamefont {Hanson}}]{Pompili2021a}%
  \BibitemOpen
  \bibfield  {author} {\bibinfo {author} {\bibfnamefont {M.}~\bibnamefont {Pompili}}, \bibinfo {author} {\bibfnamefont {S.~L.}\ \bibnamefont {Hermans}}, \bibinfo {author} {\bibfnamefont {S.}~\bibnamefont {Baier}}, \bibinfo {author} {\bibfnamefont {H.~K.}\ \bibnamefont {Beukers}}, \bibinfo {author} {\bibfnamefont {P.~C.}\ \bibnamefont {Humphreys}}, \bibinfo {author} {\bibfnamefont {R.~N.}\ \bibnamefont {Schouten}}, \bibinfo {author} {\bibfnamefont {R.~F.}\ \bibnamefont {Vermeulen}}, \bibinfo {author} {\bibfnamefont {M.~J.}\ \bibnamefont {Tiggelman}}, \bibinfo {author} {\bibfnamefont {L.}~\bibnamefont {dos Santos~Martins}}, \bibinfo {author} {\bibfnamefont {B.}~\bibnamefont {Dirkse}}, \bibinfo {author} {\bibfnamefont {S.}~\bibnamefont {Wehner}},\ and\ \bibinfo {author} {\bibfnamefont {R.}~\bibnamefont {Hanson}},\ }\bibfield  {title} {\bibinfo {title} {{Realization of a multinode quantum network of remote solid-state qubits}},\ }\href {https://doi.org/10.1126/science.abg1919} {\bibfield  {journal} {\bibinfo
  {journal} {Science}\ }\textbf {\bibinfo {volume} {372}},\ \bibinfo {pages} {259} (\bibinfo {year} {2021})}\BibitemShut {NoStop}%
\bibitem [{\citenamefont {Knaut}\ \emph {et~al.}(2024)\citenamefont {Knaut}, \citenamefont {Suleymanzade}, \citenamefont {Wei}, \citenamefont {Assumpcao}, \citenamefont {Stas}, \citenamefont {Huan}, \citenamefont {Machielse}, \citenamefont {Knall}, \citenamefont {Sutula}, \citenamefont {Baranes}, \citenamefont {Sinclair}, \citenamefont {De-Eknamkul}, \citenamefont {Levonian}, \citenamefont {Bhaskar}, \citenamefont {Park}, \citenamefont {Lon{\v c}ar},\ and\ \citenamefont {Lukin}}]{Knaut2023entanglement}%
  \BibitemOpen
  \bibfield  {author} {\bibinfo {author} {\bibfnamefont {C.~M.}\ \bibnamefont {Knaut}}, \bibinfo {author} {\bibfnamefont {A.}~\bibnamefont {Suleymanzade}}, \bibinfo {author} {\bibfnamefont {Y.-C.}\ \bibnamefont {Wei}}, \bibinfo {author} {\bibfnamefont {D.~R.}\ \bibnamefont {Assumpcao}}, \bibinfo {author} {\bibfnamefont {P.-J.}\ \bibnamefont {Stas}}, \bibinfo {author} {\bibfnamefont {Y.~Q.}\ \bibnamefont {Huan}}, \bibinfo {author} {\bibfnamefont {B.}~\bibnamefont {Machielse}}, \bibinfo {author} {\bibfnamefont {E.~N.}\ \bibnamefont {Knall}}, \bibinfo {author} {\bibfnamefont {M.}~\bibnamefont {Sutula}}, \bibinfo {author} {\bibfnamefont {G.}~\bibnamefont {Baranes}}, \bibinfo {author} {\bibfnamefont {N.}~\bibnamefont {Sinclair}}, \bibinfo {author} {\bibfnamefont {C.}~\bibnamefont {De-Eknamkul}}, \bibinfo {author} {\bibfnamefont {D.~S.}\ \bibnamefont {Levonian}}, \bibinfo {author} {\bibfnamefont {M.~K.}\ \bibnamefont {Bhaskar}}, \bibinfo {author} {\bibfnamefont {H.}~\bibnamefont {Park}}, \bibinfo {author}
  {\bibfnamefont {M.}~\bibnamefont {Lon{\v c}ar}},\ and\ \bibinfo {author} {\bibfnamefont {M.~D.}\ \bibnamefont {Lukin}},\ }\bibfield  {title} {\bibinfo {title} {Entanglement of nanophotonic quantum memory nodes in a telecom network},\ }\href@noop {} {\bibfield  {journal} {\bibinfo  {journal} {Nature}\ }\textbf {\bibinfo {volume} {629}},\ \bibinfo {pages} {573} (\bibinfo {year} {2024})}\BibitemShut {NoStop}%
\bibitem [{\citenamefont {Drmota}\ \emph {et~al.}(2024)\citenamefont {Drmota}, \citenamefont {Nadlinger}, \citenamefont {Main}, \citenamefont {Nichol}, \citenamefont {Ainley}, \citenamefont {Leichtle}, \citenamefont {Mantri}, \citenamefont {Kashefi}, \citenamefont {Srinivas}, \citenamefont {Araneda}, \citenamefont {Ballance},\ and\ \citenamefont {Lucas}}]{Drmota2023verifiable}%
  \BibitemOpen
  \bibfield  {author} {\bibinfo {author} {\bibfnamefont {P.}~\bibnamefont {Drmota}}, \bibinfo {author} {\bibfnamefont {D.~P.}\ \bibnamefont {Nadlinger}}, \bibinfo {author} {\bibfnamefont {D.}~\bibnamefont {Main}}, \bibinfo {author} {\bibfnamefont {B.~C.}\ \bibnamefont {Nichol}}, \bibinfo {author} {\bibfnamefont {E.~M.}\ \bibnamefont {Ainley}}, \bibinfo {author} {\bibfnamefont {D.}~\bibnamefont {Leichtle}}, \bibinfo {author} {\bibfnamefont {A.}~\bibnamefont {Mantri}}, \bibinfo {author} {\bibfnamefont {E.}~\bibnamefont {Kashefi}}, \bibinfo {author} {\bibfnamefont {R.}~\bibnamefont {Srinivas}}, \bibinfo {author} {\bibfnamefont {G.}~\bibnamefont {Araneda}}, \bibinfo {author} {\bibfnamefont {C.~J.}\ \bibnamefont {Ballance}},\ and\ \bibinfo {author} {\bibfnamefont {D.~M.}\ \bibnamefont {Lucas}},\ }\bibfield  {title} {\bibinfo {title} {Verifiable blind quantum computing with trapped ions and single photons},\ }\href {https://doi.org/10.1103/PhysRevLett.132.150604} {\bibfield  {journal} {\bibinfo  {journal} {Phys.
  Rev. Lett.}\ }\textbf {\bibinfo {volume} {132}},\ \bibinfo {pages} {150604} (\bibinfo {year} {2024})}\BibitemShut {NoStop}%
\bibitem [{\citenamefont {Higginbottom}\ \emph {et~al.}(2022)\citenamefont {Higginbottom}, \citenamefont {Kurkjian}, \citenamefont {Chartrand}, \citenamefont {MacQuarrie}, \citenamefont {Klein}, \citenamefont {Lee-Hone}, \citenamefont {Stacho}, \citenamefont {Bowness}, \citenamefont {Bergeron}, \citenamefont {DeAbreu}, \citenamefont {Brunelle}, \citenamefont {Harrigan}, \citenamefont {Kanaganayagam}, \citenamefont {Kazemi}, \citenamefont {Marsden}, \citenamefont {Richards}, \citenamefont {Stott}, \citenamefont {Roorda}, \citenamefont {Morse}, \citenamefont {Thewalt},\ and\ \citenamefont {Simmons}}]{Higginbottom2022}%
  \BibitemOpen
  \bibfield  {author} {\bibinfo {author} {\bibfnamefont {D.~B.}\ \bibnamefont {Higginbottom}}, \bibinfo {author} {\bibfnamefont {A.~T.~K.}\ \bibnamefont {Kurkjian}}, \bibinfo {author} {\bibfnamefont {C.}~\bibnamefont {Chartrand}}, \bibinfo {author} {\bibfnamefont {E.~R.}\ \bibnamefont {MacQuarrie}}, \bibinfo {author} {\bibfnamefont {J.~R.}\ \bibnamefont {Klein}}, \bibinfo {author} {\bibfnamefont {N.~R.}\ \bibnamefont {Lee-Hone}}, \bibinfo {author} {\bibfnamefont {J.}~\bibnamefont {Stacho}}, \bibinfo {author} {\bibfnamefont {C.}~\bibnamefont {Bowness}}, \bibinfo {author} {\bibfnamefont {L.}~\bibnamefont {Bergeron}}, \bibinfo {author} {\bibfnamefont {A.}~\bibnamefont {DeAbreu}}, \bibinfo {author} {\bibfnamefont {N.~A.}\ \bibnamefont {Brunelle}}, \bibinfo {author} {\bibfnamefont {S.~R.}\ \bibnamefont {Harrigan}}, \bibinfo {author} {\bibfnamefont {J.}~\bibnamefont {Kanaganayagam}}, \bibinfo {author} {\bibfnamefont {M.}~\bibnamefont {Kazemi}}, \bibinfo {author} {\bibfnamefont {D.~W.}\ \bibnamefont {Marsden}},
  \bibinfo {author} {\bibfnamefont {T.~S.}\ \bibnamefont {Richards}}, \bibinfo {author} {\bibfnamefont {L.~A.}\ \bibnamefont {Stott}}, \bibinfo {author} {\bibfnamefont {S.}~\bibnamefont {Roorda}}, \bibinfo {author} {\bibfnamefont {K.~J.}\ \bibnamefont {Morse}}, \bibinfo {author} {\bibfnamefont {M.~L.~W.}\ \bibnamefont {Thewalt}},\ and\ \bibinfo {author} {\bibfnamefont {S.}~\bibnamefont {Simmons}},\ }\bibfield  {title} {\bibinfo {title} {{Optical observation of single spins in silicon}},\ }\href {https://doi.org/10.1038/s41586-022-04821-y} {\bibfield  {journal} {\bibinfo  {journal} {Nature}\ }\textbf {\bibinfo {volume} {607}},\ \bibinfo {pages} {266} (\bibinfo {year} {2022})}\BibitemShut {NoStop}%
\bibitem [{\citenamefont {DeAbreu}\ \emph {et~al.}(2023)\citenamefont {DeAbreu}, \citenamefont {Bowness}, \citenamefont {Alizadeh}, \citenamefont {Chartrand}, \citenamefont {Brunelle}, \citenamefont {MacQuarrie}, \citenamefont {Lee-Hone}, \citenamefont {Ruether}, \citenamefont {Kazemi}, \citenamefont {Kurkjian}, \citenamefont {Roorda}, \citenamefont {Abrosimov}, \citenamefont {Pohl}, \citenamefont {Thewalt}, \citenamefont {Higginbottom},\ and\ \citenamefont {Simmons}}]{DeAbreu2023Waveguide-integratedCentres}%
  \BibitemOpen
  \bibfield  {author} {\bibinfo {author} {\bibfnamefont {A.}~\bibnamefont {DeAbreu}}, \bibinfo {author} {\bibfnamefont {C.}~\bibnamefont {Bowness}}, \bibinfo {author} {\bibfnamefont {A.}~\bibnamefont {Alizadeh}}, \bibinfo {author} {\bibfnamefont {C.}~\bibnamefont {Chartrand}}, \bibinfo {author} {\bibfnamefont {N.~A.}\ \bibnamefont {Brunelle}}, \bibinfo {author} {\bibfnamefont {E.~R.}\ \bibnamefont {MacQuarrie}}, \bibinfo {author} {\bibfnamefont {N.~R.}\ \bibnamefont {Lee-Hone}}, \bibinfo {author} {\bibfnamefont {M.}~\bibnamefont {Ruether}}, \bibinfo {author} {\bibfnamefont {M.}~\bibnamefont {Kazemi}}, \bibinfo {author} {\bibfnamefont {A.~T.~K.}\ \bibnamefont {Kurkjian}}, \bibinfo {author} {\bibfnamefont {S.}~\bibnamefont {Roorda}}, \bibinfo {author} {\bibfnamefont {N.~V.}\ \bibnamefont {Abrosimov}}, \bibinfo {author} {\bibfnamefont {H.-J.}\ \bibnamefont {Pohl}}, \bibinfo {author} {\bibfnamefont {M.~L.~W.}\ \bibnamefont {Thewalt}}, \bibinfo {author} {\bibfnamefont {D.~B.}\ \bibnamefont {Higginbottom}},\ and\
  \bibinfo {author} {\bibfnamefont {S.}~\bibnamefont {Simmons}},\ }\bibfield  {title} {\bibinfo {title} {{Waveguide-integrated silicon T centres}},\ }\href {https://doi.org/10.1364/oe.482008} {\bibfield  {journal} {\bibinfo  {journal} {Optics Express}\ }\textbf {\bibinfo {volume} {31}},\ \bibinfo {pages} {15045} (\bibinfo {year} {2023})}\BibitemShut {NoStop}%
\bibitem [{\citenamefont {Feng}\ \emph {et~al.}(2022)\citenamefont {Feng}, \citenamefont {Feng}, \citenamefont {Feng}, \citenamefont {Zhang}, \citenamefont {Wang}, \citenamefont {Wang}, \citenamefont {Zhou}, \citenamefont {Qiang}, \citenamefont {Guo}, \citenamefont {Guo}, \citenamefont {Guo}, \citenamefont {Ren}, \citenamefont {Ren},\ and\ \citenamefont {Ren}}]{Feng2022}%
  \BibitemOpen
  \bibfield  {author} {\bibinfo {author} {\bibfnamefont {L.}~\bibnamefont {Feng}}, \bibinfo {author} {\bibfnamefont {L.}~\bibnamefont {Feng}}, \bibinfo {author} {\bibfnamefont {L.}~\bibnamefont {Feng}}, \bibinfo {author} {\bibfnamefont {M.}~\bibnamefont {Zhang}}, \bibinfo {author} {\bibfnamefont {J.}~\bibnamefont {Wang}}, \bibinfo {author} {\bibfnamefont {J.}~\bibnamefont {Wang}}, \bibinfo {author} {\bibfnamefont {X.}~\bibnamefont {Zhou}}, \bibinfo {author} {\bibfnamefont {X.}~\bibnamefont {Qiang}}, \bibinfo {author} {\bibfnamefont {G.}~\bibnamefont {Guo}}, \bibinfo {author} {\bibfnamefont {G.}~\bibnamefont {Guo}}, \bibinfo {author} {\bibfnamefont {G.}~\bibnamefont {Guo}}, \bibinfo {author} {\bibfnamefont {X.}~\bibnamefont {Ren}}, \bibinfo {author} {\bibfnamefont {X.}~\bibnamefont {Ren}},\ and\ \bibinfo {author} {\bibfnamefont {X.}~\bibnamefont {Ren}},\ }\bibfield  {title} {\bibinfo {title} {{Silicon photonic devices for scalable quantum information applications}},\ }\href {https://doi.org/10.1364/PRJ.464808}
  {\bibfield  {journal} {\bibinfo  {journal} {Photonics Research}\ }\textbf {\bibinfo {volume} {10}},\ \bibinfo {pages} {A135} (\bibinfo {year} {2022})}\BibitemShut {NoStop}%
\bibitem [{\citenamefont {White}\ \emph {et~al.}(2020)\citenamefont {White}, \citenamefont {Lukin}, \citenamefont {Guidry}, \citenamefont {Trivedi}, \citenamefont {Morioka}, \citenamefont {Babin}, \citenamefont {Kaiser}, \citenamefont {Ul-Hassan}, \citenamefont {Son}, \citenamefont {Ohshima}, \citenamefont {Vasireddy}, \citenamefont {Nasr}, \citenamefont {Nanni}, \citenamefont {Wrachtrup},\ and\ \citenamefont {Vučković}}]{White_2020}%
  \BibitemOpen
  \bibfield  {author} {\bibinfo {author} {\bibfnamefont {A.~D.}\ \bibnamefont {White}}, \bibinfo {author} {\bibfnamefont {D.~M.}\ \bibnamefont {Lukin}}, \bibinfo {author} {\bibfnamefont {M.~A.}\ \bibnamefont {Guidry}}, \bibinfo {author} {\bibfnamefont {R.}~\bibnamefont {Trivedi}}, \bibinfo {author} {\bibfnamefont {N.}~\bibnamefont {Morioka}}, \bibinfo {author} {\bibfnamefont {C.}~\bibnamefont {Babin}}, \bibinfo {author} {\bibfnamefont {F.}~\bibnamefont {Kaiser}}, \bibinfo {author} {\bibfnamefont {J.}~\bibnamefont {Ul-Hassan}}, \bibinfo {author} {\bibfnamefont {N.~T.}\ \bibnamefont {Son}}, \bibinfo {author} {\bibfnamefont {T.}~\bibnamefont {Ohshima}}, \bibinfo {author} {\bibfnamefont {P.}~\bibnamefont {Vasireddy}}, \bibinfo {author} {\bibfnamefont {M.}~\bibnamefont {Nasr}}, \bibinfo {author} {\bibfnamefont {E.}~\bibnamefont {Nanni}}, \bibinfo {author} {\bibfnamefont {J.}~\bibnamefont {Wrachtrup}},\ and\ \bibinfo {author} {\bibfnamefont {J.}~\bibnamefont {Vučković}},\ }\bibfield  {title} {\bibinfo {title}
  {Static and dynamic stark tuning of the silicon vacancy in silicon carbide},\ }in\ \href@noop {} {\emph {\bibinfo {booktitle} {Conference on Lasers and Electro-Optics (CLEO)}}}\ (\bibinfo {year} {2020})\ pp.\ \bibinfo {pages} {1--2}\BibitemShut {NoStop}%
\bibitem [{\citenamefont {Anderson}\ \emph {et~al.}(2022)\citenamefont {Anderson}, \citenamefont {Glen}, \citenamefont {Zeledon}, \citenamefont {Bourassa}, \citenamefont {Jin}, \citenamefont {Zhu}, \citenamefont {Vorwerk}, \citenamefont {Crook}, \citenamefont {Abe}, \citenamefont {Ul-Hassan}, \citenamefont {Ohshima}, \citenamefont {Son}, \citenamefont {Galli},\ and\ \citenamefont {Awschalom}}]{Anderson2022_singleshot}%
  \BibitemOpen
  \bibfield  {author} {\bibinfo {author} {\bibfnamefont {C.~P.}\ \bibnamefont {Anderson}}, \bibinfo {author} {\bibfnamefont {E.~O.}\ \bibnamefont {Glen}}, \bibinfo {author} {\bibfnamefont {C.}~\bibnamefont {Zeledon}}, \bibinfo {author} {\bibfnamefont {A.}~\bibnamefont {Bourassa}}, \bibinfo {author} {\bibfnamefont {Y.}~\bibnamefont {Jin}}, \bibinfo {author} {\bibfnamefont {Y.}~\bibnamefont {Zhu}}, \bibinfo {author} {\bibfnamefont {C.}~\bibnamefont {Vorwerk}}, \bibinfo {author} {\bibfnamefont {A.~L.}\ \bibnamefont {Crook}}, \bibinfo {author} {\bibfnamefont {H.}~\bibnamefont {Abe}}, \bibinfo {author} {\bibfnamefont {J.}~\bibnamefont {Ul-Hassan}}, \bibinfo {author} {\bibfnamefont {T.}~\bibnamefont {Ohshima}}, \bibinfo {author} {\bibfnamefont {N.~T.}\ \bibnamefont {Son}}, \bibinfo {author} {\bibfnamefont {G.}~\bibnamefont {Galli}},\ and\ \bibinfo {author} {\bibfnamefont {D.~D.}\ \bibnamefont {Awschalom}},\ }\bibfield  {title} {\bibinfo {title} {Five-second coherence of a single spin with single-shot readout in
  silicon carbide},\ }\href {https://doi.org/10.1126/sciadv.abm5912} {\bibfield  {journal} {\bibinfo  {journal} {Science Advances}\ }\textbf {\bibinfo {volume} {8}},\ \bibinfo {pages} {eabm5912} (\bibinfo {year} {2022})}\BibitemShut {NoStop}%
\bibitem [{\citenamefont {Barrett}\ and\ \citenamefont {Kok}(2005)}]{Barrett2005}%
  \BibitemOpen
  \bibfield  {author} {\bibinfo {author} {\bibfnamefont {S.~D.}\ \bibnamefont {Barrett}}\ and\ \bibinfo {author} {\bibfnamefont {P.}~\bibnamefont {Kok}},\ }\bibfield  {title} {\bibinfo {title} {{Efficient high-fidelity quantum computation using matter qubits and linear optics}},\ }\href {https://doi.org/10.1103/PhysRevA.71.060310} {\bibfield  {journal} {\bibinfo  {journal} {Physical Review A}\ }\textbf {\bibinfo {volume} {71}},\ \bibinfo {pages} {060310} (\bibinfo {year} {2005})}\BibitemShut {NoStop}%
\bibitem [{\citenamefont {Davies}(1989)}]{davies1989}%
  \BibitemOpen
  \bibfield  {author} {\bibinfo {author} {\bibfnamefont {G.}~\bibnamefont {Davies}},\ }\bibfield  {title} {\bibinfo {title} {The optical properties of luminescence centres in silicon},\ }\href {https://doi.org/https://doi.org/10.1016/0370-1573(89)90064-1} {\bibfield  {journal} {\bibinfo  {journal} {Physics Reports}\ }\textbf {\bibinfo {volume} {176}},\ \bibinfo {pages} {83} (\bibinfo {year} {1989})}\BibitemShut {NoStop}%
\bibitem [{\citenamefont {Chartrand}\ \emph {et~al.}(2018)\citenamefont {Chartrand}, \citenamefont {Bergeron}, \citenamefont {Morse}, \citenamefont {Riemann}, \citenamefont {Abrosimov}, \citenamefont {Becker}, \citenamefont {Pohl}, \citenamefont {Simmons},\ and\ \citenamefont {Thewalt}}]{Chartrand2018}%
  \BibitemOpen
  \bibfield  {author} {\bibinfo {author} {\bibfnamefont {C.}~\bibnamefont {Chartrand}}, \bibinfo {author} {\bibfnamefont {L.}~\bibnamefont {Bergeron}}, \bibinfo {author} {\bibfnamefont {K.~J.}\ \bibnamefont {Morse}}, \bibinfo {author} {\bibfnamefont {H.}~\bibnamefont {Riemann}}, \bibinfo {author} {\bibfnamefont {N.~V.}\ \bibnamefont {Abrosimov}}, \bibinfo {author} {\bibfnamefont {P.}~\bibnamefont {Becker}}, \bibinfo {author} {\bibfnamefont {H.-J.}\ \bibnamefont {Pohl}}, \bibinfo {author} {\bibfnamefont {S.}~\bibnamefont {Simmons}},\ and\ \bibinfo {author} {\bibfnamefont {M.~L.~W.}\ \bibnamefont {Thewalt}},\ }\bibfield  {title} {\bibinfo {title} {{Highly enriched Si 28 reveals remarkable optical linewidths and fine structure for well-known damage centers}},\ }\href {https://doi.org/10.1103/PhysRevB.98.195201} {\bibfield  {journal} {\bibinfo  {journal} {Physical Review B}\ }\textbf {\bibinfo {volume} {98}},\ \bibinfo {pages} {195201} (\bibinfo {year} {2018})}\BibitemShut {NoStop}%
\bibitem [{\citenamefont {Redjem}\ \emph {et~al.}(2020)\citenamefont {Redjem}, \citenamefont {Durand}, \citenamefont {Herzig}, \citenamefont {Benali}, \citenamefont {Pezzagna}, \citenamefont {Meijer}, \citenamefont {Kuznetsov}, \citenamefont {Nguyen}, \citenamefont {Cueff}, \citenamefont {G{\'{e}}rard}, \citenamefont {Robert-Philip}, \citenamefont {Gil}, \citenamefont {Caliste}, \citenamefont {Pochet}, \citenamefont {Abbarchi}, \citenamefont {Jacques}, \citenamefont {Dr{\'{e}}au},\ and\ \citenamefont {Cassabois}}]{Redjem2020}%
  \BibitemOpen
  \bibfield  {author} {\bibinfo {author} {\bibfnamefont {W.}~\bibnamefont {Redjem}}, \bibinfo {author} {\bibfnamefont {A.}~\bibnamefont {Durand}}, \bibinfo {author} {\bibfnamefont {T.}~\bibnamefont {Herzig}}, \bibinfo {author} {\bibfnamefont {A.}~\bibnamefont {Benali}}, \bibinfo {author} {\bibfnamefont {S.}~\bibnamefont {Pezzagna}}, \bibinfo {author} {\bibfnamefont {J.}~\bibnamefont {Meijer}}, \bibinfo {author} {\bibfnamefont {A.~Y.}\ \bibnamefont {Kuznetsov}}, \bibinfo {author} {\bibfnamefont {H.~S.}\ \bibnamefont {Nguyen}}, \bibinfo {author} {\bibfnamefont {S.}~\bibnamefont {Cueff}}, \bibinfo {author} {\bibfnamefont {J.-M.}\ \bibnamefont {G{\'{e}}rard}}, \bibinfo {author} {\bibfnamefont {I.}~\bibnamefont {Robert-Philip}}, \bibinfo {author} {\bibfnamefont {B.}~\bibnamefont {Gil}}, \bibinfo {author} {\bibfnamefont {D.}~\bibnamefont {Caliste}}, \bibinfo {author} {\bibfnamefont {P.}~\bibnamefont {Pochet}}, \bibinfo {author} {\bibfnamefont {M.}~\bibnamefont {Abbarchi}}, \bibinfo {author} {\bibfnamefont
  {V.}~\bibnamefont {Jacques}}, \bibinfo {author} {\bibfnamefont {A.}~\bibnamefont {Dr{\'{e}}au}},\ and\ \bibinfo {author} {\bibfnamefont {G.}~\bibnamefont {Cassabois}},\ }\bibfield  {title} {\bibinfo {title} {{Single artificial atoms in silicon emitting at telecom wavelengths}},\ }\href {https://doi.org/10.1038/s41928-020-00499-0} {\bibfield  {journal} {\bibinfo  {journal} {Nature Electronics}\ }\textbf {\bibinfo {volume} {3}},\ \bibinfo {pages} {738–743} (\bibinfo {year} {2020})}\BibitemShut {NoStop}%
\bibitem [{\citenamefont {Udvarhelyi}\ \emph {et~al.}(2021)\citenamefont {Udvarhelyi}, \citenamefont {Somogyi}, \citenamefont {Thiering},\ and\ \citenamefont {Gali}}]{Udvarhelyi2021}%
  \BibitemOpen
  \bibfield  {author} {\bibinfo {author} {\bibfnamefont {P.}~\bibnamefont {Udvarhelyi}}, \bibinfo {author} {\bibfnamefont {B.}~\bibnamefont {Somogyi}}, \bibinfo {author} {\bibfnamefont {G.}~\bibnamefont {Thiering}},\ and\ \bibinfo {author} {\bibfnamefont {A.}~\bibnamefont {Gali}},\ }\bibfield  {title} {\bibinfo {title} {{Identification of a Telecom Wavelength Single Photon Emitter in Silicon}},\ }\href {https://doi.org/10.1103/PHYSREVLETT.127.196402/FIGURES/2/MEDIUM} {\bibfield  {journal} {\bibinfo  {journal} {Physical Review Letters}\ }\textbf {\bibinfo {volume} {127}},\ \bibinfo {pages} {196402} (\bibinfo {year} {2021})}\BibitemShut {NoStop}%
\bibitem [{\citenamefont {Ivanov}\ \emph {et~al.}(2022)\citenamefont {Ivanov}, \citenamefont {Simoni}, \citenamefont {Lee}, \citenamefont {Liu}, \citenamefont {Jhuria}, \citenamefont {Redjem}, \citenamefont {Zhiyenbayev}, \citenamefont {Papapanos}, \citenamefont {Qarony}, \citenamefont {Kant\'e}, \citenamefont {Persaud}, \citenamefont {Schenkel},\ and\ \citenamefont {Tan}}]{Ivanov2022}%
  \BibitemOpen
  \bibfield  {author} {\bibinfo {author} {\bibfnamefont {V.}~\bibnamefont {Ivanov}}, \bibinfo {author} {\bibfnamefont {J.}~\bibnamefont {Simoni}}, \bibinfo {author} {\bibfnamefont {Y.}~\bibnamefont {Lee}}, \bibinfo {author} {\bibfnamefont {W.}~\bibnamefont {Liu}}, \bibinfo {author} {\bibfnamefont {K.}~\bibnamefont {Jhuria}}, \bibinfo {author} {\bibfnamefont {W.}~\bibnamefont {Redjem}}, \bibinfo {author} {\bibfnamefont {Y.}~\bibnamefont {Zhiyenbayev}}, \bibinfo {author} {\bibfnamefont {C.}~\bibnamefont {Papapanos}}, \bibinfo {author} {\bibfnamefont {W.}~\bibnamefont {Qarony}}, \bibinfo {author} {\bibfnamefont {B.}~\bibnamefont {Kant\'e}}, \bibinfo {author} {\bibfnamefont {A.}~\bibnamefont {Persaud}}, \bibinfo {author} {\bibfnamefont {T.}~\bibnamefont {Schenkel}},\ and\ \bibinfo {author} {\bibfnamefont {L.~Z.}\ \bibnamefont {Tan}},\ }\bibfield  {title} {\bibinfo {title} {Effect of localization on photoluminescence and zero-field splitting of silicon color centers},\ }\href
  {https://doi.org/10.1103/PhysRevB.106.134107} {\bibfield  {journal} {\bibinfo  {journal} {Phys. Rev. B}\ }\textbf {\bibinfo {volume} {106}},\ \bibinfo {pages} {134107} (\bibinfo {year} {2022})}\BibitemShut {NoStop}%
\bibitem [{\citenamefont {Baron}\ \emph {et~al.}(2022)\citenamefont {Baron}, \citenamefont {Durand}, \citenamefont {Udvarhelyi}, \citenamefont {Herzig}, \citenamefont {Khoury}, \citenamefont {Pezzagna}, \citenamefont {Meijer}, \citenamefont {Robert-Philip}, \citenamefont {Abbarchi}, \citenamefont {Hartmann}, \citenamefont {Mazzocchi}, \citenamefont {G{\'{e}}rard}, \citenamefont {Gali}, \citenamefont {Jacques}, \citenamefont {Cassabois},\ and\ \citenamefont {Dr{\'{e}}au}}]{Baron2022}%
  \BibitemOpen
  \bibfield  {author} {\bibinfo {author} {\bibfnamefont {Y.}~\bibnamefont {Baron}}, \bibinfo {author} {\bibfnamefont {A.}~\bibnamefont {Durand}}, \bibinfo {author} {\bibfnamefont {P.}~\bibnamefont {Udvarhelyi}}, \bibinfo {author} {\bibfnamefont {T.}~\bibnamefont {Herzig}}, \bibinfo {author} {\bibfnamefont {M.}~\bibnamefont {Khoury}}, \bibinfo {author} {\bibfnamefont {S.}~\bibnamefont {Pezzagna}}, \bibinfo {author} {\bibfnamefont {J.}~\bibnamefont {Meijer}}, \bibinfo {author} {\bibfnamefont {I.}~\bibnamefont {Robert-Philip}}, \bibinfo {author} {\bibfnamefont {M.}~\bibnamefont {Abbarchi}}, \bibinfo {author} {\bibfnamefont {J.-M.}\ \bibnamefont {Hartmann}}, \bibinfo {author} {\bibfnamefont {V.}~\bibnamefont {Mazzocchi}}, \bibinfo {author} {\bibfnamefont {J.-M.}\ \bibnamefont {G{\'{e}}rard}}, \bibinfo {author} {\bibfnamefont {A.}~\bibnamefont {Gali}}, \bibinfo {author} {\bibfnamefont {V.}~\bibnamefont {Jacques}}, \bibinfo {author} {\bibfnamefont {G.}~\bibnamefont {Cassabois}},\ and\ \bibinfo {author}
  {\bibfnamefont {A.}~\bibnamefont {Dr{\'{e}}au}},\ }\bibfield  {title} {\bibinfo {title} {{Detection of Single W-Centers in Silicon}},\ }\href {https://doi.org/10.1021/acsphotonics.2c00336} {\bibfield  {journal} {\bibinfo  {journal} {ACS Photonics}\ }\textbf {\bibinfo {volume} {9}},\ \bibinfo {pages} {2337} (\bibinfo {year} {2022})}\BibitemShut {NoStop}%
\bibitem [{\citenamefont {Safonov}\ \emph {et~al.}(1995)\citenamefont {Safonov}, \citenamefont {Lightowlers},\ and\ \citenamefont {Davies}}]{Safonov1995}%
  \BibitemOpen
  \bibfield  {author} {\bibinfo {author} {\bibfnamefont {A.~N.}\ \bibnamefont {Safonov}}, \bibinfo {author} {\bibfnamefont {E.~C.}\ \bibnamefont {Lightowlers}},\ and\ \bibinfo {author} {\bibfnamefont {G.}~\bibnamefont {Davies}},\ }\bibfield  {title} {\bibinfo {title} {{Carbon-hydrogen deep level luminescence centre in silicon responsible for the T-line}},\ }\href {https://doi.org/10.4028/www.scientific.net/msf.196-201.909} {\bibfield  {journal} {\bibinfo  {journal} {Materials Science Forum}\ }\textbf {\bibinfo {volume} {196-201}},\ \bibinfo {pages} {909} (\bibinfo {year} {1995})}\BibitemShut {NoStop}%
\bibitem [{\citenamefont {Safonov}\ and\ \citenamefont {Lightowlers}(1999)}]{Safonov1999a}%
  \BibitemOpen
  \bibfield  {author} {\bibinfo {author} {\bibfnamefont {A.~N.}\ \bibnamefont {Safonov}}\ and\ \bibinfo {author} {\bibfnamefont {E.~C.}\ \bibnamefont {Lightowlers}},\ }\bibfield  {title} {\bibinfo {title} {{Photoluminescence characterisation of hydrogen-related centres in silicon}},\ }\href {http://linkinghub.elsevier.com/retrieve/pii/S0921510798002724 papers3://publication/doi/10.1016/S0921-5107(98)00272-4} {\bibfield  {journal} {\bibinfo  {journal} {Materials Science and Engineering: B}\ }\textbf {\bibinfo {volume} {58}},\ \bibinfo {pages} {39} (\bibinfo {year} {1999})}\BibitemShut {NoStop}%
\bibitem [{\citenamefont {Kambs}\ and\ \citenamefont {Becher}(2018)}]{Kambs2018}%
  \BibitemOpen
  \bibfield  {author} {\bibinfo {author} {\bibfnamefont {B.}~\bibnamefont {Kambs}}\ and\ \bibinfo {author} {\bibfnamefont {C.}~\bibnamefont {Becher}},\ }\bibfield  {title} {\bibinfo {title} {{Limitations on the indistinguishability of photons from remote solid state sources}},\ }\href {https://doi.org/10.1088/1367-2630/aaea99} {\bibfield  {journal} {\bibinfo  {journal} {New Journal of Physics}\ }\textbf {\bibinfo {volume} {20}},\ \bibinfo {pages} {115003} (\bibinfo {year} {2018})}\BibitemShut {NoStop}%
\bibitem [{\citenamefont {Reiserer}(2022)}]{Cavity_enhanced_QN}%
  \BibitemOpen
  \bibfield  {author} {\bibinfo {author} {\bibfnamefont {A.}~\bibnamefont {Reiserer}},\ }\bibfield  {title} {\bibinfo {title} {Colloquium: Cavity-enhanced quantum network nodes},\ }\href {https://doi.org/10.1103/RevModPhys.94.041003} {\bibfield  {journal} {\bibinfo  {journal} {Rev. Mod. Phys.}\ }\textbf {\bibinfo {volume} {94}},\ \bibinfo {pages} {041003} (\bibinfo {year} {2022})}\BibitemShut {NoStop}%
\bibitem [{\citenamefont {Safonov}\ \emph {et~al.}(1996)\citenamefont {Safonov}, \citenamefont {Lightowlers}, \citenamefont {Davies}, \citenamefont {Leary}, \citenamefont {Jones},\ and\ \citenamefont {{\"{O}}berg}}]{Safonov1996b}%
  \BibitemOpen
  \bibfield  {author} {\bibinfo {author} {\bibfnamefont {A.~N.}\ \bibnamefont {Safonov}}, \bibinfo {author} {\bibfnamefont {E.~C.}\ \bibnamefont {Lightowlers}}, \bibinfo {author} {\bibfnamefont {G.}~\bibnamefont {Davies}}, \bibinfo {author} {\bibfnamefont {P.}~\bibnamefont {Leary}}, \bibinfo {author} {\bibfnamefont {R.}~\bibnamefont {Jones}},\ and\ \bibinfo {author} {\bibfnamefont {S.}~\bibnamefont {{\"{O}}berg}},\ }\bibfield  {title} {\bibinfo {title} {{Interstitial-Carbon Hydrogen Interaction in Silicon}},\ }\href {https://link.aps.org/doi/10.1103/PhysRevLett.77.4812 papers3://publication/doi/10.1103/PhysRevLett.77.4812} {\bibfield  {journal} {\bibinfo  {journal} {Phys. Rev. Lett.}\ }\textbf {\bibinfo {volume} {77}},\ \bibinfo {pages} {4812} (\bibinfo {year} {1996})}\BibitemShut {NoStop}%
\bibitem [{\citenamefont {Brink}\ \emph {et~al.}(1962)\citenamefont {Brink}, \citenamefont {Brink},\ and\ \citenamefont {Satchler}}]{brink1962angular}%
  \BibitemOpen
  \bibfield  {author} {\bibinfo {author} {\bibfnamefont {D.}~\bibnamefont {Brink}}, \bibinfo {author} {\bibfnamefont {D.}~\bibnamefont {Brink}},\ and\ \bibinfo {author} {\bibfnamefont {G.}~\bibnamefont {Satchler}},\ }\href {https://books.google.ca/books?id=rssNAQAAIAAJ} {\emph {\bibinfo {title} {Angular Momentum}}},\ Oxford library of the physical sciences\ (\bibinfo  {publisher} {Clarendon Press},\ \bibinfo {year} {1962})\BibitemShut {NoStop}%
\bibitem [{\citenamefont {Bir}\ and\ \citenamefont {Pikus}(1974)}]{bir1974symmetry}%
  \BibitemOpen
  \bibfield  {author} {\bibinfo {author} {\bibfnamefont {G.}~\bibnamefont {Bir}}\ and\ \bibinfo {author} {\bibfnamefont {G.}~\bibnamefont {Pikus}},\ }\href {https://books.google.ca/books?id=38m2QgAACAAJ} {\emph {\bibinfo {title} {Symmetry and Strain-induced Effects in Semiconductors}}},\ A Halsted Press book\ (\bibinfo  {publisher} {Wiley},\ \bibinfo {year} {1974})\BibitemShut {NoStop}%
\bibitem [{\citenamefont {Bir}\ \emph {et~al.}(1963{\natexlab{a}})\citenamefont {Bir}, \citenamefont {Butikov},\ and\ \citenamefont {Pikus}}]{BIR1963_1}%
  \BibitemOpen
  \bibfield  {author} {\bibinfo {author} {\bibfnamefont {G.}~\bibnamefont {Bir}}, \bibinfo {author} {\bibfnamefont {E.}~\bibnamefont {Butikov}},\ and\ \bibinfo {author} {\bibfnamefont {G.}~\bibnamefont {Pikus}},\ }\bibfield  {title} {\bibinfo {title} {Spin and combined resonance on acceptor centres in ge and si type crystals—i: Paramagnetic resonance in strained and unstrained crystals},\ }\href {https://doi.org/https://doi.org/10.1016/0022-3697(63)90086-6} {\bibfield  {journal} {\bibinfo  {journal} {Journal of Physics and Chemistry of Solids}\ }\textbf {\bibinfo {volume} {24}},\ \bibinfo {pages} {1467} (\bibinfo {year} {1963}{\natexlab{a}})}\BibitemShut {NoStop}%
\bibitem [{\citenamefont {Bhattacharjee}\ and\ \citenamefont {Rodriguez}(1972)}]{Bhattacharjee_1972}%
  \BibitemOpen
  \bibfield  {author} {\bibinfo {author} {\bibfnamefont {A.~K.}\ \bibnamefont {Bhattacharjee}}\ and\ \bibinfo {author} {\bibfnamefont {S.}~\bibnamefont {Rodriguez}},\ }\bibfield  {title} {\bibinfo {title} {Group-theoretical study of the zeeman effect of acceptors in silicon and germanium},\ }\href {https://doi.org/10.1103/PhysRevB.6.3836} {\bibfield  {journal} {\bibinfo  {journal} {Phys. Rev. B}\ }\textbf {\bibinfo {volume} {6}},\ \bibinfo {pages} {3836} (\bibinfo {year} {1972})}\BibitemShut {NoStop}%
\bibitem [{\citenamefont {Bir}\ \emph {et~al.}(1963{\natexlab{b}})\citenamefont {Bir}, \citenamefont {Butikov},\ and\ \citenamefont {Pikus}}]{BIR1963_2}%
  \BibitemOpen
  \bibfield  {author} {\bibinfo {author} {\bibfnamefont {G.}~\bibnamefont {Bir}}, \bibinfo {author} {\bibfnamefont {E.}~\bibnamefont {Butikov}},\ and\ \bibinfo {author} {\bibfnamefont {G.}~\bibnamefont {Pikus}},\ }\bibfield  {title} {\bibinfo {title} {Spin and combined resonance on acceptor centres in ge and si type crystals—ii: The effect of the electrical field and relaxation time},\ }\href {https://doi.org/https://doi.org/10.1016/0022-3697(63)90087-8} {\bibfield  {journal} {\bibinfo  {journal} {Journal of Physics and Chemistry of Solids}\ }\textbf {\bibinfo {volume} {24}},\ \bibinfo {pages} {1475} (\bibinfo {year} {1963}{\natexlab{b}})}\BibitemShut {NoStop}%
\bibitem [{\citenamefont {{A A Kaplyanskii}}(1964)}]{AAKaplyanskii1964}%
  \BibitemOpen
  \bibfield  {author} {\bibinfo {author} {\bibnamefont {{A A Kaplyanskii}}},\ }\bibfield  {title} {\bibinfo {title} {{Noncubic Centers in Cubic Crystals and Their Piezospectroscopic Study}},\ }\href@noop {} {\bibfield  {journal} {\bibinfo  {journal} {Optika i spectroskopija}\ }\textbf {\bibinfo {volume} {16}},\ \bibinfo {pages} {602} (\bibinfo {year} {1964})}\BibitemShut {NoStop}%
\bibitem [{\citenamefont {Tamarat}\ \emph {et~al.}(2006)\citenamefont {Tamarat}, \citenamefont {Gaebel}, \citenamefont {Rabeau}, \citenamefont {Khan}, \citenamefont {Greentree}, \citenamefont {Wilson}, \citenamefont {Hollenberg}, \citenamefont {Prawer}, \citenamefont {Hemmer}, \citenamefont {Jelezko},\ and\ \citenamefont {Wrachtrup}}]{PhysRevLett.97.083002}%
  \BibitemOpen
  \bibfield  {author} {\bibinfo {author} {\bibfnamefont {P.}~\bibnamefont {Tamarat}}, \bibinfo {author} {\bibfnamefont {T.}~\bibnamefont {Gaebel}}, \bibinfo {author} {\bibfnamefont {J.~R.}\ \bibnamefont {Rabeau}}, \bibinfo {author} {\bibfnamefont {M.}~\bibnamefont {Khan}}, \bibinfo {author} {\bibfnamefont {A.~D.}\ \bibnamefont {Greentree}}, \bibinfo {author} {\bibfnamefont {H.}~\bibnamefont {Wilson}}, \bibinfo {author} {\bibfnamefont {L.~C.~L.}\ \bibnamefont {Hollenberg}}, \bibinfo {author} {\bibfnamefont {S.}~\bibnamefont {Prawer}}, \bibinfo {author} {\bibfnamefont {P.}~\bibnamefont {Hemmer}}, \bibinfo {author} {\bibfnamefont {F.}~\bibnamefont {Jelezko}},\ and\ \bibinfo {author} {\bibfnamefont {J.}~\bibnamefont {Wrachtrup}},\ }\bibfield  {title} {\bibinfo {title} {Stark shift control of single optical centers in diamond},\ }\href {https://doi.org/10.1103/PhysRevLett.97.083002} {\bibfield  {journal} {\bibinfo  {journal} {Phys. Rev. Lett.}\ }\textbf {\bibinfo {volume} {97}},\ \bibinfo {pages} {083002}
  (\bibinfo {year} {2006})}\BibitemShut {NoStop}%
\bibitem [{\citenamefont {Kaplyanskii}(1967)}]{Kaplyanskii1967NONCUBICCI}%
  \BibitemOpen
  \bibfield  {author} {\bibinfo {author} {\bibfnamefont {A.~A.}\ \bibnamefont {Kaplyanskii}},\ }\bibfield  {title} {\bibinfo {title} {Noncubic centers in cubic crystals and their spectra in external fields},\ }\href@noop {} {\bibfield  {journal} {\bibinfo  {journal} {Le Journal De Physique Colloques}\ }\textbf {\bibinfo {volume} {28}} (\bibinfo {year} {1967})}\BibitemShut {NoStop}%
\bibitem [{\citenamefont {Powell}(2010)}]{Powell2010SymmetryGT}%
  \BibitemOpen
  \bibfield  {author} {\bibinfo {author} {\bibfnamefont {R.~C.}\ \bibnamefont {Powell}},\ }\href@noop {} {\emph {\bibinfo {title} {Symmetry, group theory, and the physical properties of crystals}}},\ Vol.\ \bibinfo {volume} {824}\ (\bibinfo  {publisher} {Springer},\ \bibinfo {year} {2010})\BibitemShut {NoStop}%
\bibitem [{\citenamefont {Bonin}\ and\ \citenamefont {Kresin}(1997)}]{bonin1997electric}%
  \BibitemOpen
  \bibfield  {author} {\bibinfo {author} {\bibfnamefont {K.}~\bibnamefont {Bonin}}\ and\ \bibinfo {author} {\bibfnamefont {V.}~\bibnamefont {Kresin}},\ }\href {https://books.google.ca/books?id=1E8M117QOFEC} {\emph {\bibinfo {title} {Electric-dipole Polarizabilities of Atoms, Molecules, and Clusters}}}\ (\bibinfo  {publisher} {World Scientific},\ \bibinfo {year} {1997})\BibitemShut {NoStop}%
\bibitem [{\citenamefont {Becker}\ \emph {et~al.}(2010)\citenamefont {Becker}, \citenamefont {Pohl}, \citenamefont {Riemann},\ and\ \citenamefont {Abrosimov}}]{Chai_becker2010enrichment}%
  \BibitemOpen
  \bibfield  {author} {\bibinfo {author} {\bibfnamefont {P.}~\bibnamefont {Becker}}, \bibinfo {author} {\bibfnamefont {H.-J.}\ \bibnamefont {Pohl}}, \bibinfo {author} {\bibfnamefont {H.}~\bibnamefont {Riemann}},\ and\ \bibinfo {author} {\bibfnamefont {N.}~\bibnamefont {Abrosimov}},\ }\bibfield  {title} {\bibinfo {title} {Enrichment of silicon for a better kilogram},\ }\href@noop {} {\bibfield  {journal} {\bibinfo  {journal} {physica status solidi (a)}\ }\textbf {\bibinfo {volume} {207}},\ \bibinfo {pages} {49} (\bibinfo {year} {2010})}\BibitemShut {NoStop}%
\bibitem [{\citenamefont {Higginbottom}\ \emph {et~al.}(2023)\citenamefont {Higginbottom}, \citenamefont {Asadi}, \citenamefont {Chartrand}, \citenamefont {Ji}, \citenamefont {Bergeron}, \citenamefont {Thewalt}, \citenamefont {Simon},\ and\ \citenamefont {Simmons}}]{Higginbottom2023memory}%
  \BibitemOpen
  \bibfield  {author} {\bibinfo {author} {\bibfnamefont {D.~B.}\ \bibnamefont {Higginbottom}}, \bibinfo {author} {\bibfnamefont {F.~K.}\ \bibnamefont {Asadi}}, \bibinfo {author} {\bibfnamefont {C.}~\bibnamefont {Chartrand}}, \bibinfo {author} {\bibfnamefont {J.-W.}\ \bibnamefont {Ji}}, \bibinfo {author} {\bibfnamefont {L.}~\bibnamefont {Bergeron}}, \bibinfo {author} {\bibfnamefont {M.~L.}\ \bibnamefont {Thewalt}}, \bibinfo {author} {\bibfnamefont {C.}~\bibnamefont {Simon}},\ and\ \bibinfo {author} {\bibfnamefont {S.}~\bibnamefont {Simmons}},\ }\bibfield  {title} {\bibinfo {title} {{Memory and transduction prospects for silicon T center devices}},\ }\href {https://doi.org/10.1103/PRXQuantum.4.020308} {\bibfield  {journal} {\bibinfo  {journal} {PRX Quantum}\ }\textbf {\bibinfo {volume} {4}},\ \bibinfo {pages} {020308} (\bibinfo {year} {2023})}\BibitemShut {NoStop}%
\bibitem [{\citenamefont {Johnston}\ \emph {et~al.}(2024)\citenamefont {Johnston}, \citenamefont {Felix-Rendon}, \citenamefont {Wong},\ and\ \citenamefont {Chen}}]{Johnston2023cavitycoupled}%
  \BibitemOpen
  \bibfield  {author} {\bibinfo {author} {\bibfnamefont {A.}~\bibnamefont {Johnston}}, \bibinfo {author} {\bibfnamefont {U.}~\bibnamefont {Felix-Rendon}}, \bibinfo {author} {\bibfnamefont {Y.-E.}\ \bibnamefont {Wong}},\ and\ \bibinfo {author} {\bibfnamefont {S.}~\bibnamefont {Chen}},\ }\bibfield  {title} {\bibinfo {title} {Cavity-coupled telecom atomic source in silicon},\ }\href@noop {} {\bibfield  {journal} {\bibinfo  {journal} {Nature Communications}\ }\textbf {\bibinfo {volume} {15}},\ \bibinfo {pages} {2350} (\bibinfo {year} {2024})}\BibitemShut {NoStop}%
\bibitem [{\citenamefont {Islam}\ \emph {et~al.}(2023)\citenamefont {Islam}, \citenamefont {Lee}, \citenamefont {Harper}, \citenamefont {Rahaman}, \citenamefont {Zhao}, \citenamefont {Vij},\ and\ \citenamefont {Waks}}]{Islam2023cavityenhanced}%
  \BibitemOpen
  \bibfield  {author} {\bibinfo {author} {\bibfnamefont {F.}~\bibnamefont {Islam}}, \bibinfo {author} {\bibfnamefont {C.-M.}\ \bibnamefont {Lee}}, \bibinfo {author} {\bibfnamefont {S.}~\bibnamefont {Harper}}, \bibinfo {author} {\bibfnamefont {M.~H.}\ \bibnamefont {Rahaman}}, \bibinfo {author} {\bibfnamefont {Y.}~\bibnamefont {Zhao}}, \bibinfo {author} {\bibfnamefont {N.~K.}\ \bibnamefont {Vij}},\ and\ \bibinfo {author} {\bibfnamefont {E.}~\bibnamefont {Waks}},\ }\bibfield  {title} {\bibinfo {title} {Cavity-enhanced emission from a silicon t center},\ }\href@noop {} {\bibfield  {journal} {\bibinfo  {journal} {Nano Letters}\ }\textbf {\bibinfo {volume} {24}},\ \bibinfo {pages} {319} (\bibinfo {year} {2023})}\BibitemShut {NoStop}%
\bibitem [{\citenamefont {Raha}\ \emph {et~al.}(2020)\citenamefont {Raha}, \citenamefont {Chen}, \citenamefont {Phenicie}, \citenamefont {Ourari}, \citenamefont {Dibos},\ and\ \citenamefont {Thompson}}]{Raha2020opticalquantum}%
  \BibitemOpen
  \bibfield  {author} {\bibinfo {author} {\bibfnamefont {M.}~\bibnamefont {Raha}}, \bibinfo {author} {\bibfnamefont {S.}~\bibnamefont {Chen}}, \bibinfo {author} {\bibfnamefont {C.~M.}\ \bibnamefont {Phenicie}}, \bibinfo {author} {\bibfnamefont {S.}~\bibnamefont {Ourari}}, \bibinfo {author} {\bibfnamefont {A.~M.}\ \bibnamefont {Dibos}},\ and\ \bibinfo {author} {\bibfnamefont {J.~D.}\ \bibnamefont {Thompson}},\ }\bibfield  {title} {\bibinfo {title} {{Optical quantum nondemolition measurement of a single rare earth ion qubit}},\ }\href {https://doi.org/10.1038/s41467-020-15138-7} {\bibfield  {journal} {\bibinfo  {journal} {Nature Communications}\ }\textbf {\bibinfo {volume} {11}},\ \bibinfo {pages} {1} (\bibinfo {year} {2020})}\BibitemShut {NoStop}%
\bibitem [{\citenamefont {Schmidgall}\ \emph {et~al.}(2018)\citenamefont {Schmidgall}, \citenamefont {Chakravarthi}, \citenamefont {Gould}, \citenamefont {Christen}, \citenamefont {Hestroffer}, \citenamefont {Hatami},\ and\ \citenamefont {Fu}}]{strak_integrated}%
  \BibitemOpen
  \bibfield  {author} {\bibinfo {author} {\bibfnamefont {E.~R.}\ \bibnamefont {Schmidgall}}, \bibinfo {author} {\bibfnamefont {S.}~\bibnamefont {Chakravarthi}}, \bibinfo {author} {\bibfnamefont {M.}~\bibnamefont {Gould}}, \bibinfo {author} {\bibfnamefont {I.~R.}\ \bibnamefont {Christen}}, \bibinfo {author} {\bibfnamefont {K.}~\bibnamefont {Hestroffer}}, \bibinfo {author} {\bibfnamefont {F.}~\bibnamefont {Hatami}},\ and\ \bibinfo {author} {\bibfnamefont {K.-M.~C.}\ \bibnamefont {Fu}},\ }\bibfield  {title} {\bibinfo {title} {Frequency control of single quantum emitters in integrated photonic circuits},\ }\href {https://doi.org/10.1021/acs.nanolett.7b04717} {\bibfield  {journal} {\bibinfo  {journal} {Nano Letters}\ }\textbf {\bibinfo {volume} {18}},\ \bibinfo {pages} {1175} (\bibinfo {year} {2018})}\BibitemShut {NoStop}%
\bibitem [{\citenamefont {Anderson}\ \emph {et~al.}(2019)\citenamefont {Anderson}, \citenamefont {Bourassa}, \citenamefont {Miao}, \citenamefont {Wolfowicz}, \citenamefont {Mintun}, \citenamefont {Crook}, \citenamefont {Abe}, \citenamefont {Ul~Hassan}, \citenamefont {Son}, \citenamefont {Ohshima} \emph {et~al.}}]{pin_anderson2019electrical}%
  \BibitemOpen
  \bibfield  {author} {\bibinfo {author} {\bibfnamefont {C.~P.}\ \bibnamefont {Anderson}}, \bibinfo {author} {\bibfnamefont {A.}~\bibnamefont {Bourassa}}, \bibinfo {author} {\bibfnamefont {K.~C.}\ \bibnamefont {Miao}}, \bibinfo {author} {\bibfnamefont {G.}~\bibnamefont {Wolfowicz}}, \bibinfo {author} {\bibfnamefont {P.~J.}\ \bibnamefont {Mintun}}, \bibinfo {author} {\bibfnamefont {A.~L.}\ \bibnamefont {Crook}}, \bibinfo {author} {\bibfnamefont {H.}~\bibnamefont {Abe}}, \bibinfo {author} {\bibfnamefont {J.}~\bibnamefont {Ul~Hassan}}, \bibinfo {author} {\bibfnamefont {N.~T.}\ \bibnamefont {Son}}, \bibinfo {author} {\bibfnamefont {T.}~\bibnamefont {Ohshima}}, \emph {et~al.},\ }\bibfield  {title} {\bibinfo {title} {Electrical and optical control of single spins integrated in scalable semiconductor devices},\ }\href@noop {} {\bibfield  {journal} {\bibinfo  {journal} {Science}\ }\textbf {\bibinfo {volume} {366}},\ \bibinfo {pages} {1225} (\bibinfo {year} {2019})}\BibitemShut {NoStop}%
\bibitem [{\citenamefont {Tomm}\ \emph {et~al.}(2021)\citenamefont {Tomm}, \citenamefont {Javadi}, \citenamefont {Antoniadis}, \citenamefont {Najer}, \citenamefont {Löbl}, \citenamefont {Korsch}, \citenamefont {Schott}, \citenamefont {Valentin}, \citenamefont {Wieck},\ and\ \citenamefont {Warburton}}]{QD_pn}%
  \BibitemOpen
  \bibfield  {author} {\bibinfo {author} {\bibfnamefont {N.}~\bibnamefont {Tomm}}, \bibinfo {author} {\bibfnamefont {A.}~\bibnamefont {Javadi}}, \bibinfo {author} {\bibfnamefont {N.~O.}\ \bibnamefont {Antoniadis}}, \bibinfo {author} {\bibfnamefont {D.}~\bibnamefont {Najer}}, \bibinfo {author} {\bibfnamefont {M.~C.}\ \bibnamefont {Löbl}}, \bibinfo {author} {\bibfnamefont {A.~R.}\ \bibnamefont {Korsch}}, \bibinfo {author} {\bibfnamefont {R.}~\bibnamefont {Schott}}, \bibinfo {author} {\bibfnamefont {S.~R.}\ \bibnamefont {Valentin}}, \bibinfo {author} {\bibfnamefont {A.~D.}\ \bibnamefont {Wieck}},\ and\ \bibinfo {author} {\bibfnamefont {A.~L. . R.~J.}\ \bibnamefont {Warburton}},\ }\bibfield  {title} {\bibinfo {title} {A bright and fast source of coherent single photons},\ }\href@noop {} {\bibfield  {journal} {\bibinfo  {journal} {Nat. Nanotechnol}\ }\textbf {\bibinfo {volume} {16}},\ \bibinfo {pages} {399} (\bibinfo {year} {2021})}\BibitemShut {NoStop}%
\bibitem [{\citenamefont {Hall}(1967)}]{Hall_1967}%
  \BibitemOpen
  \bibfield  {author} {\bibinfo {author} {\bibfnamefont {J.~J.}\ \bibnamefont {Hall}},\ }\bibfield  {title} {\bibinfo {title} {Electronic effects in the elastic constants of $n$-type silicon},\ }\href {https://doi.org/10.1103/PhysRev.161.756} {\bibfield  {journal} {\bibinfo  {journal} {Phys. Rev.}\ }\textbf {\bibinfo {volume} {161}},\ \bibinfo {pages} {756} (\bibinfo {year} {1967})}\BibitemShut {NoStop}%
\bibitem [{\citenamefont {Kresse}\ and\ \citenamefont {Furthm\"uller}(1996)}]{G.Kresse-PRB96}%
  \BibitemOpen
  \bibfield  {author} {\bibinfo {author} {\bibfnamefont {G.}~\bibnamefont {Kresse}}\ and\ \bibinfo {author} {\bibfnamefont {J.}~\bibnamefont {Furthm\"uller}},\ }\bibfield  {title} {\bibinfo {title} {Efficient iterative schemes for ab initio total-energy calculations using a plane-wave basis set},\ }\href {https://doi.org/10.1103/PhysRevB.54.11169} {\bibfield  {journal} {\bibinfo  {journal} {Phys. Rev. B}\ }\textbf {\bibinfo {volume} {54}},\ \bibinfo {pages} {11169} (\bibinfo {year} {1996})}\BibitemShut {NoStop}%
\bibitem [{\citenamefont {Kresse}\ and\ \citenamefont {Furthm{\"u}ller}(1996)}]{G.Kresse-CMS96}%
  \BibitemOpen
  \bibfield  {author} {\bibinfo {author} {\bibfnamefont {G.}~\bibnamefont {Kresse}}\ and\ \bibinfo {author} {\bibfnamefont {J.}~\bibnamefont {Furthm{\"u}ller}},\ }\bibfield  {title} {\bibinfo {title} {Efficiency of ab-initio total energy calculations for metals and semiconductors using a plane-wave basis set},\ }\href {https://doi.org/https://doi.org/10.1016/0927-0256(96)00008-0} {\bibfield  {journal} {\bibinfo  {journal} {Computational Materials Science}\ }\textbf {\bibinfo {volume} {6}},\ \bibinfo {pages} {15} (\bibinfo {year} {1996})}\BibitemShut {NoStop}%
\bibitem [{\citenamefont {Bl\"ochl}(1994)}]{P.E.Blochl-PRB94}%
  \BibitemOpen
  \bibfield  {author} {\bibinfo {author} {\bibfnamefont {P.~E.}\ \bibnamefont {Bl\"ochl}},\ }\bibfield  {title} {\bibinfo {title} {Projector augmented-wave method},\ }\href {https://doi.org/10.1103/PhysRevB.50.17953} {\bibfield  {journal} {\bibinfo  {journal} {Phys. Rev. B}\ }\textbf {\bibinfo {volume} {50}},\ \bibinfo {pages} {17953} (\bibinfo {year} {1994})}\BibitemShut {NoStop}%
\bibitem [{\citenamefont {Perdew}\ \emph {et~al.}(1996)\citenamefont {Perdew}, \citenamefont {Burke},\ and\ \citenamefont {Ernzerhof}}]{J.Perdew-PRL96}%
  \BibitemOpen
  \bibfield  {author} {\bibinfo {author} {\bibfnamefont {J.~P.}\ \bibnamefont {Perdew}}, \bibinfo {author} {\bibfnamefont {K.}~\bibnamefont {Burke}},\ and\ \bibinfo {author} {\bibfnamefont {M.}~\bibnamefont {Ernzerhof}},\ }\bibfield  {title} {\bibinfo {title} {Generalized gradient approximation made simple},\ }\href {https://doi.org/10.1103/PhysRevLett.77.3865} {\bibfield  {journal} {\bibinfo  {journal} {Phys. Rev. Lett.}\ }\textbf {\bibinfo {volume} {77}},\ \bibinfo {pages} {3865} (\bibinfo {year} {1996})}\BibitemShut {NoStop}%
\bibitem [{\citenamefont {Heyd}\ \emph {et~al.}(2003)\citenamefont {Heyd}, \citenamefont {Scuseria},\ and\ \citenamefont {Ernzerhof}}]{Heyd2003}%
  \BibitemOpen
  \bibfield  {author} {\bibinfo {author} {\bibfnamefont {J.}~\bibnamefont {Heyd}}, \bibinfo {author} {\bibfnamefont {G.~E.}\ \bibnamefont {Scuseria}},\ and\ \bibinfo {author} {\bibfnamefont {M.}~\bibnamefont {Ernzerhof}},\ }\bibfield  {title} {\bibinfo {title} {Hybrid functionals based on a screened coulomb potential},\ }\href {https://doi.org/10.1063/1.1564060} {\bibfield  {journal} {\bibinfo  {journal} {The Journal of Chemical Physics}\ }\textbf {\bibinfo {volume} {118}},\ \bibinfo {pages} {8207} (\bibinfo {year} {2003})}\BibitemShut {NoStop}%
\bibitem [{\citenamefont {Jackson}(1975)}]{Jackson_1975}%
  \BibitemOpen
  \bibfield  {author} {\bibinfo {author} {\bibfnamefont {J.~D.}\ \bibnamefont {Jackson}},\ }\href@noop {} {\emph {\bibinfo {title} {Classical electrodynamics}}},\ \bibinfo {edition} {2nd}\ ed.\ (\bibinfo  {publisher} {J. Wiley},\ \bibinfo {year} {1975})\BibitemShut {NoStop}%
\bibitem [{\citenamefont {Frisch}\ \emph {et~al.}(2016)\citenamefont {Frisch}, \citenamefont {Trucks}, \citenamefont {Schlegel}, \citenamefont {Scuseria}, \citenamefont {Robb}, \citenamefont {Cheeseman}, \citenamefont {Scalmani}, \citenamefont {Barone}, \citenamefont {Petersson}, \citenamefont {Nakatsuji}, \citenamefont {Li}, \citenamefont {Caricato}, \citenamefont {Marenich}, \citenamefont {Bloino}, \citenamefont {Janesko}, \citenamefont {Gomperts}, \citenamefont {Mennucci}, \citenamefont {Hratchian}, \citenamefont {Ortiz}, \citenamefont {Izmaylov}, \citenamefont {Sonnenberg}, \citenamefont {Williams-Young}, \citenamefont {Ding}, \citenamefont {Lipparini}, \citenamefont {Egidi}, \citenamefont {Goings}, \citenamefont {Peng}, \citenamefont {Petrone}, \citenamefont {Henderson}, \citenamefont {Ranasinghe}, \citenamefont {Zakrzewski}, \citenamefont {Gao}, \citenamefont {Rega}, \citenamefont {Zheng}, \citenamefont {Liang}, \citenamefont {Hada}, \citenamefont {Ehara}, \citenamefont {Toyota}, \citenamefont {Fukuda},
  \citenamefont {Hasegawa}, \citenamefont {Ishida}, \citenamefont {Nakajima}, \citenamefont {Honda}, \citenamefont {Kitao}, \citenamefont {Nakai}, \citenamefont {Vreven}, \citenamefont {Throssell}, \citenamefont {Montgomery}, \citenamefont {Peralta}, \citenamefont {Ogliaro}, \citenamefont {Bearpark}, \citenamefont {Heyd}, \citenamefont {Brothers}, \citenamefont {Kudin}, \citenamefont {Staroverov}, \citenamefont {Keith}, \citenamefont {Kobayashi}, \citenamefont {Normand}, \citenamefont {Raghavachari}, \citenamefont {Rendell}, \citenamefont {Burant}, \citenamefont {Iyengar}, \citenamefont {Tomasi}, \citenamefont {Cossi}, \citenamefont {Millam}, \citenamefont {Klene}, \citenamefont {Adamo}, \citenamefont {Cammi}, \citenamefont {Ochterski}, \citenamefont {Martin}, \citenamefont {Morokuma}, \citenamefont {Farkas}, \citenamefont {Foresman},\ and\ \citenamefont {Fox}}]{Gaussian16}%
  \BibitemOpen
  \bibfield  {author} {\bibinfo {author} {\bibfnamefont {M.~J.}\ \bibnamefont {Frisch}}, \bibinfo {author} {\bibfnamefont {G.~W.}\ \bibnamefont {Trucks}}, \bibinfo {author} {\bibfnamefont {H.~B.}\ \bibnamefont {Schlegel}}, \bibinfo {author} {\bibfnamefont {G.~E.}\ \bibnamefont {Scuseria}}, \bibinfo {author} {\bibfnamefont {M.~A.}\ \bibnamefont {Robb}}, \bibinfo {author} {\bibfnamefont {J.~R.}\ \bibnamefont {Cheeseman}}, \bibinfo {author} {\bibfnamefont {G.}~\bibnamefont {Scalmani}}, \bibinfo {author} {\bibfnamefont {V.}~\bibnamefont {Barone}}, \bibinfo {author} {\bibfnamefont {G.~A.}\ \bibnamefont {Petersson}}, \bibinfo {author} {\bibfnamefont {H.}~\bibnamefont {Nakatsuji}}, \bibinfo {author} {\bibfnamefont {X.}~\bibnamefont {Li}}, \bibinfo {author} {\bibfnamefont {M.}~\bibnamefont {Caricato}}, \bibinfo {author} {\bibfnamefont {A.~V.}\ \bibnamefont {Marenich}}, \bibinfo {author} {\bibfnamefont {J.}~\bibnamefont {Bloino}}, \bibinfo {author} {\bibfnamefont {B.~G.}\ \bibnamefont {Janesko}}, \bibinfo {author}
  {\bibfnamefont {R.}~\bibnamefont {Gomperts}}, \bibinfo {author} {\bibfnamefont {B.}~\bibnamefont {Mennucci}}, \bibinfo {author} {\bibfnamefont {H.~P.}\ \bibnamefont {Hratchian}}, \bibinfo {author} {\bibfnamefont {J.~V.}\ \bibnamefont {Ortiz}}, \bibinfo {author} {\bibfnamefont {A.~F.}\ \bibnamefont {Izmaylov}}, \bibinfo {author} {\bibfnamefont {J.~L.}\ \bibnamefont {Sonnenberg}}, \bibinfo {author} {\bibfnamefont {D.}~\bibnamefont {Williams-Young}}, \bibinfo {author} {\bibfnamefont {F.}~\bibnamefont {Ding}}, \bibinfo {author} {\bibfnamefont {F.}~\bibnamefont {Lipparini}}, \bibinfo {author} {\bibfnamefont {F.}~\bibnamefont {Egidi}}, \bibinfo {author} {\bibfnamefont {J.}~\bibnamefont {Goings}}, \bibinfo {author} {\bibfnamefont {B.}~\bibnamefont {Peng}}, \bibinfo {author} {\bibfnamefont {A.}~\bibnamefont {Petrone}}, \bibinfo {author} {\bibfnamefont {T.}~\bibnamefont {Henderson}}, \bibinfo {author} {\bibfnamefont {D.}~\bibnamefont {Ranasinghe}}, \bibinfo {author} {\bibfnamefont {V.~G.}\ \bibnamefont
  {Zakrzewski}}, \bibinfo {author} {\bibfnamefont {J.}~\bibnamefont {Gao}}, \bibinfo {author} {\bibfnamefont {N.}~\bibnamefont {Rega}}, \bibinfo {author} {\bibfnamefont {G.}~\bibnamefont {Zheng}}, \bibinfo {author} {\bibfnamefont {W.}~\bibnamefont {Liang}}, \bibinfo {author} {\bibfnamefont {M.}~\bibnamefont {Hada}}, \bibinfo {author} {\bibfnamefont {M.}~\bibnamefont {Ehara}}, \bibinfo {author} {\bibfnamefont {K.}~\bibnamefont {Toyota}}, \bibinfo {author} {\bibfnamefont {R.}~\bibnamefont {Fukuda}}, \bibinfo {author} {\bibfnamefont {J.}~\bibnamefont {Hasegawa}}, \bibinfo {author} {\bibfnamefont {M.}~\bibnamefont {Ishida}}, \bibinfo {author} {\bibfnamefont {T.}~\bibnamefont {Nakajima}}, \bibinfo {author} {\bibfnamefont {Y.}~\bibnamefont {Honda}}, \bibinfo {author} {\bibfnamefont {O.}~\bibnamefont {Kitao}}, \bibinfo {author} {\bibfnamefont {H.}~\bibnamefont {Nakai}}, \bibinfo {author} {\bibfnamefont {T.}~\bibnamefont {Vreven}}, \bibinfo {author} {\bibfnamefont {K.}~\bibnamefont {Throssell}}, \bibinfo {author}
  {\bibfnamefont {J.~A.}\ \bibnamefont {Montgomery}, \bibfnamefont {{Jr.}}}, \bibinfo {author} {\bibfnamefont {J.~E.}\ \bibnamefont {Peralta}}, \bibinfo {author} {\bibfnamefont {F.}~\bibnamefont {Ogliaro}}, \bibinfo {author} {\bibfnamefont {M.~J.}\ \bibnamefont {Bearpark}}, \bibinfo {author} {\bibfnamefont {J.~J.}\ \bibnamefont {Heyd}}, \bibinfo {author} {\bibfnamefont {E.~N.}\ \bibnamefont {Brothers}}, \bibinfo {author} {\bibfnamefont {K.~N.}\ \bibnamefont {Kudin}}, \bibinfo {author} {\bibfnamefont {V.~N.}\ \bibnamefont {Staroverov}}, \bibinfo {author} {\bibfnamefont {T.~A.}\ \bibnamefont {Keith}}, \bibinfo {author} {\bibfnamefont {R.}~\bibnamefont {Kobayashi}}, \bibinfo {author} {\bibfnamefont {J.}~\bibnamefont {Normand}}, \bibinfo {author} {\bibfnamefont {K.}~\bibnamefont {Raghavachari}}, \bibinfo {author} {\bibfnamefont {A.~P.}\ \bibnamefont {Rendell}}, \bibinfo {author} {\bibfnamefont {J.~C.}\ \bibnamefont {Burant}}, \bibinfo {author} {\bibfnamefont {S.~S.}\ \bibnamefont {Iyengar}}, \bibinfo {author}
  {\bibfnamefont {J.}~\bibnamefont {Tomasi}}, \bibinfo {author} {\bibfnamefont {M.}~\bibnamefont {Cossi}}, \bibinfo {author} {\bibfnamefont {J.~M.}\ \bibnamefont {Millam}}, \bibinfo {author} {\bibfnamefont {M.}~\bibnamefont {Klene}}, \bibinfo {author} {\bibfnamefont {C.}~\bibnamefont {Adamo}}, \bibinfo {author} {\bibfnamefont {R.}~\bibnamefont {Cammi}}, \bibinfo {author} {\bibfnamefont {J.~W.}\ \bibnamefont {Ochterski}}, \bibinfo {author} {\bibfnamefont {R.~L.}\ \bibnamefont {Martin}}, \bibinfo {author} {\bibfnamefont {K.}~\bibnamefont {Morokuma}}, \bibinfo {author} {\bibfnamefont {O.}~\bibnamefont {Farkas}}, \bibinfo {author} {\bibfnamefont {J.~B.}\ \bibnamefont {Foresman}},\ and\ \bibinfo {author} {\bibfnamefont {D.~J.}\ \bibnamefont {Fox}},\ }\href@noop {} {\bibinfo {title} {Gaussian˜16 {R}evision {C}.01}} (\bibinfo {year} {2016}),\ \bibinfo {note} {gaussian Inc. Wallingford CT}\BibitemShut {NoStop}%
\end{thebibliography}%
\end{document}